\documentclass[twocolumn,prb,tightenlines,superscriptaddress,amsmath,amssymb,amsfonts,noeprint]{revtex4-2}
\usepackage{graphicx} 
\usepackage{dcolumn} 
\usepackage{bm} 
\usepackage{float}
\usepackage{hyperref} 
\usepackage[section]{placeins}
\usepackage{epstopdf} 
\usepackage[table, dvipsnames]{xcolor} 

\usepackage{times}
\usepackage[OT1]{fontenc} 

\newcommand{\ie}{\textit{i}.\textit{e}.}

\DeclareUnicodeCharacter{2212}{-}

\begin{document}

\title{Quantum-to-classical crossover in single-electron emitter}
\author{Y. Yin}
\thanks{Author to whom correspondence should be addressed}
\email{yin80@scu.edu.cn.}
\affiliation{Department of Physics,
  Sichuan University, Chengdu, Sichuan, 610065, China}
\date{\today}

\begin{abstract}
  We investigate the temperature-driven quantum-to-classical crossover in a single-electron
  emitter. The emitter is composed of a quantum conductor and an electrode, which is coupled via an
  Ohmic contact. At zero temperature, it has been shown that a single electron can be injected
  coherently by applying an unit-charge Lorentzian pulse on the electrode. As the electrode
  temperature increases, we show that the electron emission approaches a time-dependence Poisson
  process at long times. The Poissonian character is demonstrated from the time-resolved full
  counting statistics. In the meantime, we show that the emission events remain correlated, which is
  due to the Pauli exclusion principle. The correlation is revealed from the emission rates of
  individual electrons, from which a characteristic correlation time can be extracted. The
  correlation time drops rapidly as the electrode temperature increases, indicating that correlation
  can only play a non-negligible role at short times in the high-temperature limit. By using the
  same procedure, we further show that the quantum-to-classical crossover exhibits similar features
  when the emission is driven by a Lorentzian pulse carrying two electron charge. Our results show
  how the electron emission process is affected by thermal fluctuations in a single-electron
  emitter.
\end{abstract}


\maketitle

\section{Introduction}
\label{sec1}

The on-demand control of coherent electron emission in mesoscopic conductors has attracted much
interest in recent decades \cite{glattli-2016-levit-elect, baeuerle-2018-coher-contr}. In a simple
setup, the electrons are emitted from an electrode into a conductor though an Ohmic contact, which
are driven by voltage pulses $V(t)$. For a single-channel quantum conductor, the current $I(t)$
follows instantly to $V(t)$ as $I(t) = e^2 V(t)/h$, with $e$ being the electron charge and $h$ being
the Planck constant. However, the details of the emission process can be quite different in quantum
and classical limits.

In the quantum limit, the current is due to the coherent emission of electrons or holes, which are
usually accompanied by neutral electron-hole pairs. Their wave functions are well-defined, which can
be extracted by quantum tomography methods \cite{samuelsson-2006-quant-state,
  grenier-2011-singl-elect, jullien-2014-quant-tomog-elect, bisognin-2019-quant-tomog,
  roussel21_proces_quant_signal_carried_by_elect_curren}. In particular, by using an unit-charge
Lorentzian pulse $V(t)$ with time width $W$, \ie, $V(t) = h W/[\pi e(W^2 + t^2)]$, a single electron
can be emitted on top of the Fermi sea without accompanied electron-hole pairs
\cite{keeling-2006-minim-excit, dubois-2013-minim-excit}. It has been called a leviton, whose wave
function can be given as $\psi_L(t) = \sqrt{W/(2 \pi)} /(t + i W/2)$. The emission probability
density of the leviton follows instantly to the voltage pulse as
$\left| \psi_L(t) \right|^2 = I(t)/e= eV(t)/h$, indicating an excellent synchronization between the
electron emission and driving voltage. In contrast, the current corresponds to the incoherent
emission of electrons and/or holes in the classical limit. This typically occurs at high
temperatures, when thermal fluctuations in the electrode can play an important role. The emission
events can be treated as random and independent, which follow Poisson statistics. In this limit, the
voltage pulse does not control the emission probability density of individual electrons. Instead, it
decides the overall emission rate of the emission process.

One expects that the quantum to classical crossover occurs as the temperature of the electrode
increases. Indeed, the many-body quantum state of the emitted electrons can evolve from a pure state
into a mixed one due to thermal fluctuations \cite{moskalets16_singl_elect_coher,
  moskalets18_high_temper_fusion_multiel_levit}. This can lead to a reduction of the dc shot noise
\cite{glattli16_hanbur_brown_twiss_noise_correl, bocquillon12_elect_quant_optic,
  dubois-2013-integ-fract}. But the electron anti-bunching is robust against temperature, which has
been revealed from the Hong-Ou-Mandel interference
\cite{glattli16_hanbur_brown_twiss_noise_correl}. This indicates that the quantum coherence between
different electrons is still preserved and hence the temperature-induced quantum-to-classical
crossover is incomplete. In the case of dc driving when $V(t) = V_0$, this has been clearly
demonstrated from the waiting time distribution $\mathcal{W}(\tau)$ (WTD), which gives the
distribution of time delays $\tau$ between successive electrons
\cite{brandes08_waitin_times_noise_singl_partic_trans, albert-2012-elect-waitin,
  haack14_distr_elect_waitin_times_quant_coher_conduc,
  albert14_waitin_time_distr_train_quant_elect_pulses,
  dasenbrook14_floquet_theor_elect_waitin_times}. At zero temperature, $\mathcal{W}(\tau)$ follows
the Wigner-Dyson distribution, which has a zero dip around $\tau = 0$ and a Gaussian tail as
$\tau \to +\infty$ \cite{albert-2012-elect-waitin}. As the electrode temperature $T$ increases, the
tail approaches an exponential distribution when $T$ is comparable to $e V_0/k_B$, with $k_B$ being
the Boltzmann constant. In contrast, the essential shape of $\mathcal{W}(\tau)$ remains unchanged,
especially around the zero dip \cite{haack14_distr_elect_waitin_times_quant_coher_conduc}. This
implies that the emission approaches a Poisson process at long times, but the emission events are
still correlated at short times.

However, the crossover can have a different nature when the emission is driving by the Lorentzian
pulse. First, the pulse width $W$ provides an important time scale in this case. In fact, it decides
the crossover temperature, which separates the high- and low-temperature regions for the dc shot
noise \cite{moskalets16_singl_elect_coher}. Secondly, the full counting statistics (FCS) changes
drastically as the temperature increases: In the classical limit, the emission tends to follow the
Poisson statistics. While in the quantum limit, the voltage pulse injects exactly one electron into
the quantum conductor. This make the WTD less suitable for the study of emission in this limit.
Finally, as the emission is triggered by a time-dependent voltage, it is helpful to elucidate the
relation between the electron emission and the driving voltage, which is also absent from the WTD.

\begin{figure}
  \centering
  \includegraphics[width=7.5cm]{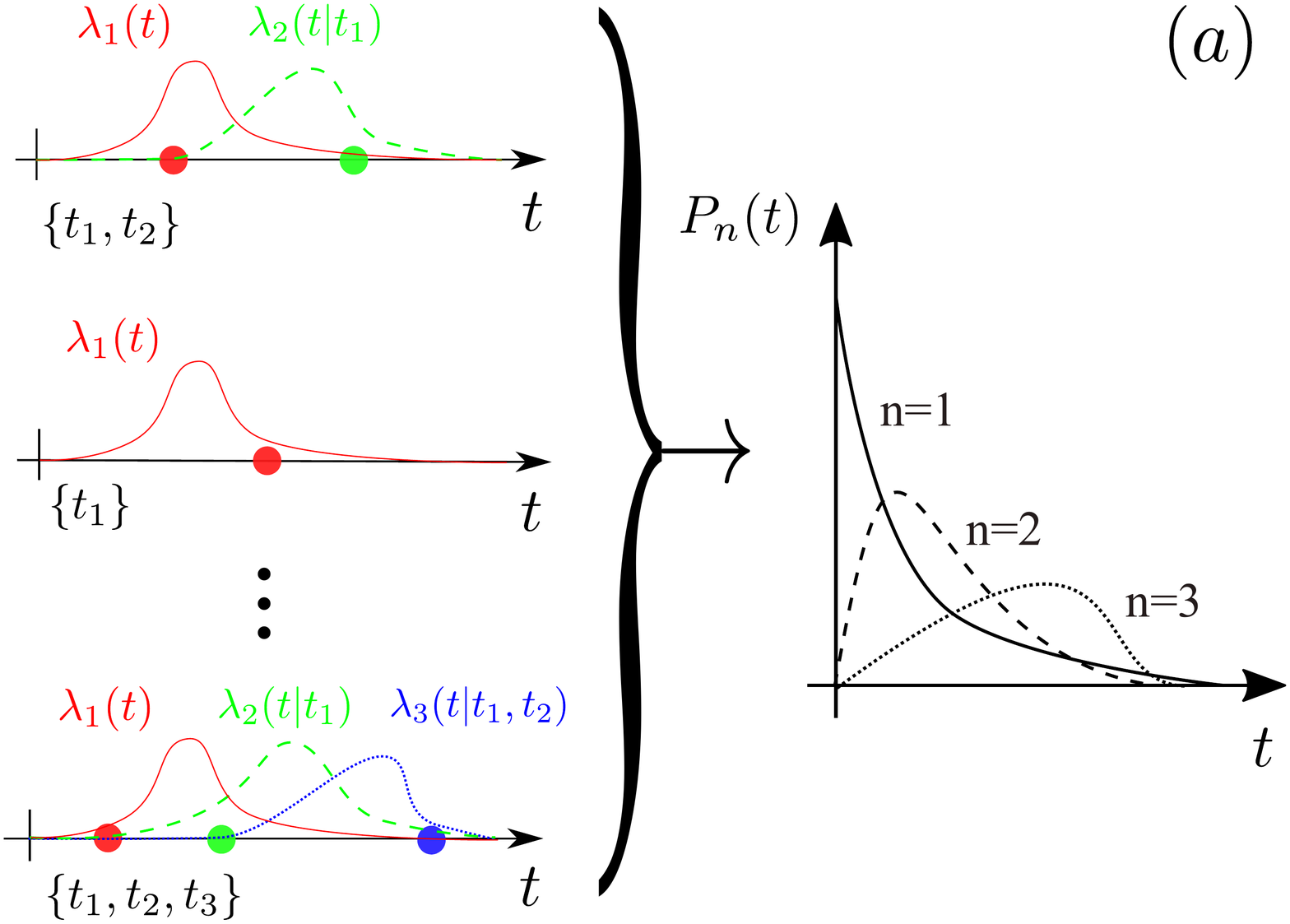}
  \includegraphics[width=8.5cm]{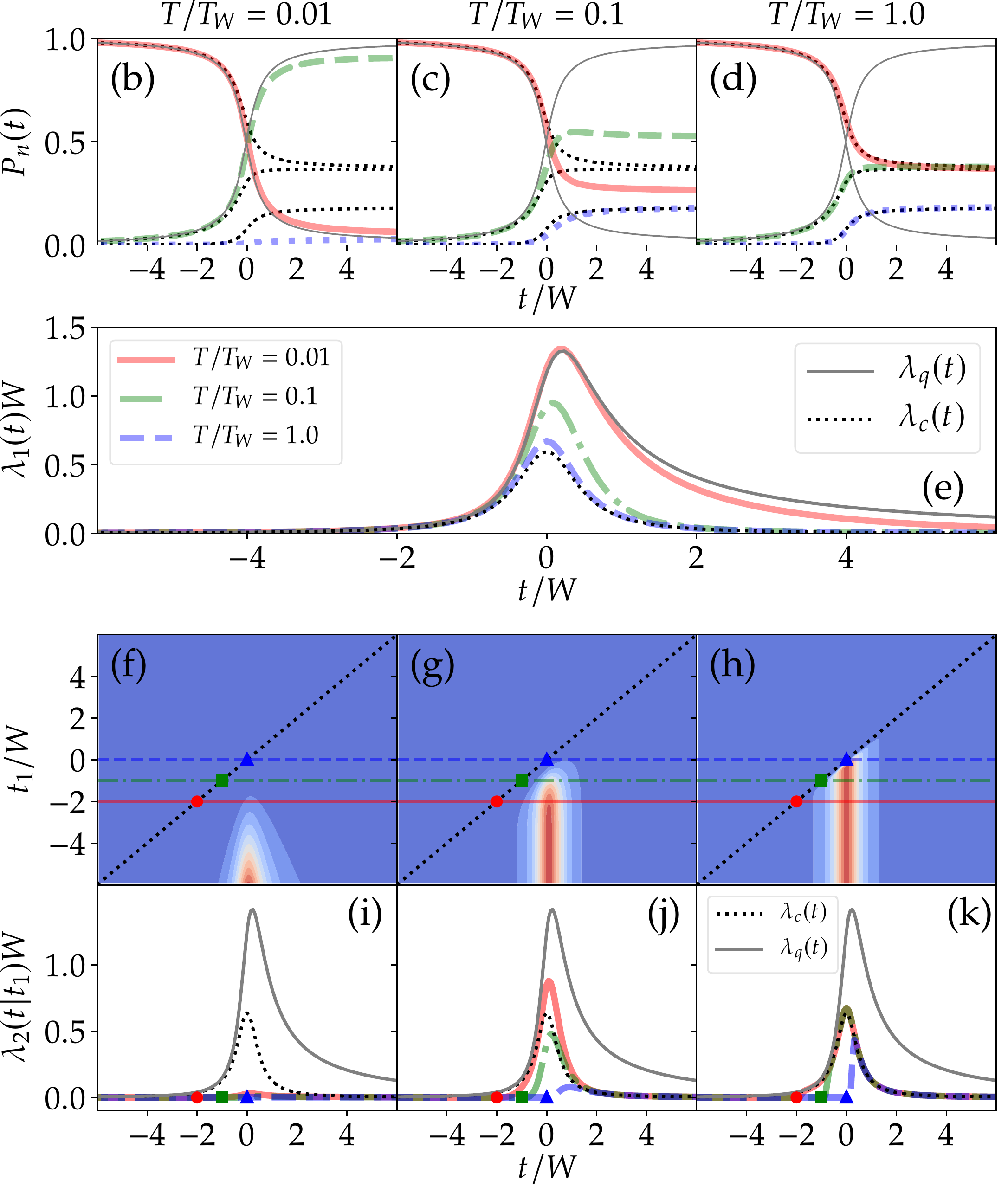}
  \caption{ (a) Illustration for the real-time electron counting measurements. A single measurement
    is represented by a sequence of random points in the time trace. The $k$-th point in each time
    trace represents the emission time of the $k$-th electron. The emission of the $k$-th electron
    is governed by the corresponding emission rate $\lambda_k(t|t_1, \dots, t_{k-1})$. In the
    figure, we mark the emission of the first, second and third electrons by the red, green and blue
    dots, which correspond to the emission rate $\lambda_1(t)$, $\lambda_2(t|t_1)$ and
    $\lambda_3(t|t_1, t_2)$, respectively. The time-resolved FCS $P_n(t)$ is obtained from the
    statistics gathered from a large number of measurements. (b-d) The time-resolved FCS $P_n(t)$ at
    different temperatures $T/T_W = 0.01$ (b), $T/T_W = 0.1$ (c) and $T/T_W = 1.0$ (d). The red
    solid, green dashed and blue dotted curves correspond to $n=0$, $1$ and $2$, respectively. The
    thin gray solid curves correspond to the quantum limit from Eq.~\eqref{s1:eq1}. The thin black
    dotted curves represent the classical limit from Eq.~\eqref{s1:eq2}. (e) The emission rate
    $\lambda_1(t)$ of the first electron at different temperatures. The gray solid and black dotted
    curves represent the quantum and classical limit for the emission rate. See text for
    details. (f-h) The emission rate $\lambda_2(t|t_1)$ of the second electron at different
    temperatures $T/T_W = 0.01$ (f), $T/T_W = 0.1$ (g) and $T/T_W = 1.0$ (h). They are shown as
    contour plots in the $t$-$t_1$ plane. (i-j) The emission rate $\lambda_2(t|t_1)$ as a function
    of $t$ for three typical $t_1$ at different temperatures. The red solid, green dash-dotted and
    blue dashed curves correspond to $t_1/W = -2.0$, $-1.0$ and $0.0$, while the red dots, green
    squares and blue triangles represent the point $t=t_1$, respectively. The classical and quantum
    limit of the emission rates $\lambda_q(t)$ and $\lambda_c(t)$ are also plotted by the gray solid
    and black dotted curves for comparison.}
  \label{fig1}
\end{figure}

To solve this problem, in this paper we discuss the quantum-to-classical crossover by combining the
time-resolved full counting statistics and electron emission rates. The time-resolved FCS $P_n(t)$
gives the probability of emitting $n$ electrons up to a given time $t$, which can be extracted from
the statistics gathered on a large number of measurements
\cite{levitov96_elect_count_statis_coher_states_elect_curren}. Each single measurement is
represented by a random time sequence $\{t_1, t_2, \dots, t_k, \dots t_n\}$, where each $t_k$ stands
for the emission time of the $k$-th electron [See Fig.~\ref{fig1}(a) for illustration]. The emission
of the $k$-th electron is governed by the corresponding emission rate
$\lambda_k(t|t_1, t_2, \dots, t_{k-1})$, which describes the expected number of electrons emitted in
a given infinitesimal time interval $[t, t+dt]$. As the emission events are generally correlated in
quantum conductors \cite{dasenbrook15_elect_waitin_times_coher_conduc_are_correl}, the emission rate
depends not only on the time $t$, but also on the emission times $\{t_1, t_2, \dots, t_{k-1}\}$ of
all previously emitted electrons.

The typical temperature-dependence of $P_n(t)$ is illustrated in Fig.~\ref{fig1}(b-d). In the
quantum limit when the temperature $T$ is very low, $P_n(t)$ can be well-approximated from the wave
function of leviton as
\begin{eqnarray}
        P_n(t) & = & \begin{cases}
                 1 - \int^t_{-\infty} d\tau \left| \psi_L(t) \right|^2,       & n=0 \\
                 \int^t_{-\infty} d\tau \left| \psi_L(t) \right|^2,       & n=1 \\
                 0, & n \ge 2
               \end{cases}
                     \label{s1:eq1}.
\end{eqnarray}
The quantum limit of $P_n(t)$ is shown by the gray solid curves in the figure. At high temperatures,
$P_n(t)$ tends to follow the classical limit, which corresponds to the time-dependent Poisson
distribution
\begin{equation}
  P_n(t) = \frac{ \left[ \int^t_{-\infty} d\tau \lambda_c(\tau) \right]^n }{n!} \exp{\left[ -\int^t_{-\infty} d\tau \lambda_c(\tau) \right]}
  \label{s1:eq2},
\end{equation}
where the emission rate $\lambda_c(t)$ is decided by the driving pulse as $\lambda_c(t) = e
V(t)/h$. The classical limit of $P_n(t)$ is plotted by the black dotted curves. The figure shows
that $P_n(t)$ evolves from the quantum limit to the classical limit as the temperatures $T$
approaches $T_W = \hbar/(k_B W)$, which providing a clear signature of the quantum-to-classical
crossover.

The behavior of $P_n(t)$ suggests that the emission rates of all the electrons should approach the
classical limit $\lambda_c(t)$ at high temperatures. We find that this only holds for the emission
rate of the first electron $\lambda_1(t)$, which is illustrated in Fig.~\ref{fig1}(e). The figure
shows that $\lambda_1(t)$ evolves from the quantum limit
$\lambda_q(t) = V(t)/[ 1 - \int^t_{-\infty} d\tau V(\tau) ]$ (gray solid curve) to the classical
limit $\lambda_c(t)$ (black dotted curve) as the temperature $T$ approaches $T_W$. However, other
emission rates do not fully follow this behavior. This can be seen from the emission rate of the
second electron $\lambda_2(t|t_1)$, which are displayed as contour plots in the $t$-$t_1$ plane
[Fig.~\ref{fig1}(f-h)]. As the emission of the second electron is the coeffect of the driving pulse
and thermal fluctuations, it remains rather small at low temperatures [Fig.~\ref{fig1}(f)]. At
moderate temperatures [Fig.~\ref{fig1}(g)], it depends significantly on both $t$ and $t_1$,
indicating strong correlations between the emission of the first and second electrons. In
particular, $\lambda_2(t|t_1)$ always drops to zero as $t$ approaches $t_1$. This is better
demonstrated in Fig.~\ref{fig1}(j), where the points $t=t_1$ for three typical $t_1$ are marked by
the red dots, blue triangles and green squares. This indicates that the emission of the second
electron is coupled to the first one due to the Pauli exclusion principle. At high temperatures
[Fig.~\ref{fig1}(h)], the correlations remain pronounced, but can only be seen at short times. We
find that $\lambda_2(t|t_1)$ tends to follow $\Theta(t-t_1)\lambda_c(t)$ in the high temperature
limit. This can be better seen from Fig.~\ref{fig1}(k), where $\lambda_c(t)$ is plotted by the black
dotted curve for comparison. These behaviors show that, as the electrode temperature increases, the
electron emission approaches a time-dependent Poisson process at long times, but the correlations
between the emission events are always preserved at short times. By using a similar procedure, we
find that similar behaviors can also be seen when the emission is driven by a Lorentzian pulse
carrying two electron charge.
 
This paper is organized as follows. In Sec.~\ref{sec2}, we introduce a procedure to calculate the
time-resolved statistics and emission rates for the electron emission in quantum conductors. In
Sec.~\ref{sec3}, we demonstrate the procedure by studying the single and two electron emissions at
zero temperature. The finite temperature effect is discussed in Sec.~\ref{sec4}. We summarize our
results in Sec.~\ref{sec5}.

\section{Time-resolved statistics of a quantum emitter}
\label{sec2}

The statistics of the electron emission process can be described by two equivalently ways. On one
hand, it can be characterized by $n$-fold delayed coincidences, which gives the joint probability
that one electron is emitted in each infinitesimal time interval $[t_i, t_i + dt]$ (with
$i = 1, 2, \dots, n$). It can be expressed as
\begin{eqnarray}
  g_n(t_1,\dots, t_n) & = & \langle a^{\dagger}(t_1) \dots a^{\dagger}(t_n) a(t_n) \dots
                            a(t_1)\rangle \nonumber\\
                      &-& \langle a^{\dagger}(t_1) \dots a^{\dagger}(t_n) a(t_n) \dots a(t_1) \rangle_0,
  \label{s2:eq10}
\end{eqnarray}
where $a^{\dagger}(t)$ and $a(t)$ represent the creation and annihilation operations of electrons,
respectively. Here the two angular brackets $\langle \dots \rangle$ and $\langle \dots \rangle_0$
denote the quantum-statistical average over the many-body electron states with and without the
emitted electrons, respectively. In doing so, one excludes the contribution from the undisturbed
Fermi sea \cite{haack12_glaub_coher_singl_elect_sourc, grenier-2011-singl-elect}. The coincidences
functions are just the diagonal part of the Glauber's correlation functions
\cite{glauber63_coher_incoh_states_radiat_field, glauber63_photon_correl,
  kelley64_theor_elect_field_measur_photoel_count, cahill99_densit_operat_fermion,
  glauber06_quant_theor_optic_coher, glauber06_nobel_lectur}, which can be extracted from the
current correlation measurements \cite{samuelsson-2006-quant-state,
  mahe10_curren_correl_deman_elect_sourc, grenier-2011-singl-elect}.

In contrast, the emission process can also be investigated by real-time electron counting techniques
\cite{gustavsson09_elect_count_quant_dots, maisi11_real_time_obser_discr_andreev_tunnel_event,
  kurzmann19_optic_detec_singl_elect_trans_dynam, ranni21_real_time_obser_cooper_pair,
  brange21_contr_emiss_time_statis_dynam}. This can be used to extract the information of electron
process in single-shot experiments. In this case, the emission process can be described by recording
each individual emission event in a time trace [See Fig.\ref{fig1}(a) for illustration]. This allows
us to represent emission events by random points in a line. One usually further assumes that two
emission events cannot occur exactly at the same time, \ie, there can only exist at most one
emission event in an arbitrary infinitesimal time interval $[t, t+dt]$. This assumption has been
proved to be valid for typical emission processes, such as photon emission in quantum optics and
neuronal spike emission in neuroscience \cite{kelley64_theor_elect_field_measur_photoel_count,
  mandel65_coher_proper_optic_field, macchi75_coinc_approac_to_stoch_point_proces,
  iwankiewicz95_dynam_mechan_system_under_random_impul}.

The emission process can be characterized by the time-resolved statistics of these events. For
example, it can be characterized by the time-dependent counting statistics $P_n(t_s, t_e)$, which
gives the probability to emit $n$ electrons in a given time interval $[t_s, t_e]$. It can also be
described by statistics of times, such as idle time distribution $\Pi_0(t_s, t_e)$, which gives the
probability that no electron is emitted in the time interval $[t_s, t_e]$. The WTD can be obtained
from $\Pi_0(t_s, t_e)$ by its second derivative \cite{albert-2012-elect-waitin}. Note that all these
quantities usually depend on two times $t_s$ and $t_e$, which can be treated as the starting and
ending times of the electron counting measurements.

It is inconvenient to describe the time-resolved statistics directly in terms of the $n$-fold
delayed coincidences $g_n(t_1, \dots, t_n)$. In contrast, the exclusive joint probability density
$f_n(t_1, \dots, t_n ; t_s, t_e )$ (with $n \ge 1$) has been introduced
\cite{kelley64_theor_elect_field_measur_photoel_count, D.J.Daley2003}. It gives the joint
probability that one electron is emitted in each infinitesimal time interval $[t_i, t_i + dt]$ (with
$i = 1, 2, \dots, n$ and all $t_i \in [t_s, t_e]$), while no other electron is emitted in the time
interval $[t_s, t_e]$. It has been shown that $f_n(t_1, \dots, t_k ; t_s, t_e )$ can be related to
$g_n(t_1, \dots, t_n)$ as
\begin{widetext}
  \begin{equation}
    f_n(t_1, \dots, t_n ; t_s, t_e) = \sum_{k=n}^{+\infty}
    \frac{(-1)^{k-n}}{(k-n)!} \int^{t_e}_{t_s} dt_{n+1} ... \int^{t_e}_{t_s} dt_k 
    g_k(t_1, \dots, t_n, t_{n+1}, \dots, t_k).
    \label{s2:eq20}
  \end{equation}
\end{widetext}
This relation is derived from the definition of $f_n(t_1, \dots, t_n ; t_s, t_e )$ and
$g_n(t_1, \dots, t_n)$, which is independent on the nature of the emitted particles
\cite{kelley64_theor_elect_field_measur_photoel_count}. So it can be applied to both Bosons and
Fermions.

All the time-resolved statistics can be obtained from $f_n(t_1, \dots, t_n ; t_s, t_e )$ in a direct
way. In particular, the FCS $P_n(t_s, t_e)$ for $n \ge 1$ can be given as
\begin{equation}
  P_n(t_s, t_e) = \frac{1}{n!} \int^{t_e}_{t_s} dt_1 \dots \int^{t_e}_{t_s} dt_n f_n(t_1, t_2, \dots, t_n ; t_s, t_e).
  \label{s2:eq30a}
\end{equation}
For $n=0$, the corresponding FCS $P_0(t_s, t_e)$, or equivalently the idle time distribution $\Pi_0(t_s, t_e)$, can be
obtained via the normalization relation
\begin{equation}
  \Pi_0(t_s, t_e) = P_0(t_s, t_e) = 1 - \sum^{+\infty}_{n=1} P_n(t_s, t_e).
  \label{s2:eq30b}
\end{equation}

Comparing to the FCS, the exclusive joint probability densities $f_n(t_1, \dots, t_n ; t_s, t_e )$
contain much detailed information on the emission process. In particular, one can extract the
emission rate of each individual electron from them \cite{snyder91_random_point_proces_time_space,
  iwankiewicz95_dynam_mechan_system_under_random_impul, D.J.Daley2003}. This can be better
understood by starting from a stationary Poisson process with emission rate $\lambda_0$. In this
case, the emission events are random and independent. The corresponding coincidence function can be
simply given as $g_k(t_1, \dots, t_n) = \lambda^n_0$. From Eq.~\eqref{s2:eq20}, one finds that
\begin{equation}
  f_n(t_1, \dots, t_n ; t_s, t_e ) = \lambda^n_0 \exp[ - \lambda_0 (t_e - t_s)],
  \label{s2:eq32}
\end{equation}
whose FCS follows the well-known Poisson statistics
\begin{equation}
  P_n(t_s, t_e) = \frac{ \left[\lambda_0 (t_e - t_s)\right]^n}{n!} \exp[ -
    \lambda_0 (t_e - t_s) ].
  \label{s2:eq31}
\end{equation}
The corresponding idle time distribution can be expressed as
$\Pi_0(t_s, t_e) = \exp[ - \lambda_0 (t_e - t_s)]$.

The physical meaning of Eq.~\eqref{s2:eq32} can be better seen by rewriting it in the form
\begin{eqnarray}
   f_n(t_1, \dots, t_n ; t_s, t_e ) & = & \exp[ - \lambda_0 (t_1 - t_s) ]
  \nonumber\\
   &&\times \prod^{n-1}_{i=1} \left\{ \lambda_0 \exp[ - \lambda_0 (t_{i+1}- t_i) ] \right\} \nonumber\\
   &&\times \lambda_0 \exp[ - \lambda_0 (t_e - t_n) ].
  \label{s2:eq33}
\end{eqnarray}
Here each $\lambda_0$ gives the emission probability of an electron in each infinitesimal time
interval $[t_i, t_i+dt]$. The exponential factors are just idle time distributions, they ensure that
no other electron can be emitted between these infinitesimal time intervals. From the above
expression, one finds that the emission rate can be obtained from $f_n(t_1, \dots, t_n ; t_s, t_e )$
as
\begin{equation}
  \lambda_0 = \frac{f_{n+1}(t_1, \dots, t_n, t_{n+1} ; t_s, t_{n+1} )}{f_n(t_1, \dots, t_n ; t_s, t_n )}.
  \label{s2:eq34}
\end{equation}
Alternatively, it can also be related to the idle time distribution as
\begin{equation}
  \lambda_0 = \frac{f_1(t_1 ; t_s, t_1 )}{\Pi_0(t_s, t_1)}.
  \label{s2:eq35}
\end{equation}

In general cases, the emission rate of an electron has a more complicated form. On one hand, it
depends explicitly on the time $t$, which is due to the time-dependence of the driving voltage. On
the other hand, it also depends on the history of the emission process, which can be induced by the
quantum coherence between electrons
\cite{dasenbrook15_elect_waitin_times_coher_conduc_are_correl}. So the emission rates are different
for different electrons. For the $n$-th electron, the corresponding emission rate can be written as
$\lambda_n(t|t_s, t_1, \dots, t_{n-1})$. It is essentially a conditional intensity, which gives the
expected number of electrons emitted in the infinitesimal time interval $[t, t+dt]$, under the
condition that there are $n-1$ electrons emitted previously in each infinitesimal time interval
$[t_i, t_i+dt]$ (with $i = 1, 2, \dots, n-1$). Here the history of the emission process is
represented by the ordered time sequence $t_1, \dots, t_{n-1}$, which satisfies
\begin{equation}
  t_s < t_1 < \dots <t_{n-1} < t.
  \label{s2:eq40}
\end{equation}


Equation~\eqref{s2:eq33} can be generalized to incorporate the history-dependence of the emission
rate. For example, $f_1(t_1 ; t_s, t_e)$ can be expressed in terms of $\lambda_1(t)$ and
$\lambda_2(t|t_1)$, which has the form
\begin{eqnarray}
  \hspace{-0.5cm}f_1(t_1 ; t_s, t_e) & = & \exp\left[- \int^{t_1}_{t_s} d\tau \lambda_1(\tau|t_s)\right] \nonumber\\
  && \times \lambda_1(t_1|t_s) \exp\left[ - \int^{t_e}_{t_1} d\tau \lambda_2(\tau|t_s, t_1) \right]. \label{s2:eq50a-1} 
\end{eqnarray}
The expression looks complicated at the first sight, but it can be understood in the similar way as
Eq.~\eqref{s2:eq33}. It is composed by three factors: 1) The exponential factor
$\exp\left[- \int^{t_1}_{t_s} d\tau \lambda_1(\tau|t_s)\right]$ is just the idle time distribution
$\Pi_0(t_s, t_1)$, it ensures that no electron can be emitted in the time interval $[t_s, t_1]$; 2)
$\lambda(t_1|t_s)$ gives the emission probability of the first electron in the infinitesimal time
interval $[t_1, t_1+dt]$; 3) The exponential factor
$\exp\left[ - \int^{t_e}_{t_1} d\tau \lambda_2(\tau|t_s, t_1) \right]$ can be treated as a
conditional idle time distribution. It guarantees that if the first electron has been emitted at the
time $t_1$, the second electron cannot be emitted before the time $t_e$. From the definition of
$f_1(t_1 ; t_s, t_e)$, one finds that the emission rate $\lambda_1(t|t_s)$ of the first electron can
be given as
\begin{equation}
  \lambda_1( t| t_s)  = \frac{f_1(t;t_s, t)}{\Pi_0(t_s, t)}.
  \label{s2:eq61}
\end{equation}

The exclusive joint probability density for arbitrary $n$ can be constructed in a similar way, which
depends on the emission rates up to the $(n+1)$-th electron. To write it in a more compact form, one
reminds that the emission times $t_i$ follows the relation given in Eq.~\eqref{s2:eq40}. Due to this
restriction, all the emission rates can be composed into a piece-wise function $\lambda^{\ast}(t)$,
which has the form
\begin{eqnarray}
  \hspace{-0.5cm}\lambda^{\ast}(t) & = & \begin{cases}
    \lambda_1(t|t_s),       & t_s <  t  \le t_1 \\
    \lambda_2(t|t_s, t_1),  & t_1 <  t  \le t_2 \\
    \vdots\\
    \lambda_{n+1}(t|t_s, t_1, \dots, t_n), & t_n <  t  \le t_e
  \end{cases}                                                      \label{s2:eq50b}                                                      
\end{eqnarray}
In doing so, $f_n(t_1, \dots, t_n ; t_s, t_e )$ can be written as \cite{D.J.Daley2003}
\begin{equation}
  f_n(t_1, t_2, \dots, t_n ; t_s, t_e) = \prod^n_{i=1} \lambda^{\ast}(t_i) \exp\left[- \int^{t_e}_{t_s} d\tau \lambda^{\ast}(\tau)\right]. \label{s2:eq50a}
\end{equation}
From this expression, one finds that the emission rate for the $n$-th electron ($n \ge 2$) can be
given as
\begin{equation}
  \lambda_n( t| t_s, t_1, \dots, t_{n-1}) = \frac{f_n(t_1, \dots, t_{n-1}, t ; t_s, t)}{f_{n-1}(t_1, \dots, t_{n-2}, t ; t_s, t )}.
  \label{s2:eq60}
\end{equation}

Generally speaking, one can calculate the FCS and emission rates from the coincidences
$g_n(t_1, \dots, t_n)$ by using
Eqs.~\eqref{s2:eq20},~\eqref{s2:eq30a},~\eqref{s2:eq30b},~\eqref{s2:eq61} and ~\eqref{s2:eq60}
. However, the calculation is rather involved in general cases. For non-interacting systems, this
can be greatly simplified, as the system can be fully decided by the corresponding first-order
Glauber correlation functions
$G(t, t') = \langle a^{\dagger}(t) a(t') \rangle - \langle a^{\dagger}(t) a(t') \rangle_0$
\cite{beenakker05_optim_spin_entan_elect_hole_pair_pump,
  cheong04_many_body_densit_matric_free_fermion, corney06_gauss_phase_space_repres_fermion,
  yin-2019-quasip-states, yue21_quasip_states_integ_fract_charg}. In a previous work, Macchi has
shown that $f_n(t_1, \dots, t_n ; t_s, t_e )$ can be calculated from $G(t, t')$ by solving the
eigenvalue equation \cite{macchi75_coinc_approac_to_stoch_point_proces}:
\begin{equation}
  \int^{t_e}_{t_s} dt' G(\tau, \tau') \varphi_{\alpha}(\tau') = \nu_{\alpha}(t_s, t_e) \varphi_{\alpha}(\tau),
  \label{s2:eq70a}
\end{equation}
with $\alpha = 1, 2, \dots$ being the index of the eigenvalues and eigenfunctions. The eigenvalue
$\nu_{\alpha}(t_s, t_e)$ satisfies $ 0 \le \nu_{\alpha} \le 1$, while the eigenfunctions $\varphi_{\alpha}(\tau)$ form
an orthonormal basis within the time interval $[t_s, t_e]$, \ie,
\begin{equation}
  \int^{t_e}_{t_s} d\tau \varphi^{\ast}_{\alpha}(\tau) \varphi_{\alpha'}(\tau) =
  \delta_{\alpha, \alpha'}.
  \label{s2:eq70b}
\end{equation}

The time-resolved FCS can be solely decided from the eigenvalues $\nu_\alpha(t_s, t_e)$, whose momentum generating
function can be given as
\begin{eqnarray}
  \Phi(\chi; t_s, t_e) & = & \sum^{+\infty}_{n=0} P_n(t_s, t_e) e^{i n \chi} \nonumber\\
                       & = & \prod_{\alpha=1}^N \left[1 - \nu_\alpha(t_s, t_e) + e^{i \chi} \nu_\alpha(t_s, t_e) \right].
  \label{s2:eq70c}
\end{eqnarray}
The corresponding idle time distribution can be given following Eqs.~\eqref{s2:eq30a} and
\eqref{s2:eq30b}, which can be written as
\begin{equation}
  \Pi_0(t_s, t_e) = \prod_{\alpha=1}^N \left[1 - \nu_\alpha(t_s, t_e) \right].
  \label{s2:eq70d}
\end{equation}

The exclusive joint probability density $f_n(t_1, \dots, t_n ; t_s, t_e )$ depends on both the eigenvalues and
eigenfunctions, which can be expressed as
\begin{eqnarray}
  && f_n(t_1, t_2, \dots, t_n ; t_s,t_e) = \Pi_0(t_s, t_e) \nonumber\\
  && \times \begin{vmatrix}
    C(t_1, t_1) & C(t_1, t_2) & \dots & C(t_1, t_n) \\
    C(t_2, t_1) & C(t_2, t_2) & \dots & C(t_2, t_n) \\
    \hdotsfor{1}                                    \\
    C(t_n, t_1) & C(t_n, t_2) & \dots & C(t_n, t_n)
    \label{s2:eq70e}
  \end{vmatrix},
\end{eqnarray}
with
\begin{equation}
  C(t, t') = \sum_{\alpha=1}^{+\infty} \frac{\nu_\alpha}{1-\nu_\alpha}
  \varphi_{\alpha}(t) \varphi^{\ast}_{\alpha}(t').
  \label{s2:eq70f}
\end{equation}
The emission rates can then be calculated from Eqs.~\eqref{s2:eq61} and~\eqref{s2:eq60}. This
provides an efficient numerical methods to evaluate the real-time electron statistics in mesoscopic
transports.

Both the time-resolved FCS and emission rates depend on the starting time $t_s$ of the
measurement. To obtain the full information of the emission process, the measurement has to start
early enough, corresponding to the limit $t_s \to -\infty$. In the following discussion, we shall
concentrate on this limit and hence omit $t_s$ in all the expressions for clarification.

\section{Zero temperature}
\label{sec3}

The procedure introduced in the above section allows one to extract both the time-resolved FCS and
emission rates of electrons in a unified manner. In this section, we demonstrate this procedure by
studying the emission of a single and two levitons at zero temperature. In these two cases, both the
FCS and emission rates can be given analytically. It helps one to better understand the relation
between the real-time electron statistics and the wave functions of emitted electrons.

\subsection{Single-leviton emitter}
\label{sec3a}

Let us first consider the simplest case, when a single Lorentzian voltage pulse is applied on the
electrode at zero temperature. In this case, the many-body state of the emitted electron can be
written as
\begin{equation}
  | \Psi \rangle = \psi_1^{\dagger} | F \rangle,
  \label{s3:eq10}
\end{equation}
with $\psi_1^{\dagger} = \int dt \psi_1(t) a^{\dagger}(t)$. Here $a^{\dagger}(t)$ represents the
electron creation operator in the time domain and $\psi_1(t) = \sqrt{W/(2 \pi)} /(t + i W/2)$ is the
wave function of a single leviton.  The corresponding first-order Glauber correlation function can
be given as
\begin{equation}
  G(t, t') = \psi^{\dagger}_1(t) \psi_1(t')
  \label{s3:eq21}.
\end{equation}

This is just the emission of a single leviton. The corresponds emission process is solely determined
by the corresponding wave function $\psi_1(t)$. Indeed, the solution of Eqs.~(\ref{s2:eq70a} -
\ref{s2:eq70f}) is trivial: The only nonzero exclusive joint density can be given as
$f_1(t_1; t) = | \psi_1(t_1) |^2$. The corresponding FCS can be described by the momentum generating
function:
\begin{equation}
  \Phi(\chi; t) = P_0(t) + P_1(t) e^{i \chi} = 1 - \nu_1(t) + e^{i \chi} \nu_1(t). 
  \label{s3:eq40}
\end{equation}
where $\nu_1(t) = \int^{t}_{-\infty} d\tau |\psi_1(\tau)|^2$ plays the role as the emission
probability of the electron. Clearly one has $P_0(t) = 1- \nu_1(t)$, $P_1(t) = \nu_1(t)$, while
other $P_n(t)$ are all zero [see also Eq.~\eqref{s1:eq1}].

\begin{figure}
  \centering
  \includegraphics[width=7.5cm]{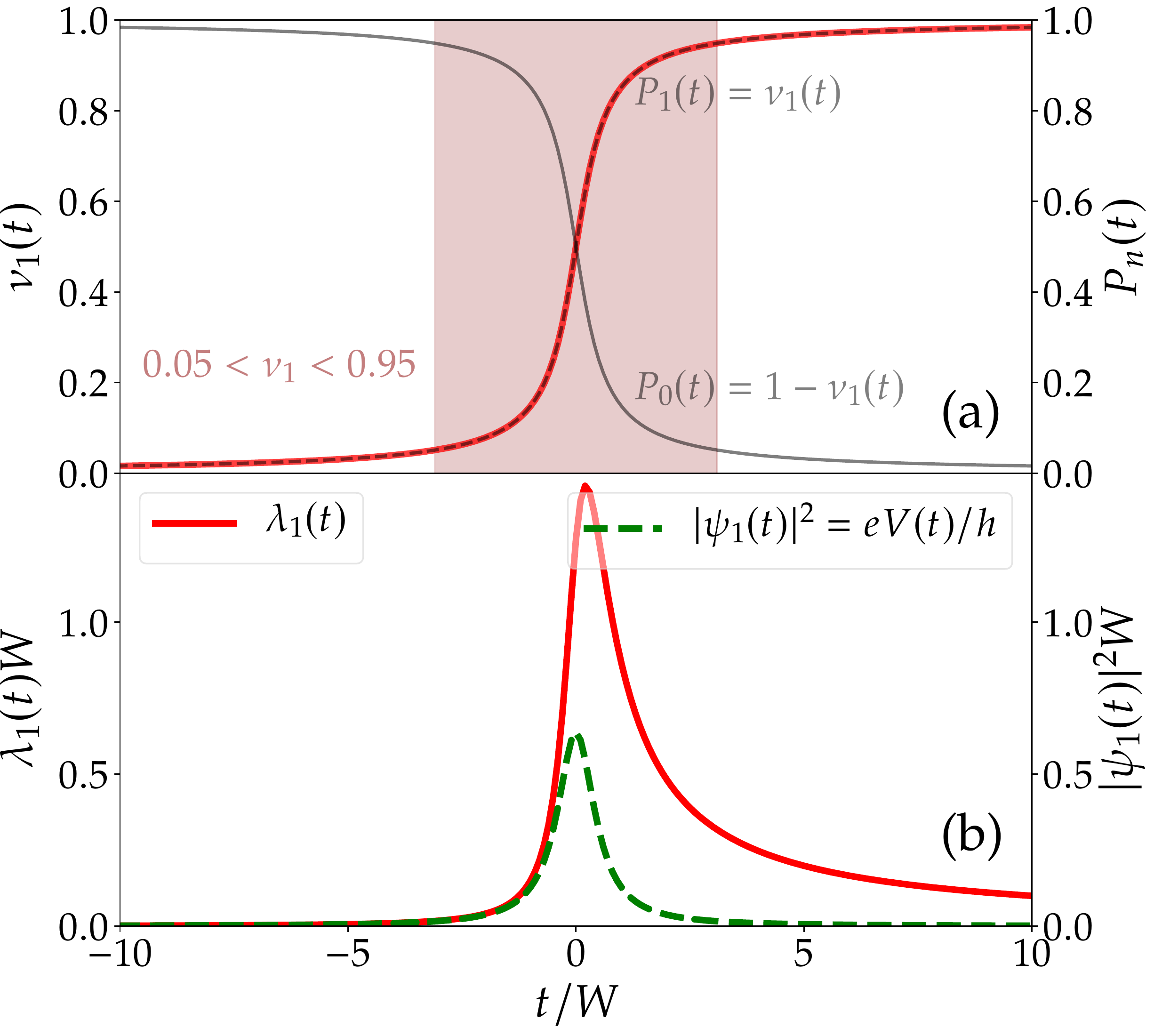}
  \caption{ (a) Emission probability $\nu_1(t)$ (red solid curve) of the electron in the case of
    single leviton emission. The red region ($t/W \in [-3.1, 3.1]$) marks the emission time window,
    which is extracted by requiring $0.05 < \nu_1(t) < 0.95$. The distribution
    $P_0(t) = 1 - \nu_1(t)$ and $P_1(t) = \nu_1(t)$ is plotted by the grey solid and black dashed
    curves for comparison. (b) The emission rate $\lambda_1(t)$ (red solid curve) of the first
    electron for the single leviton emission. The leviton wave function $\left| \psi_1(t) \right|^2$
    is plotted by the green dashed curve for comparison.}
  \label{fig2}
\end{figure}

The typical behavior of $\nu_1(t)$ as a function of normalized time $t/W$ is plotted in
Fig.~\ref{fig2}(a) by the red solid curve. The modules of the wave function $|\psi_1(t)|^2$ is also
plotted in Fig.~\ref{fig2}(b) by the green dashed curve for comparison. One finds that $\nu_1(t)$
remains smaller than $0.05$ for $t/W < -3.1$. It increases rapidly from $0.05$ to $0.95$ for
$t/W \in [-3.1, 3.1]$. This indicates that the electron is mostly likely to be emitted within this
time window, which is marked by the red region in the figure. As $t/W \to + \infty$, $\nu_1(t)$
approaches $1.0$, indicating that the voltage pulse emits exactly one electron in the long time
limit.

From the FCS, the corresponding idle time distribution can be given as
$\Pi_0(t) = 1 - \int^t_{-\infty} d\tau | \psi_1(\tau) |^2$. By using Eq.~\eqref{s2:eq60}, the
emission rate of the electron can be related to the wave function $\psi_1(t)$ as
\begin{equation}
  \lambda_1(t) = \frac{| \psi_1(t) |^2}{1 - \int^t_{-\infty} d\tau |
    \psi_1(\tau) |^2}.
  \label{s3:eq30}
\end{equation}
The typical behavior $\lambda_1(t)$ is illustrated in Fig.~\ref{fig2}(b) by the red solid curve. One
can see that, while the modules of the wave function $|\psi_1(t)|^2$ follows instantly to the
voltage pulse $V(t)$, the emission rate exhibits a different profile. It increases rapidly in the
rising edge, while decays relatively slow in the falling edge.  This shows that the emission is
switched on by the voltage pulse $V(t)$, but it does not switch off immediately when $V(t)$ drops to
zero.

Note that by using the relation $| \psi_1(t) |^2 = eV(t)/h$, the emission rate can be written as
\begin{equation}
  \lambda_1(t) = \frac{eV(t)/h}{1 - \int^t_{-\infty} d\tau eV(t)/h}.
  \label{s3:eq31}
\end{equation}
This can be treated as the quantum limit of the emission rate, which describes the emission of a
single leviton.

\subsection{Two-leviton emitter}
\label{sec3b}

Now let us turn to the two-leviton emitter, which can be realized by applying a Lorentzian pulse
carrying two electron charges. To better clarify the basic concept of the emission rate, let us
start from a more general case, when two unit-charge Lorentzian pulses are applied with the time
delay $t_c$. In this case, the many-body state represents two levitons propagating on top of the
Fermi sea, whose wave functions can be written as
$\psi_1(t) = \sqrt{W/(2 \pi)} /(t - t_c/2 + i W/2)$ and
$\psi_2(t) = \sqrt{W/(2 \pi)} /(t + t_c/2 + i W/2)$. It is convenient to write the many-body state
in the form similar to Eq.~\eqref{s3:eq10}. This can be done by constructing an orthonormal basis
from $\psi_1(t)$ and $\psi_2(t)$ via Gram-Schmidt orthonormal procedure, which gives
\begin{eqnarray}
  \psi_a(t) & = & \psi_1(t), \label{s3:eq50-1}\\
  \psi_b(t) & = & \frac{1}{D}\left[ \psi_2(t) - \langle \psi_1 | \psi_2 \rangle \psi_2(t) \right]. \label{s3:eq50-2}
\end{eqnarray}
Here the normalization constant $D$ can be given as
$D = 1 - \left| \langle \psi_1 | \psi_2 \rangle \right|^2$, with
$\langle \psi_1 | \psi_2 \rangle = \int^{+\infty}_{-\infty} d\tau \psi^{\ast}_1(\tau) \psi_2(\tau) $
being the inner product of the two wave functions. The corresponding many-body state can be given as
\begin{equation}
  | \Psi \rangle = \psi_a^{\dagger} \psi_b^{\dagger} | F \rangle, \label{s3:eq60}
\end{equation}
where $\psi_\alpha^{\dagger} = \int dt \psi_\alpha(t) a^{\dagger}(t)$ (with $\alpha = a, b$)
represents the creation operator of the two emitted electrons. The corresponding first-order Glauber
correlation function can be related to the wave functions as
\begin{equation}
  G(t, t') = \psi^{\dagger}_a(t) \psi_a(t') + \psi^{\dagger}_b(t) \psi_b(t')
  \label{s3:eq61}.
\end{equation}

By solving Eqs.~(\ref{s2:eq70a} - \ref{s2:eq70f}), one finds that only the first two exclusive joint
densities are non-zero, which can be written as \footnote{Alternatively, it can also be obtained
  from Eq.~\eqref{s2:eq20}, which can be calculated more easily in this case.}
\begin{widetext}
  \begin{eqnarray}
    f_1(t_1; t) & = & \left| \psi_a(t_1) \right|^2 \left( 1 - \int^t_{-\infty} d\tau \left| \psi_b(\tau) \right|^2 \right) + \left| \psi_b(t_1) \right|^2 \left( 1 - \int^t_{-\infty} d\tau \left| \psi_a(\tau) \right|^2 \right) \nonumber\\
                    &&\mbox{}+ 2 \operatorname{Re}\left[ \psi_a(t_1) \psi^{\ast}_b(t_1) \int^t_{-\infty} d\tau \psi_b(\tau) \psi^{\ast}_a(\tau) \right], \label{s3:eq70-1}\\
    f_2(t_1, t_2; t) & = & \left| \psi_a(t_1) \psi_b(t_2) \right|^2 + \left| \psi_a(t_2) \psi_b(t_1) \right|^2 - 2 \operatorname{Re}\left[ \psi_a(t_1) \psi_b(t_2) \psi^{\ast}_a(t_2) \psi^{\ast}_b(t_1) \right].  \label{s3:eq70-2}
  \end{eqnarray}
  It is worth noting that $f_2(t_1, t_2; t)$ is just equal to the coincidence function
  $g_2(t_1, t_2)$, which can also be written as
  \begin{equation}
    f_2(t_1, t_2; t) = g_2(t_1, t_2) = \left| \psi_a(t_1) \psi_b(t_2) - \psi_a(t_2) \psi_b(t_1) \right|^2.
    \label{s3:eq75}
  \end{equation}
  In Eqs.~\eqref{s3:eq70-1} and~\eqref{s3:eq70-2}, the first and second terms correspond to the
  incoherence contributions, while the last terms represents the contributions from two-electron
  coherence. The corresponding FCS can be described by the momentum generating function:
  \begin{equation}
    \Phi(\chi) = P_0(t) + P_1(t) e^{i \chi} + P_2(t) e^{2 i \chi} = \left[ 1 - \nu_1(t) + e^{i \chi} \nu_1(t) \right]\left[ 1 - \nu_2(t) + e^{i \chi} \nu_2(t) \right].
    \label{s3:eq80}
  \end{equation}
  It corresponds to a generalized binomial process, where two electrons attempt to emit with the
  probabilities $\nu_1(t)$ and $\nu_2(t)$, respectively. They can be related to the wave functions
  $\psi_a(t)$ and $\psi_b(t)$ as
  \begin{eqnarray}
    \nu_1(t) & = & \frac{I_a + I_b}{2} + \sqrt{ \left( \frac{I_a - I_b}{2} \right)^2 + \left| I_{ab} \right|^2 }, \label{s3:eq90-1}\\
    \nu_2(t) & = & \frac{I_a + I_b}{2} - \sqrt{ \left( \frac{I_a - I_b}{2} \right)^2 + \left| I_{ab} \right|^2 }. \label{s3:eq90-2}
  \end{eqnarray}
  The two terms $I_a = \int^t_{-\infty} d\tau \left| \psi_a(\tau) \right|^2$ and
  $I_b = \int^t_{-\infty} d\tau \left| \psi_b(\tau) \right|^2$ correspond to the incoherent
  contributions from $\psi_a(t)$ and $\psi_b(t)$, respectively. The term
  $I_{ab} = \int^t_{-\infty} d\tau \psi^{\ast}_a(\tau) \psi_b(\tau)$ represents the contributions
  due to the two-electron coherence. The idle time probability can be obtained from the FCS $P_0(t)$
  [Eq.~\eqref{s2:eq30b}], which has the form
  \begin{equation}
    \Pi_0(t) = P_0(t) = \left[ 1 - \int^t_{-\infty} d\tau \left| \psi_a(\tau) \right|^2 \right] \left[ 1 - \int^t_{-\infty} d\tau \left| \psi_b(\tau) \right|^2 \right] - \left| \int^t_{-\infty} d\tau \psi_a(\tau) \psi^{\ast}_b(\tau) \right|^2,
    \label{s3:eq100}    
  \end{equation}
  where the contributions from two-electron coherence is represented by the last term.
\end{widetext}
By substituting Eqs.~\eqref{s3:eq70-1},~\eqref{s3:eq70-2} and~\eqref{s3:eq100} into
Eqs.~\eqref{s2:eq61} and~\eqref{s2:eq60}, the emission rates of the two electrons can be given as
\begin{eqnarray}
  \lambda_1(t) & = & \frac{f_1(t| t)}{\Pi_0( t)}, \label{s3:eq101-1}\\
  \lambda_2(t|t_1) & = & \frac{f_2(t_1, t| t)}{f_1(t| t)}.   \label{s3:eq101-2}
\end{eqnarray}

\begin{figure}
  \centering
  \includegraphics[width=7.5cm]{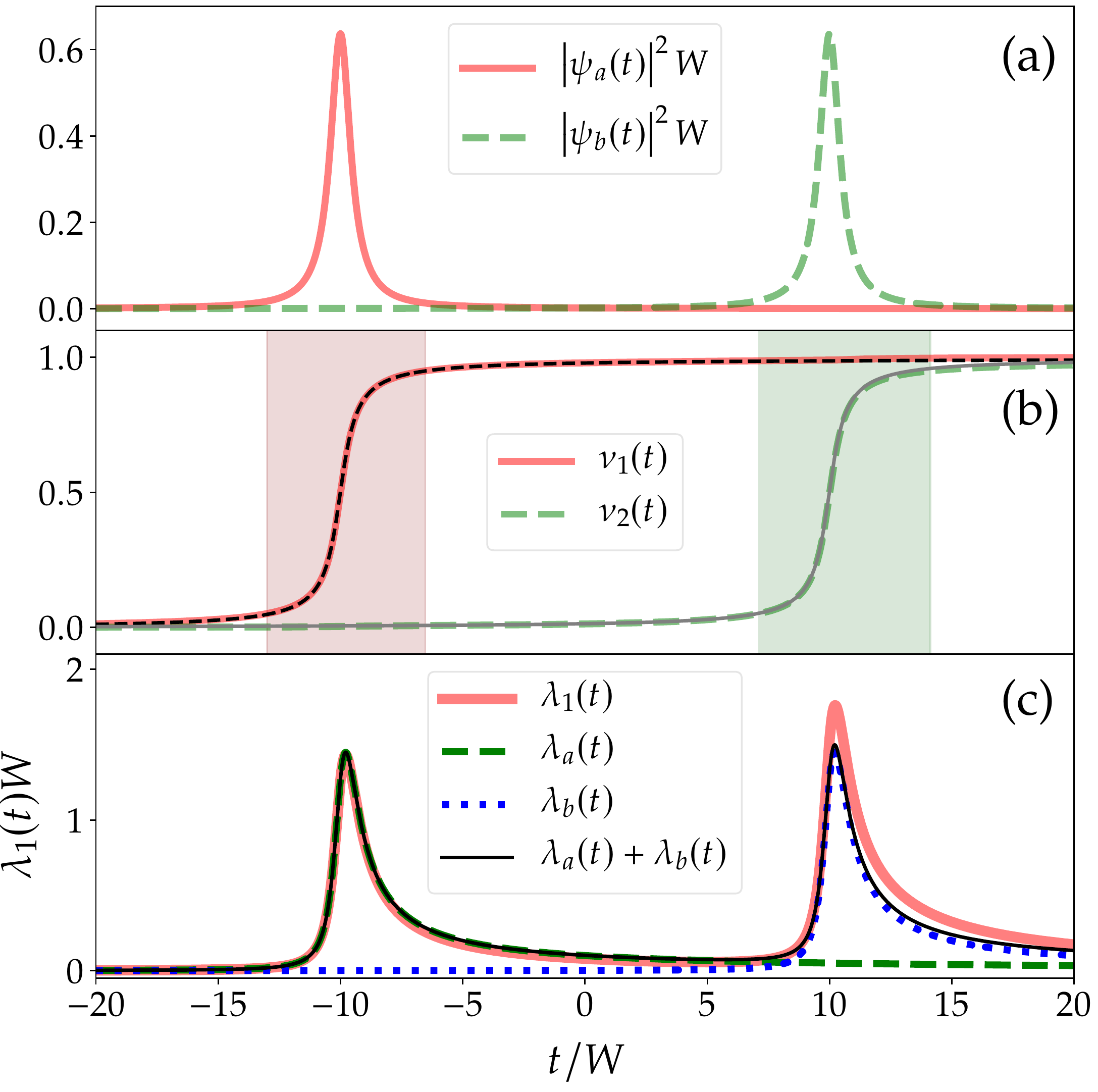}
  \caption{ (a) Wave functions of the emitted electron for $t_c/W = 20.0$. (b) Emission
    probabilities $\nu_1(t)$ (red solid curve) and $\nu_2(t)$ (green dashed curve) of the first and
    second electrons. The black dashed and gray solid curves represent the approximation from
    Eq.~\eqref{s3:eq110-1} and Eq.~\eqref{s3:eq110-2}, respectively. (c) The emission rate
    $\lambda_1(t)$ (red solid curve) of the first electron. The green dashed and blue dotted curves
    represent the contribution solely from $\psi_a(t)$ [Eq.~\eqref{s3:eq120b-1}] and $\psi_b(t)$
    [Eq.~\eqref{s3:eq120b-2}], respectively. The black solid curve represent the incoherent
    summation $\lambda_a(t) + \lambda_a(t)$.}
  \label{fig3}
\end{figure}

To better understand the physical meaning of the emission rates, let us first consider on a limit
case, when the wave functions of two levitons are well-separated from each other. In this case, the
two electrons are essentially distinguishable and one expect that their emissions can be treated as
independent events. This is demonstrated in Fig.~\ref{fig3}(a), corresponding to $t_c/W = 20.0$. In
this case, one has $\psi_a(t) \approx \psi_1(t)$ and $\psi_b(t) \approx \psi_2(t)$. The overlap
between them are rather small, indicating a negligible two-electron coherence. By dropping the
coherence contributions $I_{ab}$ in Eqs.~\eqref{s3:eq90-1} and~\eqref{s3:eq90-2}, the emission
probabilities can be well-approximated as
\begin{eqnarray}
  \nu_1(t) & \approx & \int^t_{-\infty} d\tau \left| \psi_a(\tau) \right|^2, \label{s3:eq110-1}\\
  \nu_2(t) & \approx & \int^t_{-\infty} d\tau \left| \psi_b(\tau) \right|^2. \label{s3:eq110-2}
\end{eqnarray}
The approximation is illustrated in Fig.~\ref{fig3}(b). The red solid and green dash curves
represent the exact result from Eqs.~\eqref{s3:eq90-1} and~\eqref{s3:eq90-2}, while the black dashed
and gray solid curves correspond to the approximation given in Eqs.~\eqref{s3:eq110-1}
and~\eqref{s3:eq110-2}, respectively. One can see that they agrees quite well, indicating that the
emission of the two electron are dominated by the corresponding wave functions $\psi_a(t)$ and
$\psi_b(t)$, respectively.

The independence of the emission can also be seen from the emission time windows. From
Fig.~\ref{fig3}(b), one can see that $\nu_1(t)$ [$\nu_2(t)$] increases rapidly from $0.05$ to $0.95$
as $t/W$ increases from $-12.42$ to $-5.45$ ($7.22$ to $15.47$). This indicates that the two
electrons are most likely to be emitted in two time windows $t/W \in [-12.42, -5.45]$ and
$t/W \in [7.22, 15.47]$, which are marked by the the red and green regions in
Fig.~\ref{fig3}(b). One can see that they are well-separated from each other, indicating that the
emission of the two electrons can be treated as independent events, which are distinguishable from
their emission times.

Now let us discuss the emission rate $\lambda_1(t)$ of the first electron. From
Eqs.~\eqref{s3:eq70-1},~\eqref{s3:eq100} and~\eqref{s3:eq101-1}, one finds that both $\psi_a(t)$ and
$\psi_b(t)$ can contribute to $\lambda_1(t)$. By dropping the coherence contributions [the last
terms in Eqs.~\eqref{s3:eq70-1} and~\eqref{s3:eq100} ], the emission rate of the first electron
$\lambda_1(t)$ can be approximated as
\begin{equation}
  \lambda_1(t) \approx \lambda_a(t) + \lambda_b(t),
  \label{s3:eq120a}
\end{equation}
with 
\begin{eqnarray}
  \lambda_a(t) & = & \frac{| \psi_a(t) |^2}{1 - \int^t_{-\infty} d\tau | \psi_a(\tau) |^2}, \label{s3:eq120b-1}\\
  \lambda_b(t) & = & \frac{| \psi_b(t) |^2}{1 - \int^t_{-\infty} d\tau | \psi_b(\tau) |^2}. \label{s3:eq120b-2}
\end{eqnarray}
By comparing to Eq.~\eqref{s3:eq30}, one finds that these two terms are just the emission rates
solely due to $\psi_a(t)$ and $\psi_b(t)$, respectively. The two terms can lead to two
well-separated peaks in the emission rate $\lambda_1(t)$. This can be seen from
Fig.~\ref{fig3}(c). In the figure, the red solid curve represents the exact emission rate
$\lambda_1(t)$, while the emission rates $\lambda_a(t)$ and $\lambda_b(t)$ are plotted by the green
dashed and blue dotted curves, respectively. Around the first peak ($t/W=-10.0$), one finds that the
emission rate $\lambda_1(t)$ is almost solely due to $\lambda_a(t)$. In contrast, although
$\lambda_b(t)$ dominates the emission rate around the second peak ($t/W=10.0$), the two-electron
coherence can also play a non-negligible role. As a consequence, the approximation from
Eq.~\eqref{s3:eq120a} slightly underestimates the exact emission rate, which can be seen by
comparing the red solid curve (exact emission rate) to the black solid one
[Eq.~\eqref{s3:eq120a}]. However, as the emission of the first electron is concentrated in the
emission time window of the first electron [red region in Fig.~\ref{fig3}(b)], $\lambda_1(t)$ is
essentially decided by $\lambda_a(t)$ alone.

\begin{figure}
  \centering
  \includegraphics[width=7.0cm]{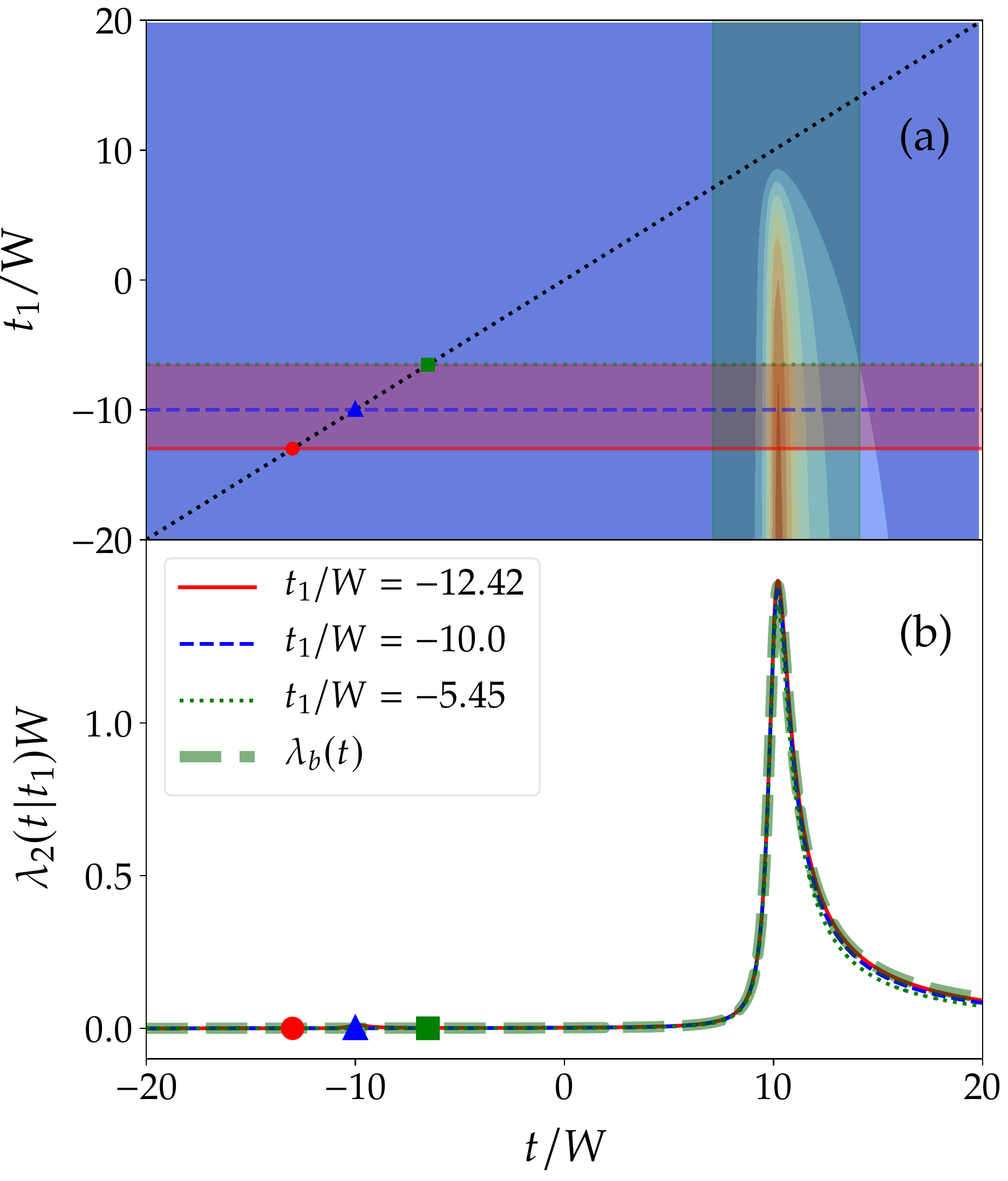}
  \caption{ (a) The emission rate $\lambda_2(t|t_1)$ as a contour plot in $t$-$t_1$ plane. It is
    only nonzero for $t_1 < t$ (lower-right side of plane. The black dashed curve represents $t_1=t$
    ). The red and green regions represent the emission time windows of the first and second
    electrons, respectively. (b) $\lambda_2(t|t_1)$ as a function of $t$ for three typical $t_1$ in
    the first emission time window. The red solid, blue dashed and green dotted curves correspond to
    $t_1/W = -12.42$, $-10.0$ and $-5.45$. The red dots, blue triangles and green squares in both
    (a) and (b) mark the corresponding points $t=t_1$.}
  \label{fig4}
\end{figure}

Now we turn to the emission rate $\lambda_2(t|t_1)$ of the second electron. By dropping the
coherence contributions [the last terms in Eqs.~\eqref{s3:eq70-1} and~\eqref{s3:eq70-2}], it can be
approximated as
\begin{widetext}
  \begin{equation}
    \lambda_2(t|t_1)  \approx \frac{ \left| \psi_a(t_1) \psi_b(t) \right|^2 +
      \left| \psi_a(t) \psi_b(t_1) \right|^2 }{ \left| \psi_a(t_1) \right|^2
      \left( 1 - \int^t_{-\infty} d\tau \left| \psi_b(\tau) \right|^2 \right) +
      \left| \psi_b(t_1) \right|^2 \left( 1 - \int^t_{-\infty} d\tau \left|
      \psi_a(\tau) \right|^2 \right) }.
    \label{s3:eq120}
  \end{equation}  
\end{widetext}
Here $t_1$ represents the emission time of the first electron, which satisfies $t_1 < t$ following
the restriction given in Eq.~\eqref{s2:eq40}. Due to the $t_1$-dependence, the behavior of
$\lambda_2(t|t_1)$ appears more complicated. This can be seen from Fig.~\ref{fig4}(a), where we plot
$\lambda_2(t|t_1)$ as a contour plot in the $t$-$t_1$ plane. Note that $\lambda_2(t|t_1)$ is only
nonzero in the lower-right side of the plane, which is due to the restriction $t_1 < t$. From the
contour plot, one finds that $\lambda_2(t|t_1)$ exhibits a strong peak around $t/W=10.0$. The
amplitudes of the peak drop to zero as $t_1$ approach $t$. It seems that the emission of the second
electron is correlated to the first one even in the absence of two-electron coherence.

However, the $t_1$-dependence is essentially negligible when $t_1$ lies in the first emission time
window, which is marked by the red region in Fig.~\ref{fig4}(a). In this case,
$\left|\psi_b(t_1)\right|^2$ is rather small [see Fig.~\ref{fig3}(a)] and can be omitted in
Eq.~\eqref{s3:eq120}. In doing so, one finds that $\lambda_2(t|t_1)$ can be written as
\begin{equation}
  \lambda_2(t|t_1) \approx \frac{ \left| \psi_b(t) \right|^2 }{ 1 - \int^t_{-\infty} d\tau \left| \psi_b(\tau) \right|^2 } = \lambda_b(t).
  \label{s3:eq130}
\end{equation}
It is just the emission rate solely from the wave function $\psi_b(t)$, which does not depends on
$t_1$. This approximation is demonstrated in Fig.~\ref{fig4}(b). In the figure, the red solid, blue
dashed and green dotted curves corresponds to $\lambda_2(t|t_1)$ with three typical $t_1$ in the
emission time window of the first electron. The thick green dashed curve represents
$\lambda_b(t)$. One can see that they agree quite well, indicating that the emission of the second
electron is dominated by $\psi_b(t)$.

The above results show that when the two pulses are well-separated in the time domain, the emission
of the two electrons can be treated as independent quantum events, whose emission rates are decided
solely by their own wave functions.

\begin{figure}
  \centering
  \includegraphics[width=7.5cm]{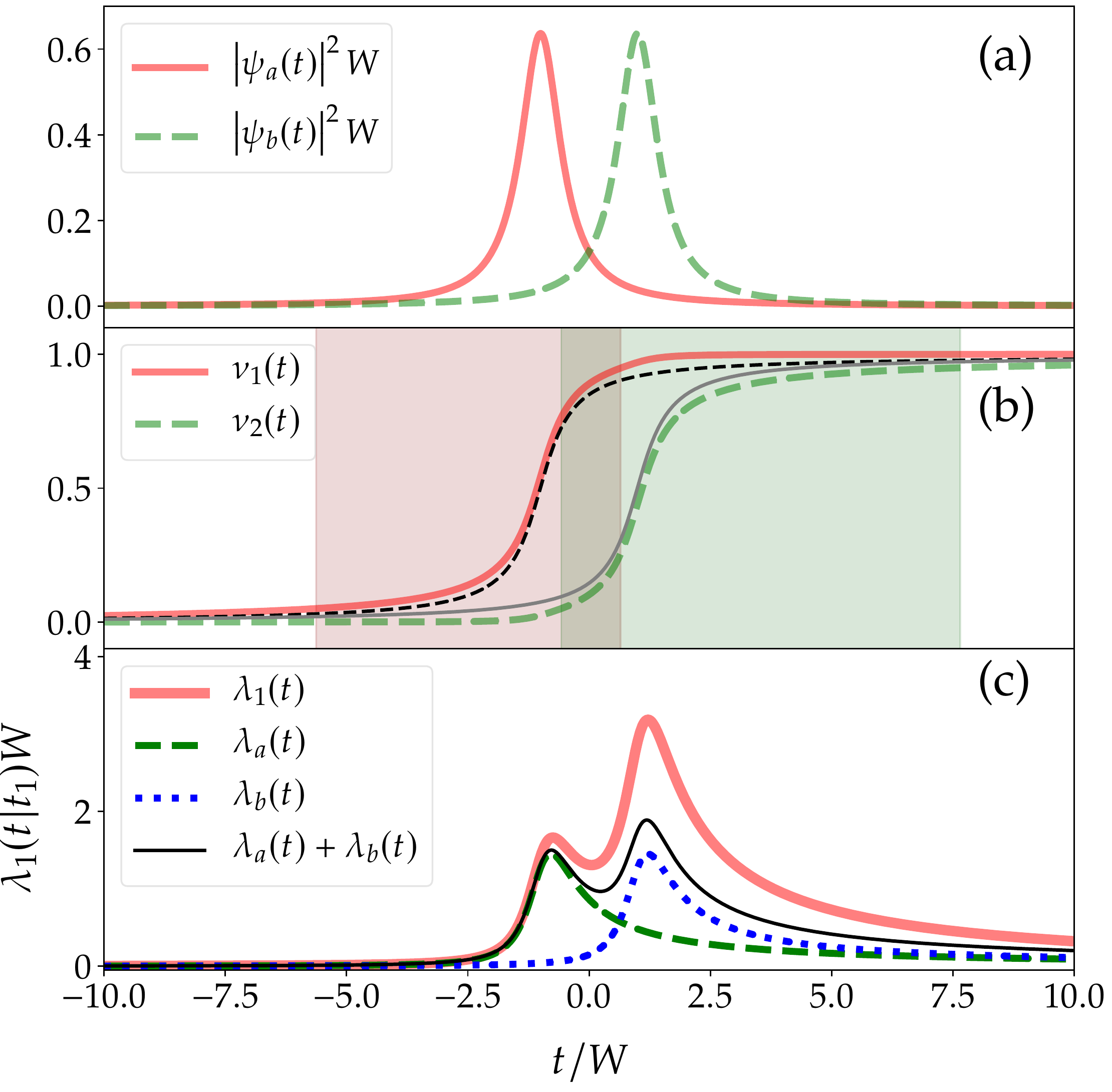}
  \caption{ The same as Fig.~\ref{fig3} but for $t_c/W = 2.0$.}
  \label{fig5}
\end{figure}

As the two pulses approach each other, the wave function of the two electrons can overlap. This is
demonstrated Fig.~\ref{fig5}(a), corresponding to $t_c/W = 2.0$. The emission probabilities
$\nu_1(t)$ and $\nu_2(t)$ are plotted in Fig.~\ref{fig5}(b) by the red solid and green dashed
curves, respectively. In this case, the two electrons are most likely to be emitted in two time
windows $t/W \in [-4.97, 0.7]$ and $t/W \in [-0.53, 9.4]$, which are marked by the red and green
regions in Fig.~\ref{fig5}(b). The two time windows can overlap with each other, indicating that
their emissions are correlated. Indeed, one finds that the approximations for the emission
probabilities Eqs.~\eqref{s3:eq110-1} (black dashed curve) and~\eqref{s3:eq110-2} (gray solid curve)
becomes inaccurate, as illustrated in Fig.~\ref{fig5}(b). This indicates that the two-electron
coherence can play a non-negligible role on the emission process.

The impact of the two-electron coherence can also be seen from the emission rate
$\lambda_1(t)$. This is demonstrated in Fig.~\ref{fig5}(c). The red solid curve represents the exact
emission rate $\lambda_1(t)$, while the green dashed and blue dotted curves correspond to the
emission rates solely from $\lambda_a(t)$ [Eq.~\eqref{s3:eq120b-1}] and $\lambda_b(t)$
[Eq.~\eqref{s3:eq120b-2}], respectively. The black solid curve represents the incoherent summation
$\lambda_a(t) + \lambda_b(t)$.  One can see that $\lambda_1(t)$ is larger than the
$\lambda_a(t) + \lambda_b(t)$, which can be attributed to the contribution from the two-electron
coherence.

\begin{figure}
  \centering
  \includegraphics[width=7.0cm]{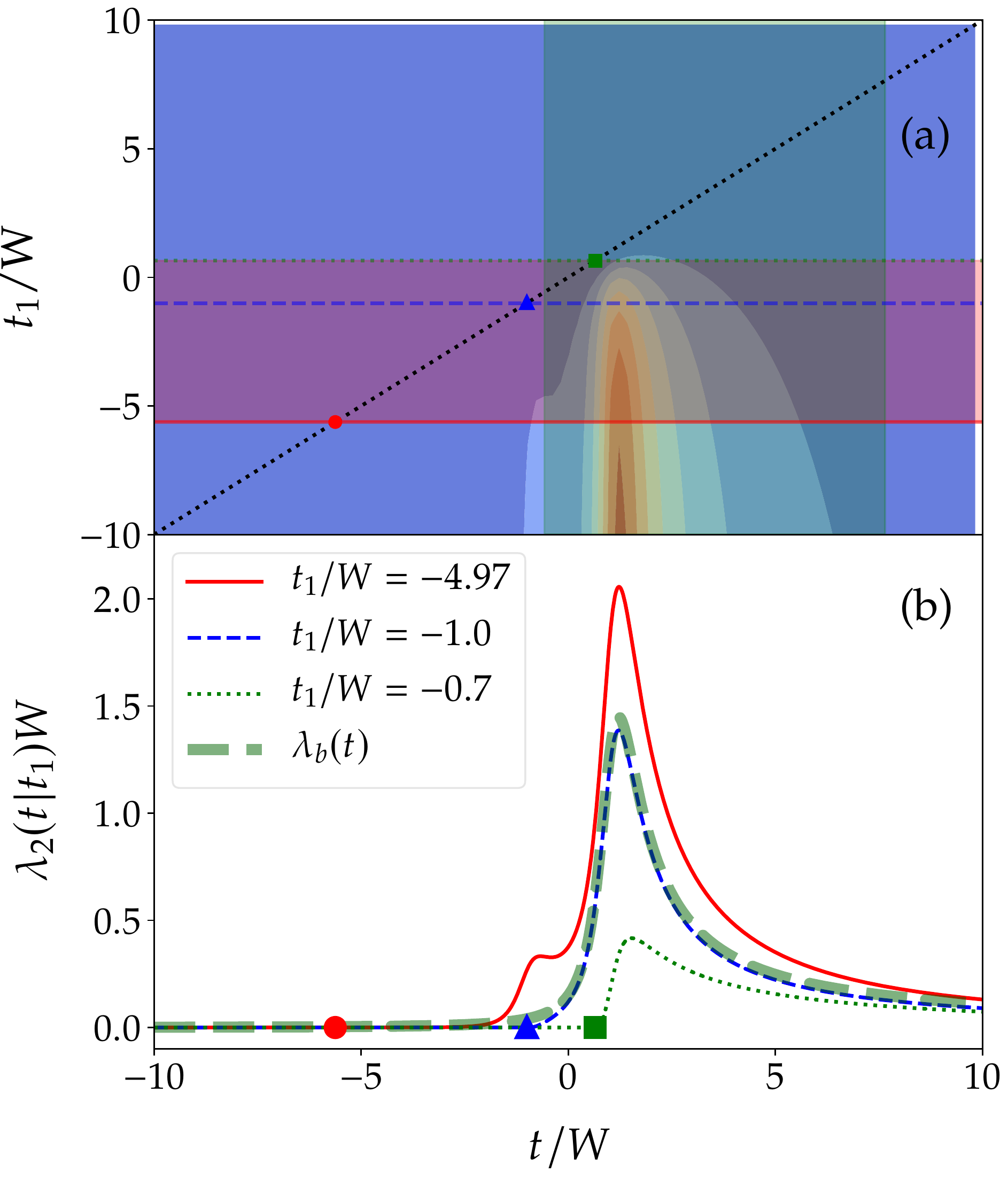}
  \caption{ The same as Fig.~\ref{fig4} but for $t_c/W=2.0$. The red solid, blue dashed and green
    dotted curves correspond to $t_1/W = -4.97$, $-1.0$ and $-0.7$.}
  \label{fig6}
\end{figure}

While the two-electron coherence enhances the emission rate $\lambda_1(t)$ of the first electron, it
plays a different role for the emission rate $\lambda_2(t|t_1)$ of the second electron. This can be
seen from Fig.~\ref{fig6}(a), where we plot $\lambda_2(t|t_1)$ as a contour plot in the $t$-$t_1$
plane. One can see that $\lambda_2(t|t_1)$ exhibits a strong peak around $t/W = 1.0$ and a small
shoulder peak around $t/W = -1.0$. The amplitudes of both peaks drop as $t_1$ approaches $t$. The
dependence remains pronounced when $t_1$ lies in the first emission time window, which is marked by
the red region in Fig.~\ref{fig6}(a). The $t_1$-dependence can be better seen from
Fig.~\ref{fig6}(b). In the figure, the red solid, blue dashed and green dotted curves corresponds to
$\lambda_2(t|t_1)$ with three typical $t_1$ in the emission time window of the first electron. The
emission rate $\lambda_b(t)$ solely due to $\psi_b(t)$ [Eq.~\eqref{s3:eq120b-2}] is plotted by the
thick green dashed curve for comparison. One finds that $\lambda_b(t)$ can only give a good
estimation of $\lambda_2(t|t_1)$ for $t_1/W = -1.0$ (blue dashed curve). In contrast,
$\lambda_2(t|t_1)$ is suppressed for $t_1/W = 0.7$ (green dotted curve) and it is enhanced for
$t_1/W = -4.97$ (red solid curve), indicating that the emission of the second electron is strongly
correlated to the first one.  In particular, $\lambda_2(t|t_1)$ always drops to zero as $t_1$
approaches $t$. This is also illustrated in Fig.~\ref{fig6}(b), where the point $t=t_1$ are marked
by the red dot ($t_1/W=-4.97$), blue triangle ($t_1/W=-1.0$) and green square ($t_1/W=0.7$),
respectively.

The behavior of $\lambda_2(t|t_1)$ can be understood in an intuitive way. When $t_1/W = -1.0$, one
has $|\psi_a(t_1)| \gg |\psi_b(t_1)|$ [see Fig.~\ref{fig5}(a)]. Then the emission of the first
electron at $t_1$ is almost solely due to $\psi_a(t_1)$. As the first electron has been emitted, the
many-body state is most likely to be collapsed to $|\Psi\rangle = \psi^{\dagger}_b|F\rangle$, which
is just a single-electron state with wave function $\psi_b(t)$. This explains why one has
$\lambda_2(t|t_1) \approx \lambda_b(t)$ for $t_1/W = -1.0$. As $t_1$ approaches $t$, the Pauli
exclusion principle prevents two electrons emit at the same time, which can greatly suppress the
emission rate $\lambda_2(t|t_1)$ when $t_1$ is sufficiently close to $t$. In particularly, it
requires $\lambda_2(t_1|t_1) = 0$. This can be seen more clearly by substituting Eq.~\eqref{s3:eq75}
into Eq.~\eqref{s3:eq101-2}. When $t_1$ is far from $t$, both $\psi_a(t)$ and $\psi_b(t)$ can
contribute to the emission of the second electron, while the suppression due to the Pauli exclusion
principle becomes less important. As a consequence, the emission rate $\lambda_2(t|t_1)$ is enhanced
comparing to $\lambda_b(t)$.

The above results show that the two-electron coherence can have a different impact on the emission
of two electrons. While it always enhances the emission rate $\lambda_1(t)$, its impact on
$\lambda_2(t|t_1)$ depends strongly on $t_1$. In particular, it greatly suppresses
$\lambda_2(t|t_1)$ when $t$ is close enough to $t_1$. This can be attributed to the Pauli exclusion
principle, indicating a strong correlation at short times.

\begin{figure}
  \centering
  \includegraphics[width=7.0cm]{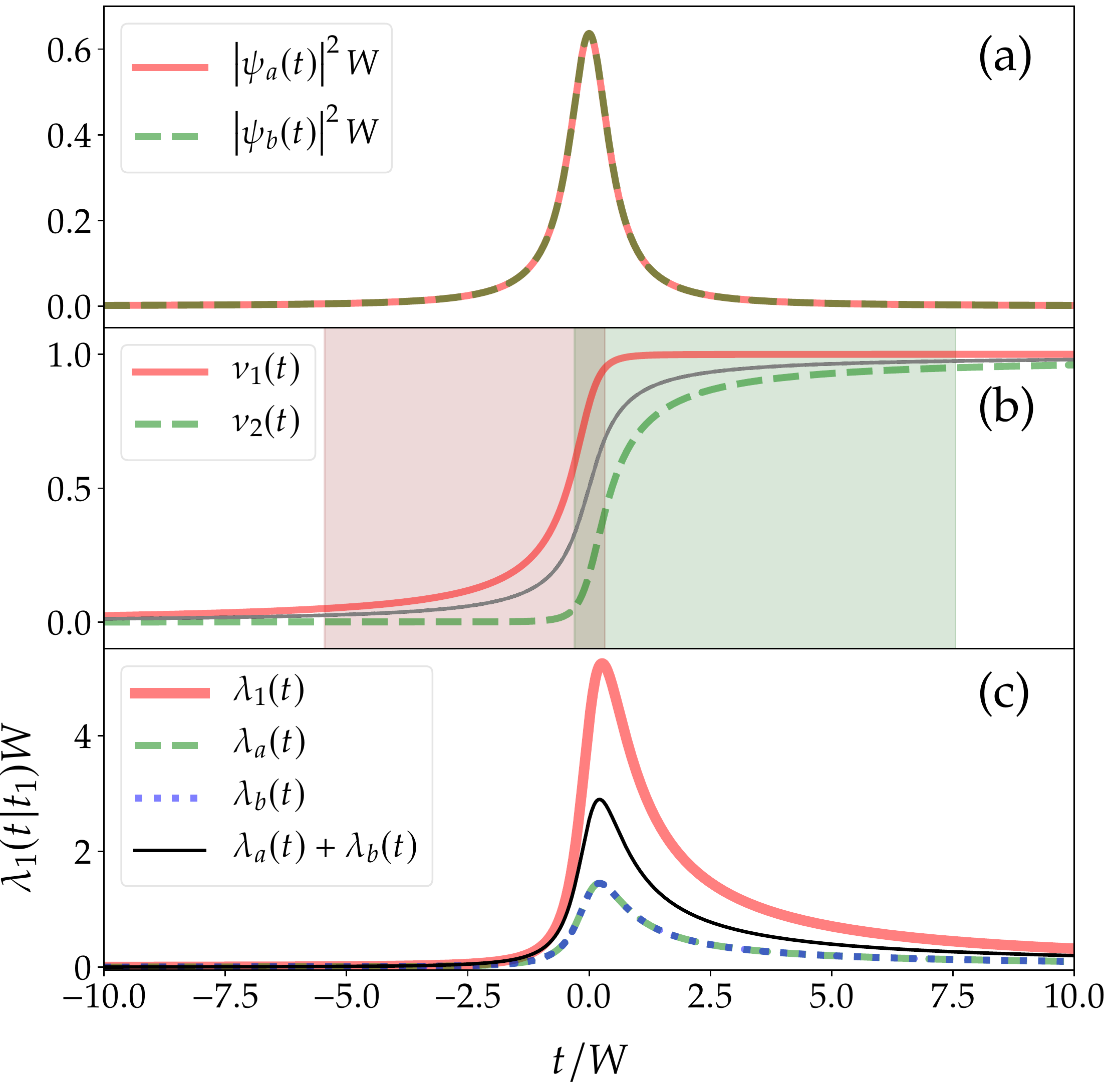}
  \caption{ The same as Fig.~\ref{fig3} but for $t_c/W = 0.0$. }
  \label{fig7}
\end{figure}

In the extreme case when $t_c/W = 0$, the two voltage pulses reduce to a single pulse carrying two
electron charges. In this case, the modules of the two wave functions $\psi_a(t)$ and $\psi_b(t)$
exhibit the same profile, which is shown in Fig.~\ref{fig7}(a). But the emission of the two
electrons are distributed in two time windows $t/W \in [-4.8, 0.32]$ (red region) and
$t/W \in [-0.3, 9.3]$ (green region), which are only slightly overlapped. This can be seen from the
corresponding emission probabilities plotted in Fig.~\ref{fig7}(b).

\begin{figure}
  \centering
  \includegraphics[width=7.0cm]{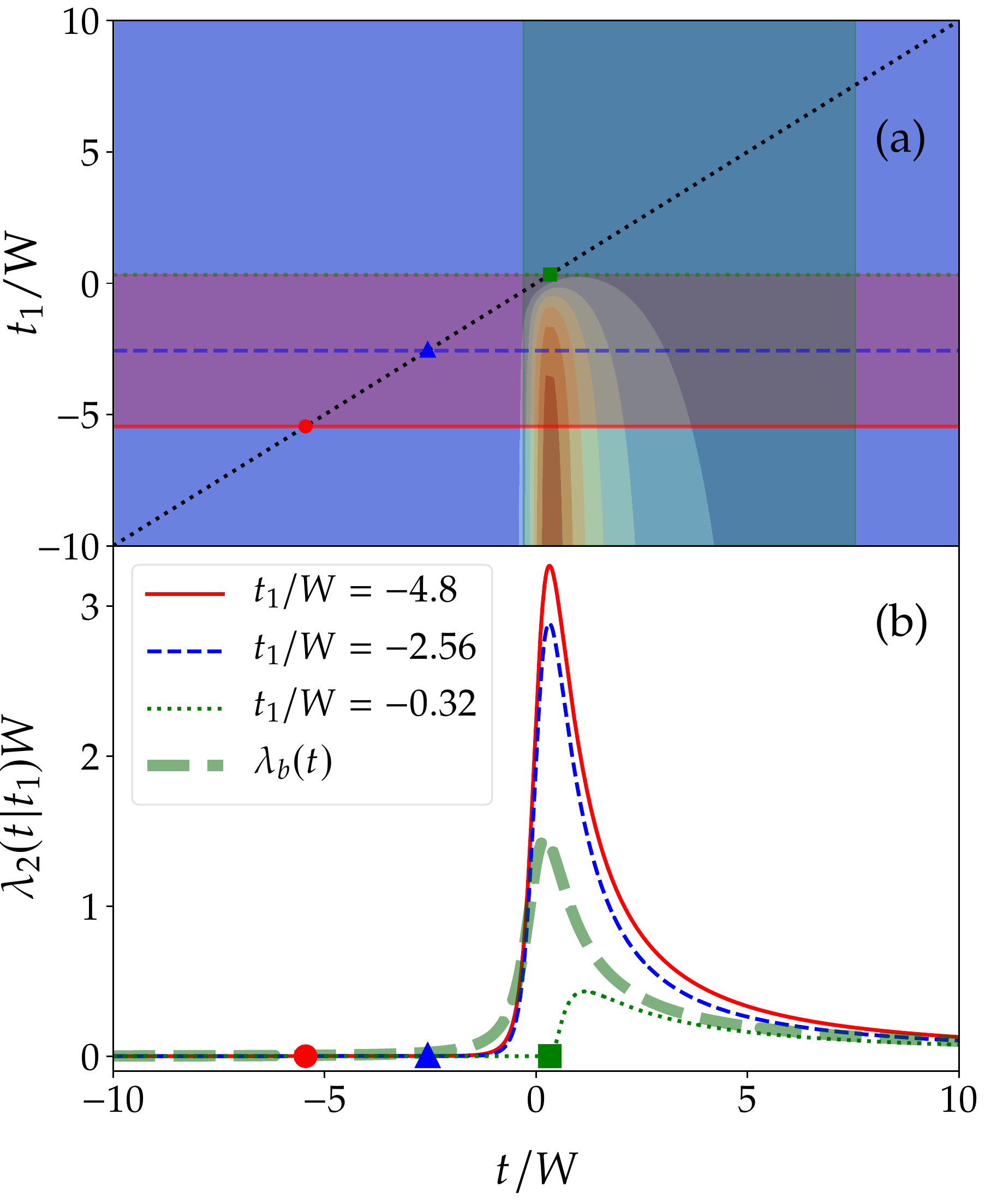}
  \caption{ The same as Fig.~\ref{fig4} but for $t_c/W=0.0$. The red solid, blue dashed and green
    dotted curves correspond to $t_1/W = -4.8$, $-2.56$ and $-0.32$.  }
  \label{fig8}
\end{figure}

In this case, the two-electron coherence can make a large contribution, which greatly enhances the
emission rate of the first electron $\lambda_1(t)$. This can be seen by comparing the red solid
curve to the black dotted one in Fig.~\ref{fig7}(c). Due to the contribution from the two-electron
coherence, the $t_1$-dependence of $\lambda_2(t|t_1)$ also becomes more pronounced, which can be
seen from Fig.~\ref{fig8}.

\begin{figure}
  \centering
  \includegraphics[width=7.5cm]{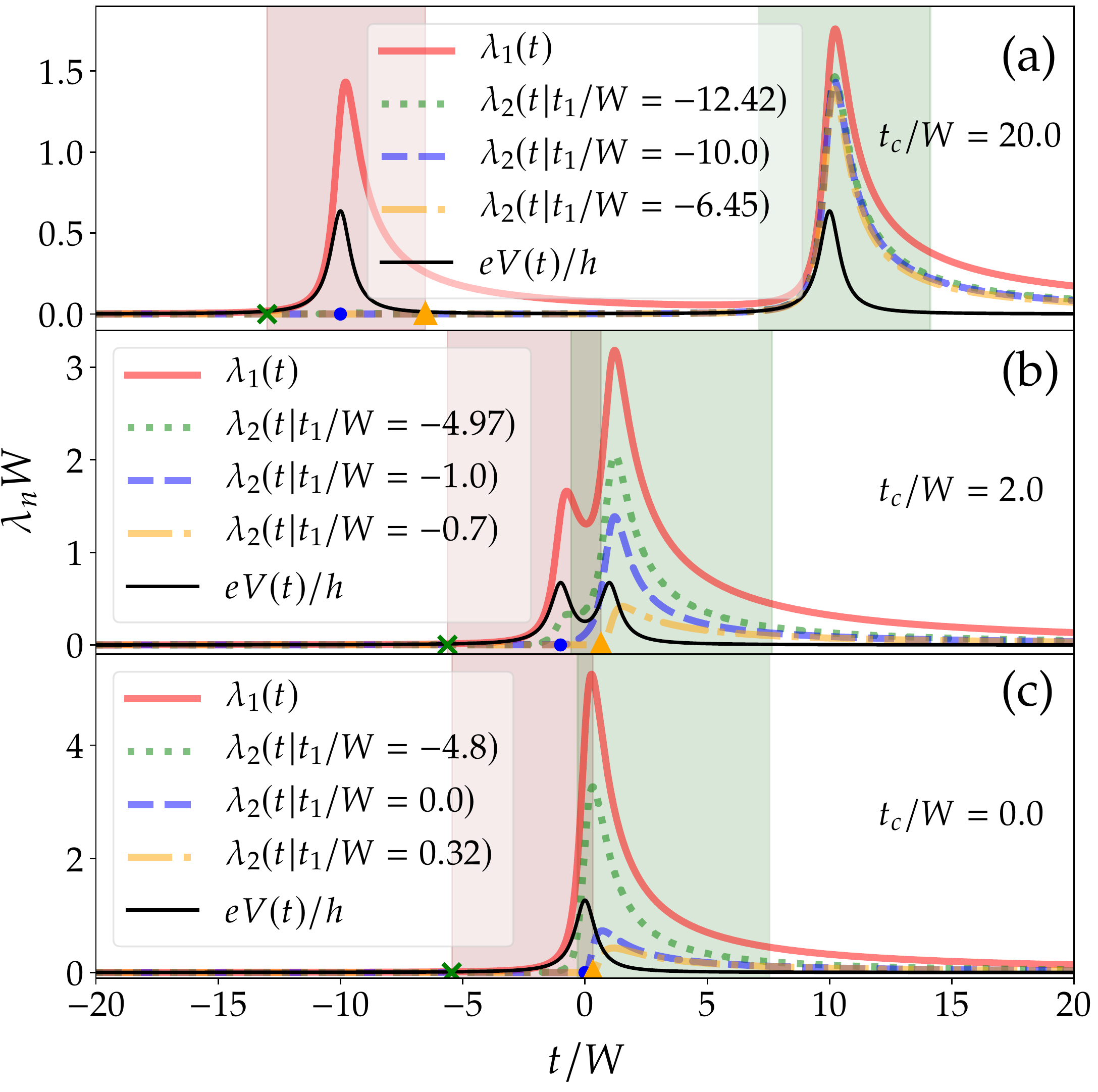}
  \caption{The relation between the emission rates $\lambda_1(t)$, $\lambda_2(t)$ and normalized
    driving pulse $eV(t)/h$ for $t_c/W=20.0$ (a), $t_c/W=2.0$ (b) and $t_c/W=0.0$ (c). The emission
    time windows for the first and second electrons are marked by the red and green regions. The red
    solid curve represents $\lambda_1(t)$. The green dotted, blue dashed and orange dash-dotted
    curves represent $\lambda_2(t|t_1)$ with three typical $t_1$, which lie in the first emission
    time window.}
  \label{fig9}
\end{figure}

In the end of this section, let us elucidate the relation between the electron emission and driving
pulse by using the emission rates. This is illustrated in Fig.~\ref{fig9}, corresponding to
$t_c/W = 20.0$ (a), $t_c/W = 2.0$ (b) and $t_c/W = 0.0$ (c). In the figure, the red solid curve
represents the emission rate of the first electron $\lambda_1(t)$. The green dotted, blue dashed and
orange dashed-dotted curves represent the emission rates of the second electron $\lambda_2(t|t_1)$
for three typical $t_1$ in the emission time window of the first electron. The normalized driving
voltage $eV(t)/h$ are plotted by the black solid curve for comparison. When the two pulses are
well-separated [Fig.~\ref{fig9}(a)], one finds that the rising edges of $\lambda_1(t)$
[$\lambda_2(t|t_1)$] follows the first [second] driving pulse. In the meantime, the emission rate
$\lambda_2(t|t_1)$ is almost independent on $t_1$. This indicates a good synchronization between the
electron emissions and driving pulses. In this case, the two electrons tend to be emitted in two
separated time windows (red and green regions), which are approximately centered around the peak of
the each pulse. As the two pulses approach each other, the emission rate $\lambda_1(t)$ is greatly
enhanced, leading to a fast rising edge. As a consequence, the first electron tends to be emitted
before the peak of the first pulse. This can be seen by comparing the red regions in
Fig.~\ref{fig9}(a-c). In contrast, the emission rate $\lambda_2(t|t_1)$ becomes $t_1$-dependent,
which is enhanced when $t_1$ is far from $t$ and greatly suppressed when $t_1$ is close to $t$. As a
consequence, the emission of the second electron tends to be emitted after the peak of the second
pulse. This can be seen by comparing the green regions in Fig.~\ref{fig9}(a-c).

\section{Finite temperatures}
\label{sec4}

Now let us discuss the impact of the finite temperature. In this case, the first-order Glauber
function can be expressed as \cite{moskalets18_high_temper_fusion_multiel_levit,
  moskalets18_singl_elect_secon_order_correl}
\begin{eqnarray}
  G(t, t') & = & \sum_i \int d\epsilon \left[ - \frac{ \partial f(\epsilon) }{\partial \epsilon} \right] \nonumber\\
           && \times \left[ \psi_i(t) e^{-i \epsilon t} \right] \left[ \psi_i(t') e^{-i \epsilon t'} \right]^{\dagger}
  \label{s4:eq10}.
\end{eqnarray}
For the single electron emission, one can choose $i = 1$ [see Eq.~\eqref{s3:eq21}], while for the
two electron emission, one can choose $i = a, b$ [see Eq.~\eqref{s3:eq61}]. Here
$f(\epsilon) = 1/\left[ 1 + e^{-\epsilon/(k_B T)} \right]$ represents the Fermi-Dirac distribution, with $T$
being the electron temperature. Note that we set the zero of energy to be the Fermi energy in the
electrode. In the calculation, we choose $T_W = \hbar/(k_B W)$ as the temperature unit.

\subsection{Single-electron emitter}
\label{sec4a}

\begin{figure}
  \centering
  \includegraphics[width=7.5cm]{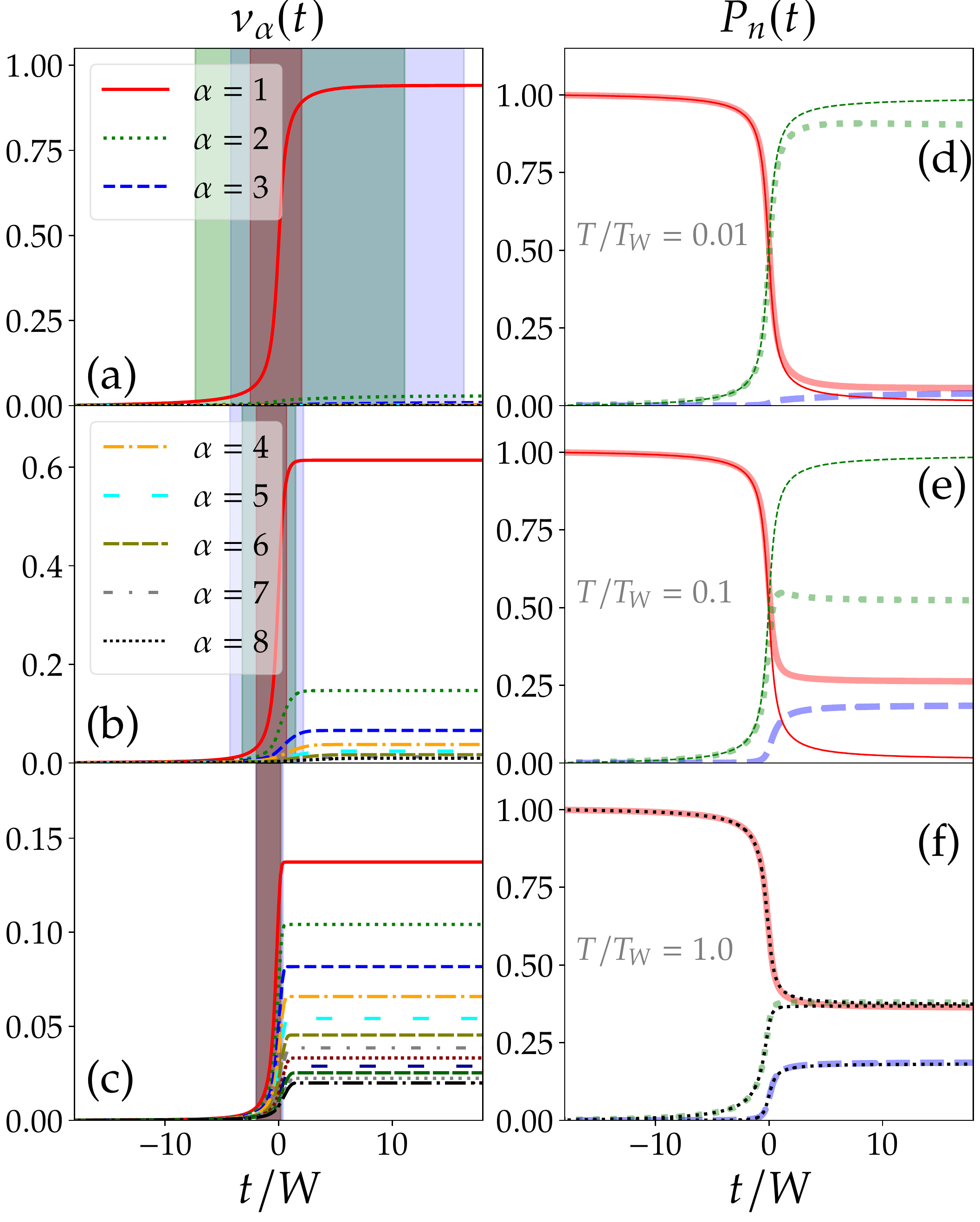}
  \caption{ (a-c) Emission probabilities $\nu_\alpha$ ($\alpha=1, 2, \dots, 8$) at different
    temperatures. The emission is driven by a single Lorentzian pulse carrying an unit electron
    charge. The red, green and blue regions correspond to the emission time windows of the first
    three electrons. (d-f) The corresponding distributions $P_n(t)$. The thick red solid, green
    dotted and blue dashed curves represent $P_0(t)$, $P_1(t)$ and $P_2(t)$, respectively. The thin
    red solid and green dashed curves in (d) and (e) represent the quantum limit of $P_0(t)$ and
    $P_1(t)$ from Eq.~\eqref{s1:eq1}. The black dashed curves in (f) represent the classical limit,
    corresponding to the Poisson process with emission rate $\lambda_c(t) = eV(t)/h$.}
  \label{fig10}
\end{figure}

Equation~\eqref{s4:eq10} corresponds to a mixed state. It indicates that a large number of electrons
can be emitted due to thermal fluctuations \cite{moskalets18_high_temper_fusion_multiel_levit}. Each
electron is emitted with the probability $p(\epsilon) = - \partial_{\epsilon} f(\epsilon)$, whose
wave function can be given as $\psi_i(t) e^{-i \epsilon t}$. As a consequence, even a single
Lorentzian pulse can trigger the emission of multiple electrons. This can be seen from the
time-resolved FCS, which is illustrated in Fig.~\ref{fig10}.  In the left column [(a-c)] of the
figure, we plot emission probabilities $\nu_\alpha(t)$ as functions of the normalized time
$t/W$. The three sub-figures (a), (b) and (c) correspond to $T/T_W = 0.01$, $T/T_W = 0.1$ and
$T/T_W = 1.0$, respectively. The corresponding distributions $P_n(t)$ are plotted in the right
column [(d-f)]. For better visualization, we only show the first three distributions $P_0(t)$,
$P_1(t)$ and $P_2(t)$, which are plotted by the thick red solid, green dotted and blue dashed
curves, respectively.

At low temperatures, the emission probability $\nu_1(t)$ plays the dominant role. For
$T/T_W = 0.01$, it can reach the maximum value $0.94$ in the long time limit $t \to +\infty$. This
is illustrated by the red solid curve in Fig.~\ref{fig10}(a). It increases from $5\%$ to $95 \%$ of
its maximum value for $t/W \in [-2.5, 2.03]$, indicating that the emission is concentrated in this
time window (marked by the red region). In contrast, other emission probabilities are rather
small. In Fig.~\ref{fig10}(a), one can only identify $\nu_2(t)$ (green dotted curve) and $\nu_3(t)$
(blue dashed curve), which merely reach $0.03$ and $0.01$ in the long time limit. The corresponding
emission time windows are $t/W \in [-7.34, 11.1]$ and $t/W \in [-4.22, 16.33]$, which are marked by
the green and blue regions in the figure. The two time windows are much wider than the emission time
window of the first electron, indicating that the emission of the second and third electrons are
less concentrated in the time domain. This results show that the emission due to thermal
fluctuations is weak at this temperature. In fact, the corresponding distribution $P_n(t)$ can still
be well-approximated by the quantum limit from Eq.~\eqref{s1:eq1}, which can be seen from
Fig.~\ref{fig10}(d).

As the temperature increases to $T/T_W = 0.1$, the emission probability $\nu_1(t)$ is suppressed,
which can only reach $0.61$ in the long time limit [red solid curve in Fig.~\ref{fig10}(b)]. It
increases from $5\%$ to $95 \%$ of its maximum value for $t/W \in [-1.95, 0.7]$ [red region in
Fig.~\ref{fig10}(b)]. Comparing to Fig.~\ref{fig10}(a), one can see that that the corresponding
emission time window is narrowed. In the meantime, the emission probabilities $\nu_2(t)$ and
$\nu_3(t)$ are enhanced, which can reach $0.15$ and $0.07$ in the long time limit. Their emission
time windows become $t/W \in [-3.2, 1.48]$ and $t/W \in [-4.3, 2.19]$, which are also narrowed
significantly. Comparing to Fig.~\ref{fig10}(a), one also finds that other emission probabilities
$\nu_\alpha(t)$ ($\alpha = 4, 5, \dots, 8$) can have non-negligible contributions. This makes the
distribution $P_n(t)$ departs from the quantum limit, as illustrated in Fig.~\ref{fig10}(e).

As the temperature $T$ further increases, more emission probabilities can contribute. This is
illustrated in Fig.~\ref{fig10}(c), corresponding to $T/T_W = 1.0$. From the figure, one finds that
the emission is not dominated by a single probability. Instead, there exits a large number of
emission probabilities, which are comparable in magnitude. This makes the corresponding distribution
$P_n(t)$ approaches the time-dependent Poisson distribution from Eq.~\eqref{s1:eq2}. This can be
seen from Fig.~\ref{fig10}(f), where the Poisson distribution is plotted by the black dotted
curves. At this temperature, the emission time windows of the first three electrons are almost
completely overlapped, which can be seen by comparing the red, green and blue regions in
Fig.~\ref{fig10}(c). This suggests that all the electron tends to be emitted in the same time window
at high temperatures.

\begin{figure}
  \centering
  \includegraphics[width=6.5cm]{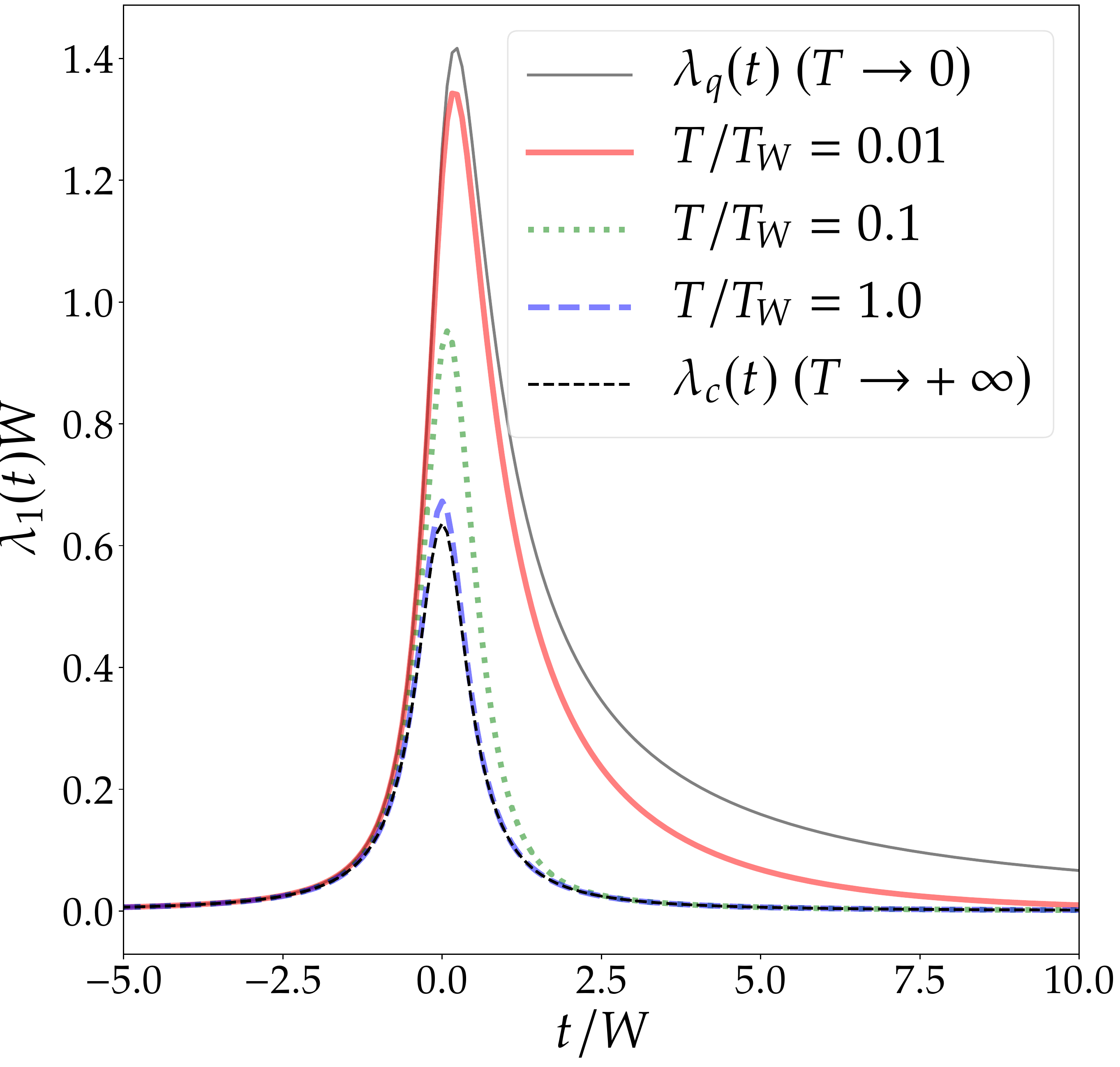}
  \caption{ Emission rate $\lambda_1(t)$ of the first electron at different temperatures. The gray
    solid curve corresponds to the quantum limit from Eq.~\eqref{s3:eq31}, while the black dashed
    curve corresponds to the classical limit $\lambda_c(t)=eV(t)/h$. }
  \label{fig11}
\end{figure}

The above results indicate that the system can be treated as a Poisson emitter at high temperatures,
whose emission rate follows the driving pulse as $\lambda_c(t) = eV(t)/h$. We find that the emission
rate $\lambda_1(t)$ of the first electron does evolve from the quantum limit $\lambda_q(t)$ from
Eq.~\eqref{s3:eq31} (gray solid curve) to the classical limit $\lambda_c(t)$ (black dotted curve) as
the temperature $T$ increases from zero to $T_W$, which is plotted in Fig.~\ref{fig11}.

\begin{figure}
  \centering
  \hspace{-1.0cm}\includegraphics[width=9cm]{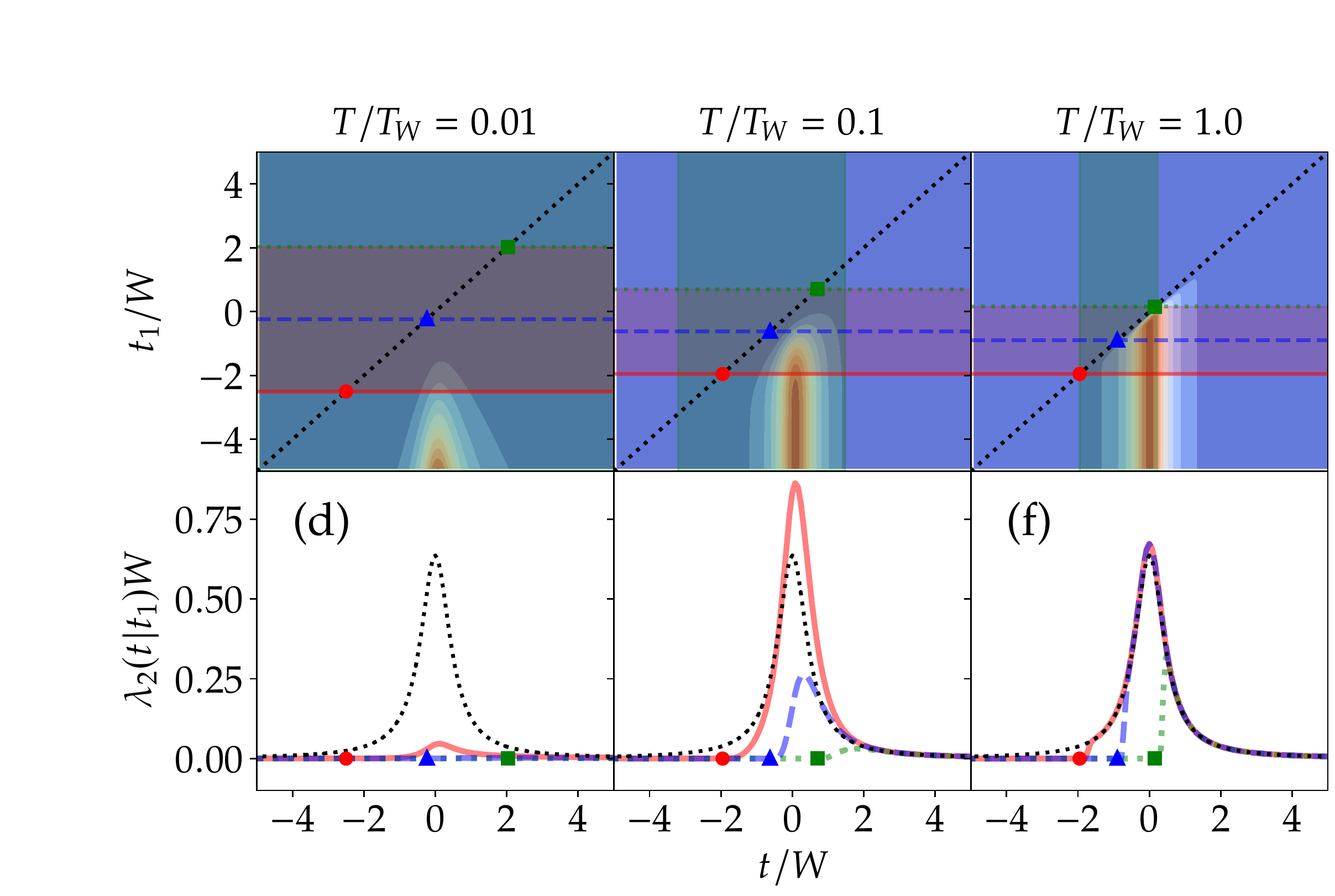}
  \caption{ (a-c) Emission rate $\lambda_2(t|t_1)$ at different temperatures, which are plotted as
    contour plots in the $t$-$t_1$ plane. The red and green regions mark the emission time windows
    of the first and second electrons. (d-f) $\lambda_2(t|t_1)$ as a functions of $t$ at different
    temperatures for three typical $t_1$ lies in the emission time windows of the first electron. In
    (d), the red solid, blue dashed and green dotted curves correspond to $t_1/W=-2.5$, $-0.23$ and
    $2.03$, respectively. In (e), the red solid, blue dashed and green dotted curves correspond to
    $t_1/W=-1.95$, $-0.63$ and $0.7$, respectively. In (e), the red solid, blue dashed and green
    dotted curves correspond to $t_1/W=-1.95$, $-0.9$ and $0.16$, respectively. The point $t=t_1$
    are marked by the red dots, blue triangles and green squares. The black dotted curves represent
    the classical limit $\lambda_c(t) = eV(t)/h$.}
  \label{fig12}
\end{figure}

However, the Poissonian character from $P_n(t)$ and $\lambda_1(t)$ does not mean that the emission
events are completely uncorrelated. This can be seen from the emission rate $\lambda_2(t|t_1)$ of
the second electron, which is illustrated in Fig.~\ref{fig12}. In the figure, we compare
$\lambda_2(t|t_1)$ at three typical temperatures $T/T_W = 0.01$ (a), $T/T_W = 0.1$ (b) and
$T/T_W = 1.0$ (c), which are plotted as contour plots in the $t$-$t_1$ plane. The emission time
windows for the first and second two electrons are also shown by the red and green regions in the
$t$-$t_1$ plane. As the emission of the second electron is the coeffect of the driving pulse and
thermal fluctuations, it remains very weak at low temperatures. This is illustrated in
Fig.~\ref{fig12}(a), corresponding to $T/T_W=0.01$. In this case, $\lambda_2(t|t_1)$ is much smaller
than the classical limit $\lambda_c(t)$. This can be seen from Fig.~\ref{fig12}(d), where we compare
$\lambda_2(t|t_1)$ with three typical times $t_1$ to the classical limit $\lambda_c(t)$ (black
dotted curve). Note that at this temperature, the emission time window of the second electron is
rather wide. It corresponds to $t/W \in [-7.34,11.1]$, which covers the whole range of $t/W$-axis.

The emission rate $\lambda_2(t|t_1)$ increases rapidly as the temperature reaches $T/T_W=0.1$. This
can be seen by comparing Fig.~\ref{fig12}(b) to Fig.~\ref{fig12}(a). This suggests that the emission
of the second electron is enhanced by thermal fluctuations. In the meantime, it is still strongly
correlated to the emission of the first electron. This is better illustrated in Fig.~\ref{fig12}(e),
where we plot $\lambda_2(t|t_1)$ with three typical times $t_1/W=-1.95$, $-0.63$ and $0.7$ by the
red solid, blue dashed and green dotted curves, respectively. One can see that $\lambda_2(t|t_1)$
depends strongly on the value of $t_1$. In particular, it drops rapidly to zero as $t$ approaches
$t_1$, where the points corresponding to $t = t_1$ are marked by the red dot ($t_1/W=-1.95$), blue
triangle ($t_1/W=-0.63$) and green square ($t_1/W=0.7$). This indicates that the correlation due to
the Pauli exclusion principle is still preserved at this temperature. The correlation remains
pronounced at high temperatures. This is demonstrated in Fig.~\ref{fig12}(c), corresponding to
$T/T_W = 1.0$. At this temperature, one finds that $\lambda_2(t|t_1)$ largely follows $\lambda_c(t)$
when $t > t_1$, but drops to zero rapidly as $t$ approaches $t_1$. The relation between
$\lambda_2(t|t_1)$ and $\lambda_c(t)$ is better illustrated in Fig.~\ref{fig12}(f), where we plot
$\lambda_2(t|t_1)$ with three typical $t_1$ by the red solid, green dotted and blue dashed curves,
while $\lambda_c(t)$ is plotted by the black dotted curve for comparison.

\begin{figure}
  \centering
  \vspace{0.5cm}
  \includegraphics[width=7.5cm]{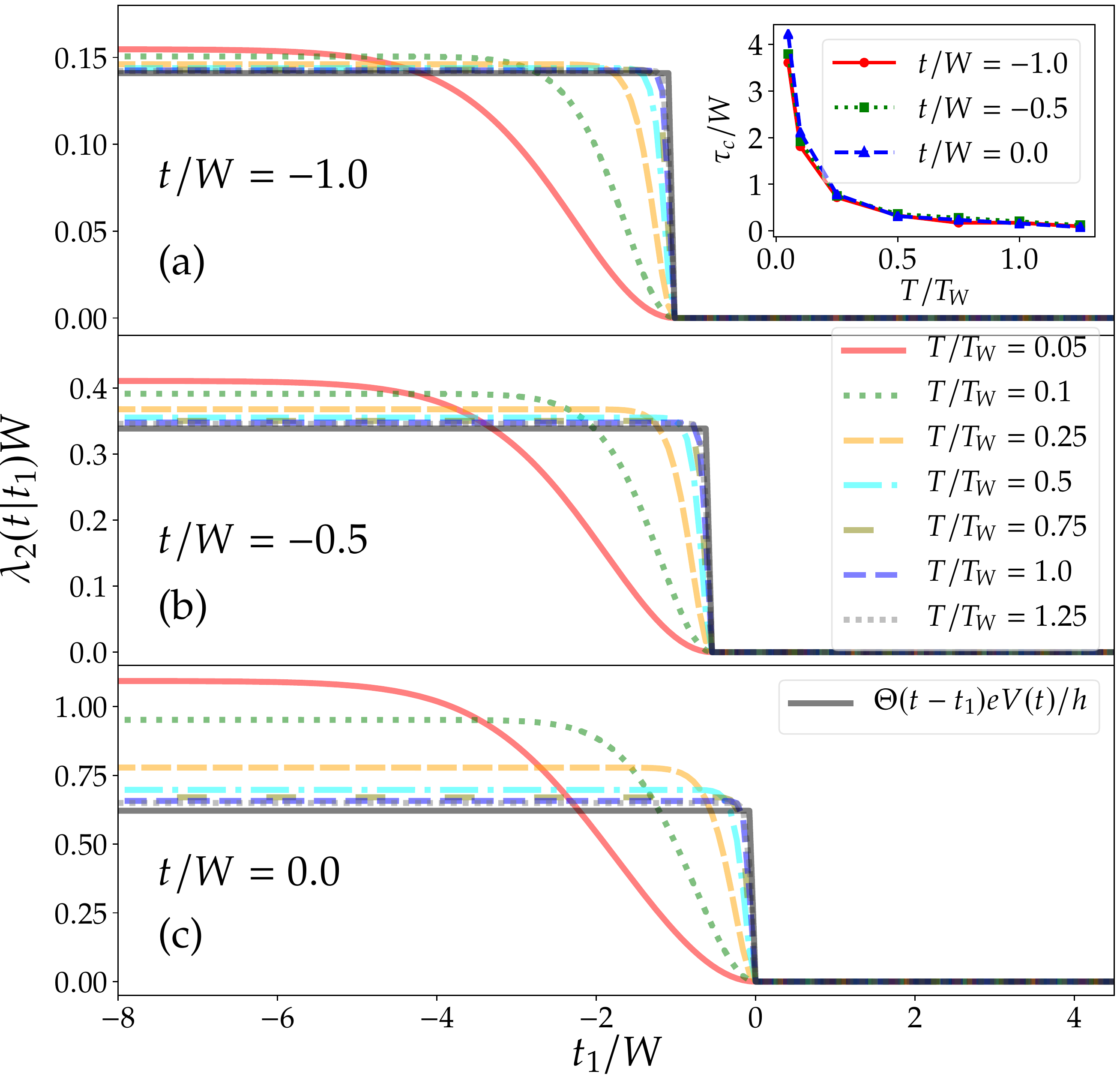}
  \caption{ Emission rate $\lambda_2(t|t_1)$ as a function of $t_1/W$ for three typical $t/W=-1.0$
    (a), $-0.5$ (b) and $0.0$ (c). Curves with different colors and line types correspond to
    different temperatures. The gray solid curve represent the classical limit
    $\Theta(t-t_1) eV(t)/h$. The inset shows the correlation times $\tau_c$ as a function of
    temperature $T/T_W$.}
  \label{fig13}
\end{figure}

Although the correlations are always present, they can only play a non-negligible role at short
times. To better seen this, we plot the emission rates $\lambda_2(t|t_1)$ as a function of $t_1/W$
in Fig.~\ref{fig13}, where subfigures (a), (b) and (c) correspond to three typical emission time
$t/W = -1.0$, $-0.5$ and $0.0$, respectively. One can see that $t_1$ can only affect the emission
rate $\lambda_2(t|t_1)$ when $t_1$ is close enough to $t$. If $t_1$ is too far from $t$, \ie,
$t - t_1 > \tau_c$, $\lambda_2(t|t_1)$ is almost a constant, indicating the absence of the
correlation at long times. Here $\tau_c$ can be regarded as a correlation time, which should depends
on both the temperature $T$ and the emission time $t$. We estimate the value of $\tau_c$ by
requiring that
\begin{equation}
  \frac{\lambda_2(t|t_1 \to -\infty) - \lambda_2(t|t_1)}{\lambda_2(t|t_1 \to -\infty)} < 0.05
\end{equation}
when $t - t_1 > \tau_c$. In doing so, we obtain $\tau_c$ as a function of normalized temperature
$T/T_W$ for three typical emission time $t$. This is shown in the inset of Fig.~\ref{fig13}, where
different curves corresponds to different $t$. One can see that $\tau_c$ for different $t$ exhibits
similar temperature dependence, which drops to zero rapidly as the temperature $T/T_W$
increases. This results show that, although the correlations are always present, they are difficult
to be observed at high temperatures due to the short correlation times. As a consequence, the
emission rate $\lambda_2(t|t_1)$ tends to follow the classical limit $\Theta(t-t_1) eV(t)/h$, which
is plotted by the gray solid curves in Fig.~\ref{fig13}.

The above results show that thermal fluctuations in the electrode can suppress the correlations,
making the emission process behaves like a time-dependence Poisson process at long times. As these
results are obtained in the case of single Lorentzian pulse, one may wonder if such conclusion still
holds in more general cases. To further confirm our conclusion, in the next subsection we consider
the case when a Lorentzian pulse carrying two electric charges are applied on the electrode.

\subsection{Two-electron emitter}
\label{sec4b}

\begin{figure}
  \centering
  \includegraphics[width=7.5cm]{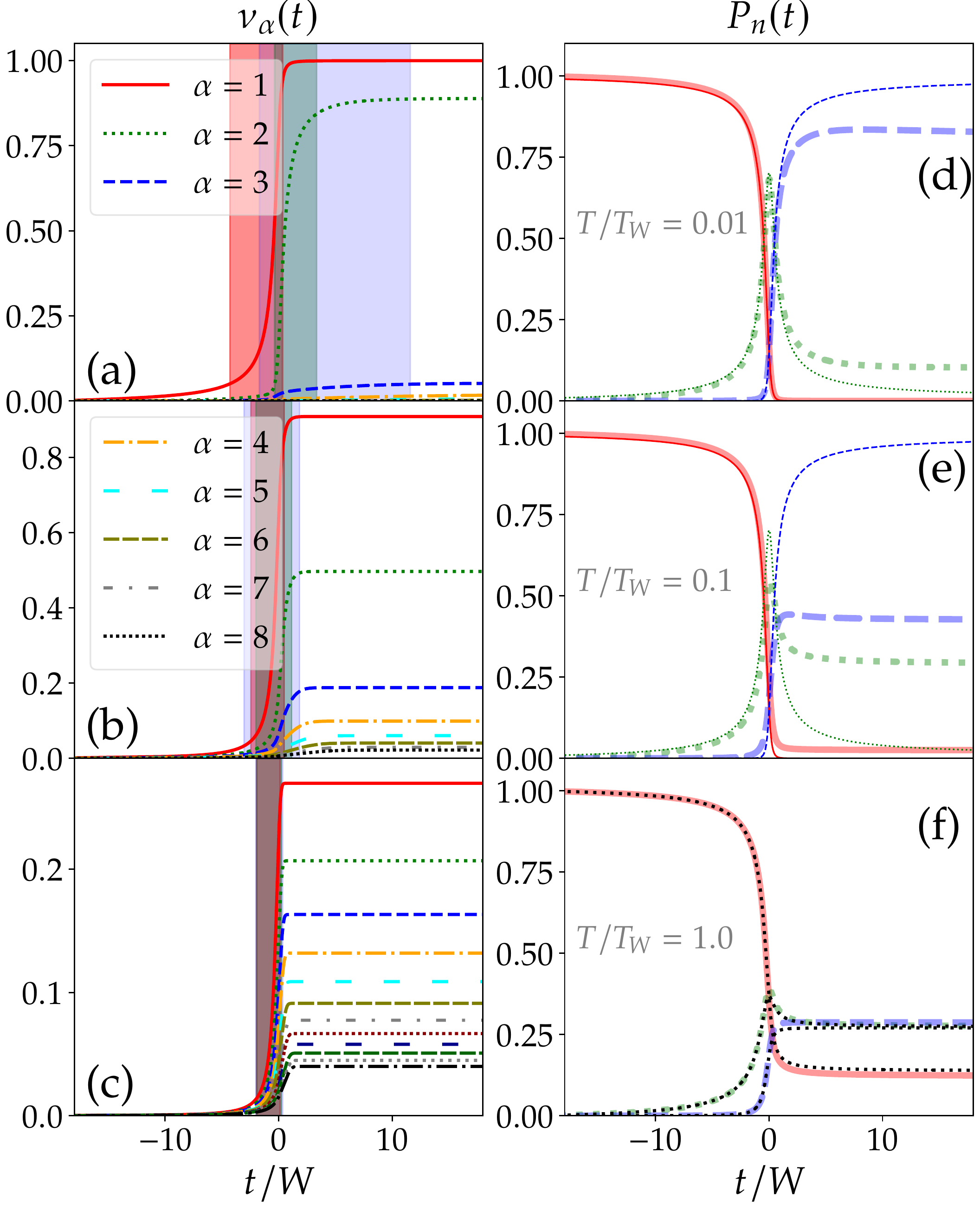}
  \caption{ The same as Fig.~\ref{fig10} but with the driving pulse carrying two electron charge. In
    this case, the quantum limit of $P_n(t)$ is obtained from Eq.~\eqref{s3:eq80}. The thin red
    solid, green dotted and blue dashed curves in (d) and (e) correspond to $P_0(t)$, $P_1(t)$ and
    $P_2(t)$, respectively.}
  \label{fig14}
\end{figure}

This corresponds to the case $t_c/W = 0$ discussed in Sec.~\ref{sec3b}. The typical behavior of the
FCS is demonstrated in Fig.~\ref{fig14}. Comparing to Fig.~\ref{fig10}, one can identify the same
features as the temperature $T$ reaches $T_W$: 1) The distribution $P_n(t)$ evolves from the quantum
limit [thin colored curves in Fig.~\ref{fig14}(d) and (e)] to the classical limit [thin dotted
curves in Fig.~\ref{fig14}(f)]. Note that in this case, the quantum limit of $P_n(t)$ is calculated
from Eq.~\eqref{s3:eq80}. There are three non-zero distributions $P_0(t)$, $P_1(t)$ and $P_2(t)$,
which are plotted by thin red solid, green dotted and blue dashed curves, respectively. 2) Multiple
electrons can be emitted due to thermal fluctuations, which tends to be emitted in the same time
window [the colored region in Fig.~\ref{fig14}(c)]. These features indicates that the
quantum-to-classical crossover exhibit same behavior as in the case of unit-charged pulse.

\begin{figure}
  \centering  
  \includegraphics[width=6.5cm]{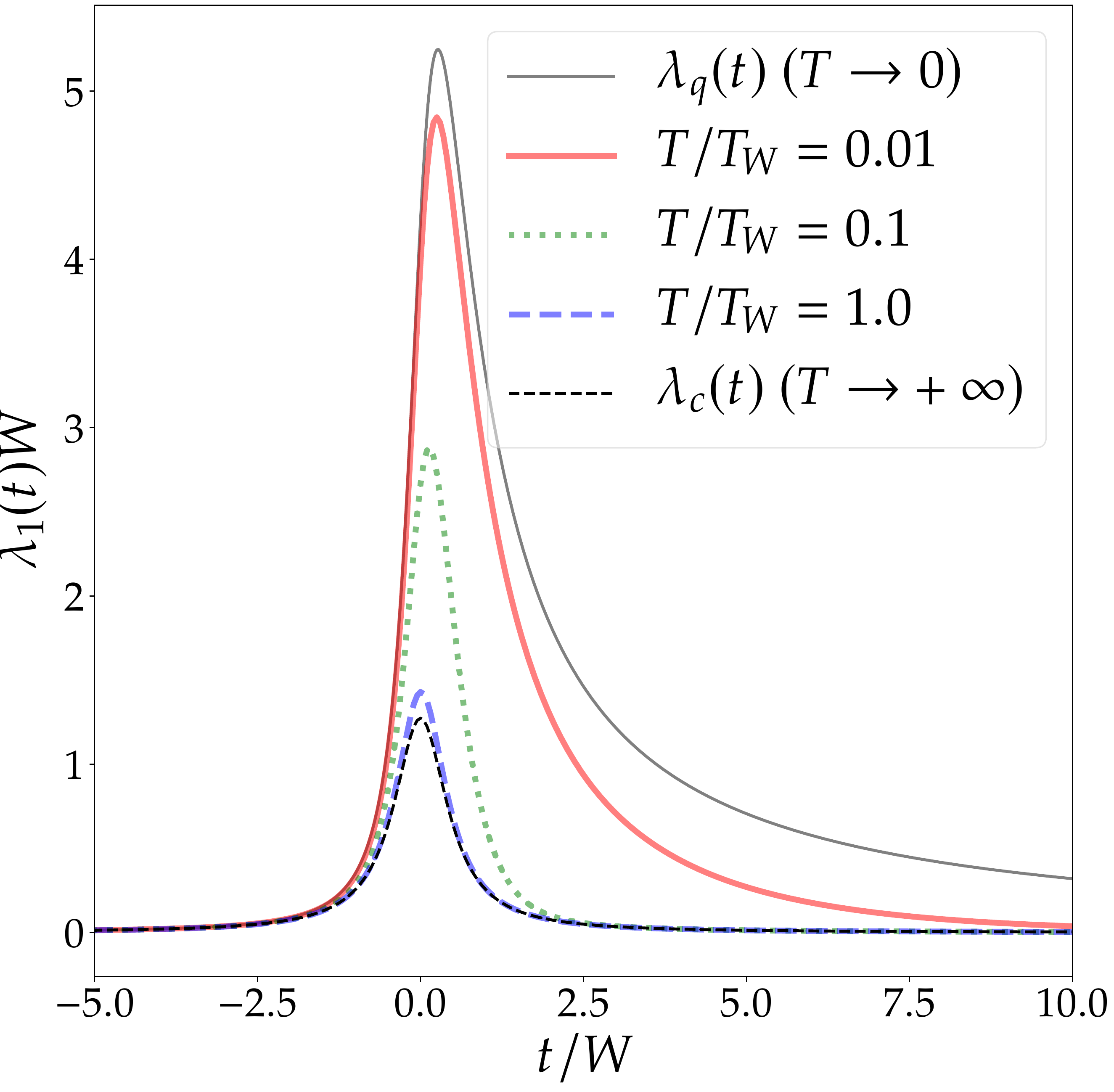}
  \caption{ The same as Fig.~\ref{fig11} but with the driving pulse carrying two electron charge. In
    this case, the classical limit $\lambda_c(t)$ still follows $eV(t)/h$, but the quantum limit
    $\lambda_q(t)$ is determined from Eq.~\eqref{s3:eq101-1}.}
  \label{fig15}
\end{figure}

\begin{figure}
  \centering
  \hspace{-1.0cm}\includegraphics[width=9cm]{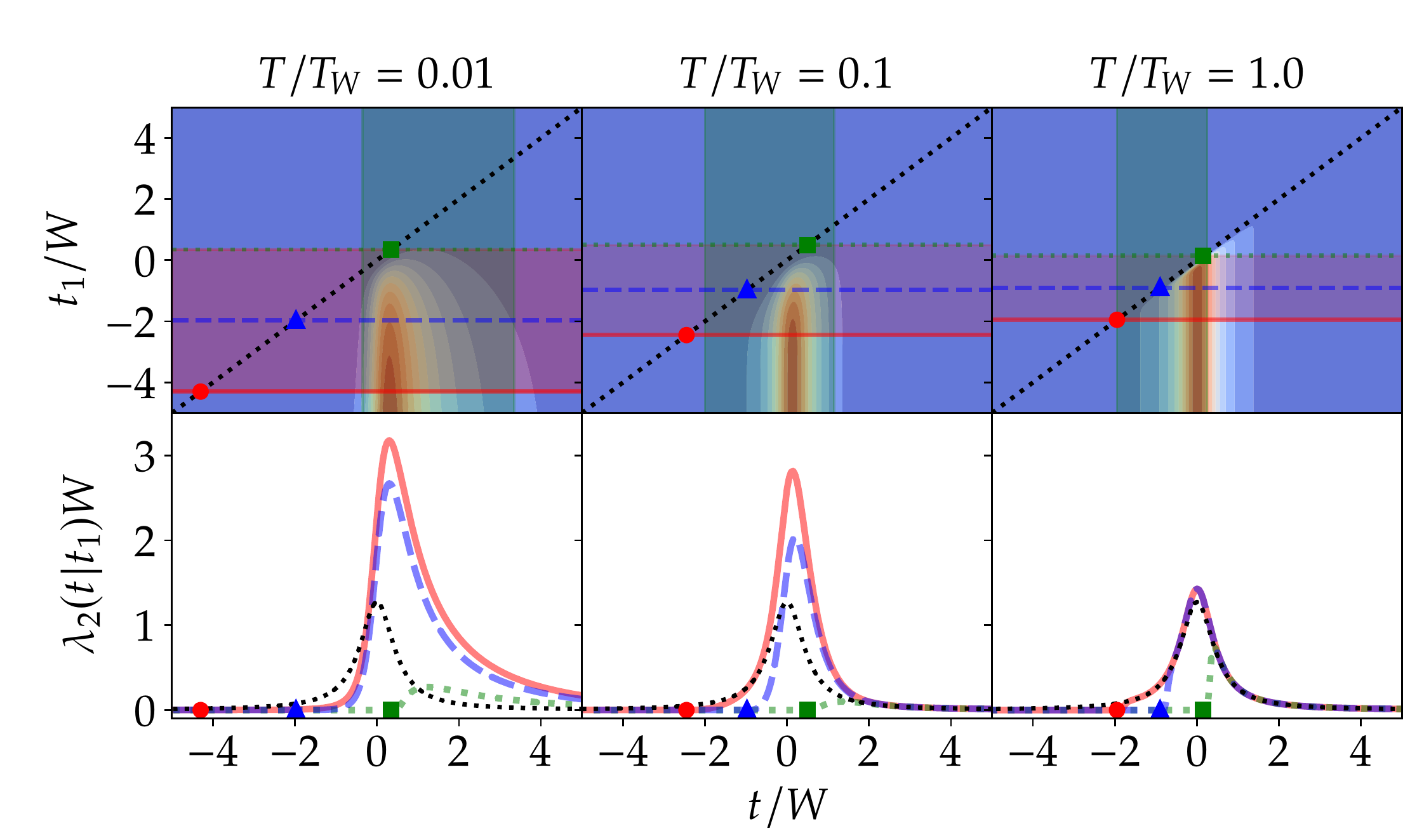}  
  \caption{ The same as Fig.~\ref{fig12} but with the driving pulse carrying two electron charge.}
  \label{fig16}
\end{figure}

\begin{figure}
  \centering
  \includegraphics[width=7.5cm]{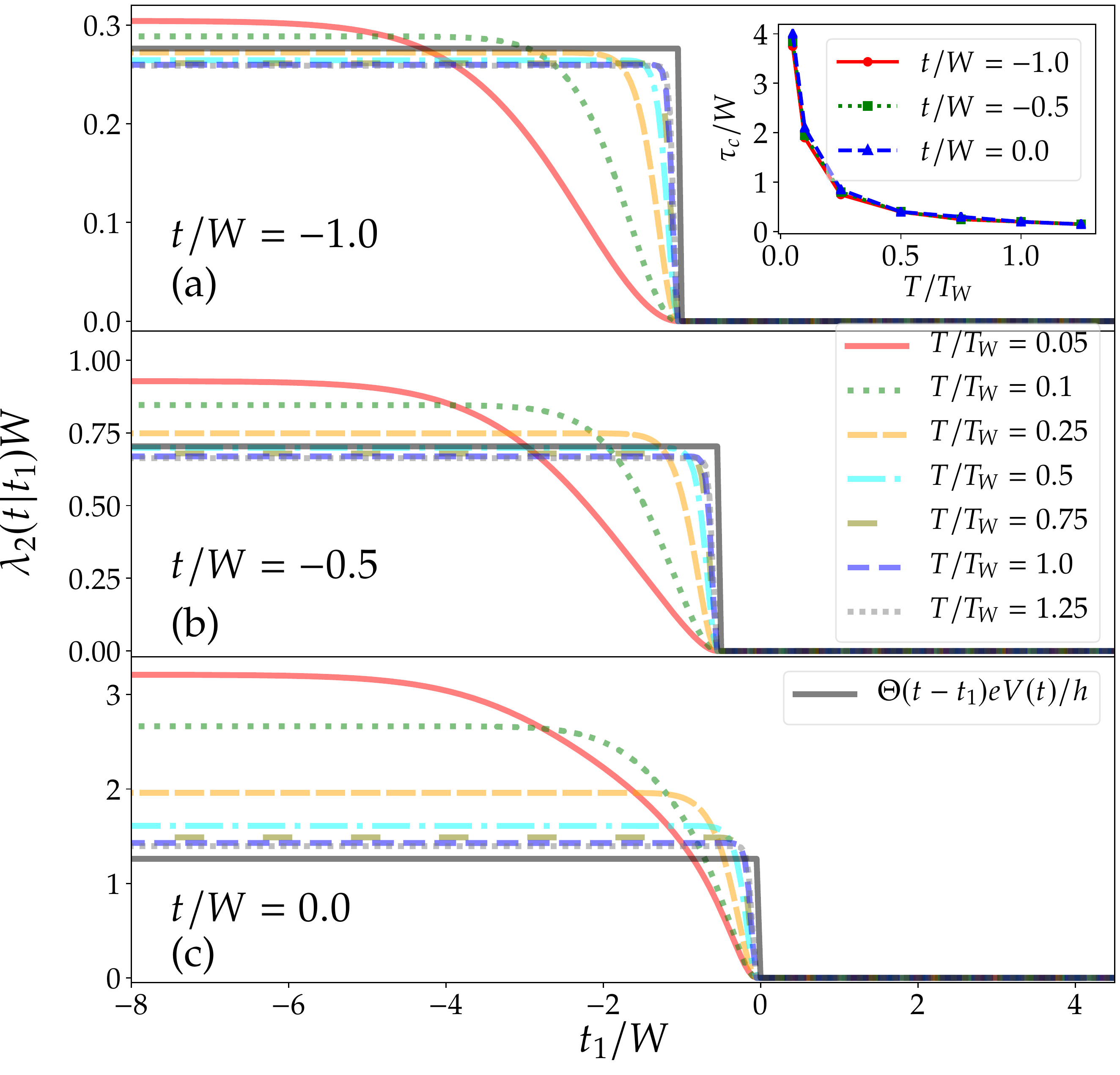}
  \caption{ The same as Fig.~\ref{fig13} but with the driving pulse carrying two electron charge. }
  \label{fig17}
\end{figure}

\begin{figure}
  \centering
  \includegraphics[width=6.5cm]{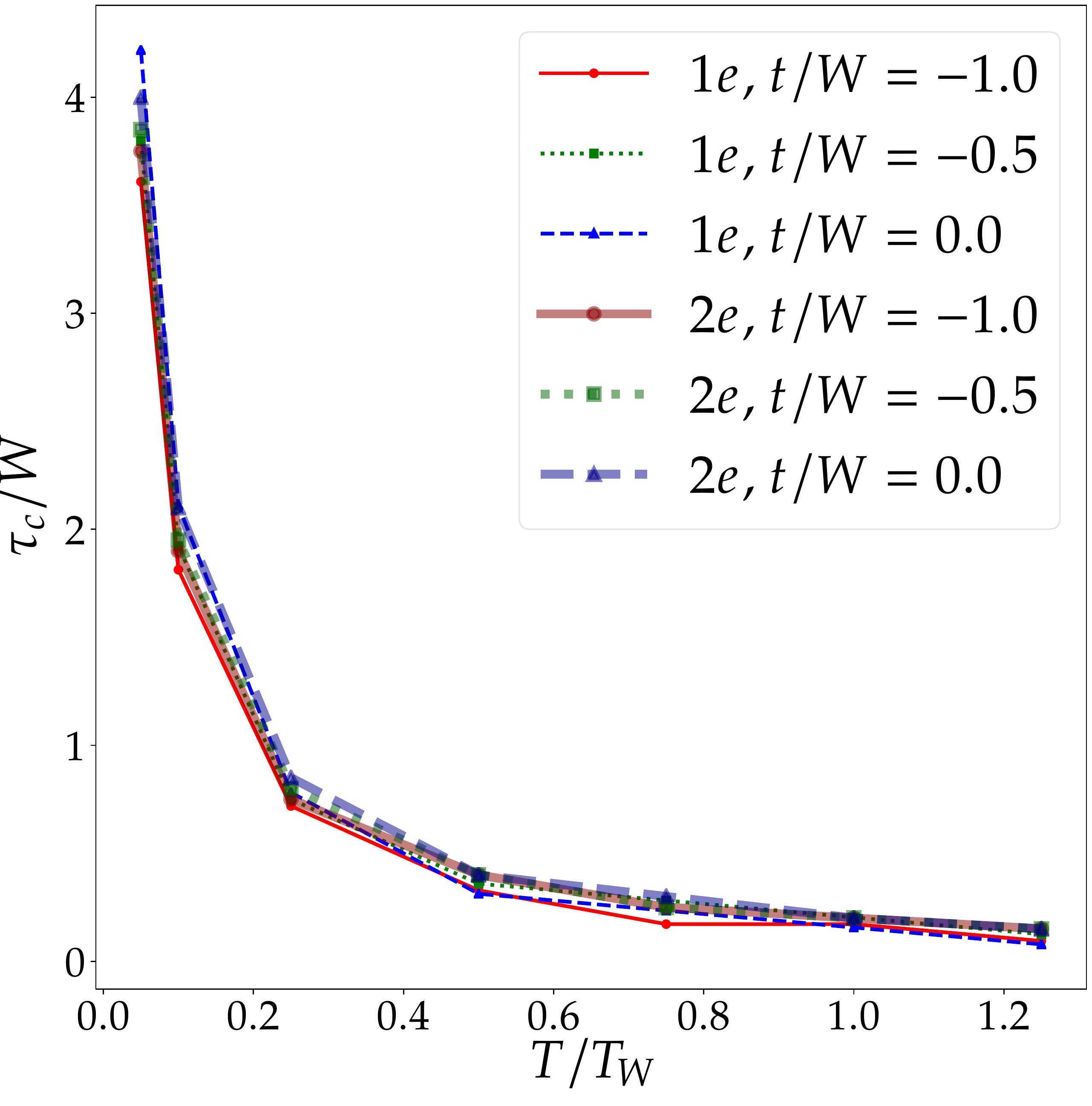}
  \caption{ The correlation time $\tau_c$ as a function of temperature $T/T_W$. Curves with the
    label ``$1e$'' represent the correlation times from the inset of Fig.~\ref{fig13}, while curves
    with the label ``$2e$" represent the correlation times from the inset of Fig.~\ref{fig17}.}
  \label{fig18}
\end{figure}

The similarity of the crossover can also be seen from the emission rates. In Fig.~\ref{fig15}. One
can see that the emission rate $\lambda_1(t)$ of the first electron evolves from the quantum limit
to the classical limit $\lambda_c(t) = eV(t)/h$ as temperature $T$ approaches $T_W$. Note that in
this case, the quantum limit $\lambda_q(t)$ is calculated from Eq.~\eqref{s3:eq101-1}. In the
meantime, the emission rate $\lambda_2(t|t_1)$ of the second electron tends to follow the classical
counterpart $\Theta(t-t_1) eV(t)/h$, which is illustrated in Fig.~\ref{fig16}
and~\ref{fig17}. Moreover, we also compare the temperature-dependence correlations time $\tau_c$ in
the two cases, which are illustrated in Fig.~\ref{fig18}. In the figure, the curves with the label
``$1e$'' represent the correlation times from the inset of Fig.~\ref{fig13}, while curves with the
label ``$2e$" represent the correlation times from the inset of Fig.~\ref{fig17}. One finds that
they tends to follow the same temperature dependence. These results indicates that the
quantum-to-classical crossover exhibit similar behavior, despite the voltage pulse carrying one or
two electron charges.

\section{Summary and Discussion}
\label{sec5}

In summary, we have investigated the quantum-to-classical crossover in a single-electron emitter,
which is driven by electrode temperature. By combining the time-resolved FCS and emission rates, we
have shown that the emission approaches a time-dependence Poisson process at long times. In
contrast, the correlation between emission events remains at short times, which is due to the Pauli
exclusion principle. This behavior is demonstrated in two cases, when the electron emission is
driven by a Lorentzian pulse carrying a single and two electron charges.

It is worth noting that the quantum-to-classical crossover has also been studied previously in
another type of single-electron emitter \cite{kashcheyevs17_class_to_quant_cross_elect_deman_emiss},
which is based on a driven localized state \cite{feve-2007-deman-coher,
  fletcher12_clock_contr_emiss_singl_elect, ubbelohde14_partit_deman_elect_pairs}. Instead of
increasing the temperature, the quantum-to-classical crossover in such emitter is realized by
manipulating the driving rate of the localized state. The quantum-to-classical crossover is
characterized via the Wigner function \cite{wigner32_quant_correc_therm_equil,
  ferraro-2013-wigner-funct}. The emitted electron from such emitter can have a energy which is far
above the Fermi level, whose Wigner function can be partially resolved from a continuous-variable
tomography techniques \cite{fletcher19_contin_variab_tomog_solit_elect}. It will be interesting to
study the quantum-to-classical crossover in such emitter by using our approach, which we will
explore in future works.

\begin{acknowledgments}
  The authors would like to thank Professor M. V. Moskalets for helpful comments and
  discussion. This work was partially supported by the National Key Research and Development Program
  of China under Grant No. 2022YFF0608302 and SCU Innovation Fund under Grant No. 2020SCUNL209.
\end{acknowledgments}




\bibliographystyle{apsrev4-2}

\begin{thebibliography}{53}%
\makeatletter
\providecommand \@ifxundefined [1]{%
 \@ifx{#1\undefined}
}%
\providecommand \@ifnum [1]{%
 \ifnum #1\expandafter \@firstoftwo
 \else \expandafter \@secondoftwo
 \fi
}%
\providecommand \@ifx [1]{%
 \ifx #1\expandafter \@firstoftwo
 \else \expandafter \@secondoftwo
 \fi
}%
\providecommand \natexlab [1]{#1}%
\providecommand \enquote  [1]{``#1''}%
\providecommand \bibnamefont  [1]{#1}%
\providecommand \bibfnamefont [1]{#1}%
\providecommand \citenamefont [1]{#1}%
\providecommand \href@noop [0]{\@secondoftwo}%
\providecommand \href [0]{\begingroup \@sanitize@url \@href}%
\providecommand \@href[1]{\@@startlink{#1}\@@href}%
\providecommand \@@href[1]{\endgroup#1\@@endlink}%
\providecommand \@sanitize@url [0]{\catcode `\\12\catcode `\$12\catcode
  `\&12\catcode `\#12\catcode `\^12\catcode `\_12\catcode `\%12\relax}%
\providecommand \@@startlink[1]{}%
\providecommand \@@endlink[0]{}%
\providecommand \url  [0]{\begingroup\@sanitize@url \@url }%
\providecommand \@url [1]{\endgroup\@href {#1}{\urlprefix }}%
\providecommand \urlprefix  [0]{URL }%
\providecommand \Eprint [0]{\href }%
\providecommand \doibase [0]{https://doi.org/}%
\providecommand \selectlanguage [0]{\@gobble}%
\providecommand \bibinfo  [0]{\@secondoftwo}%
\providecommand \bibfield  [0]{\@secondoftwo}%
\providecommand \translation [1]{[#1]}%
\providecommand \BibitemOpen [0]{}%
\providecommand \bibitemStop [0]{}%
\providecommand \bibitemNoStop [0]{.\EOS\space}%
\providecommand \EOS [0]{\spacefactor3000\relax}%
\providecommand \BibitemShut  [1]{\csname bibitem#1\endcsname}%
\let\auto@bib@innerbib\@empty
\bibitem [{\citenamefont {Glattli}\ and\ \citenamefont
  {Roulleau}(2016{\natexlab{a}})}]{glattli-2016-levit-elect}%
  \BibitemOpen
  \bibfield  {author} {\bibinfo {author} {\bibfnamefont {D.~C.}\ \bibnamefont
  {Glattli}}\ and\ \bibinfo {author} {\bibfnamefont {P.~S.}\ \bibnamefont
  {Roulleau}},\ }\href {https://doi.org/10.1002/pssb.201600650} {\bibfield
  {journal} {\bibinfo  {journal} {Phys. Status Solidi B}\ }\textbf
  {\bibinfo {volume} {254}},\ \bibinfo {pages} {1600650} (\bibinfo {year}
  {2016}{\natexlab{a}})}\BibitemShut {NoStop}%
\bibitem [{\citenamefont {B{\"a}uerle}\ \emph {et~al.}(2018)\citenamefont
  {B{\"a}uerle}, \citenamefont {Glattli}, \citenamefont {Meunier},
  \citenamefont {Portier}, \citenamefont {Roche}, \citenamefont {Roulleau},
  \citenamefont {Takada},\ and\ \citenamefont
  {Waintal}}]{baeuerle-2018-coher-contr}%
  \BibitemOpen
  \bibfield  {author} {\bibinfo {author} {\bibfnamefont {C.}~\bibnamefont
  {B{\"a}uerle}}, \bibinfo {author} {\bibfnamefont {D.~C.}\ \bibnamefont
  {Glattli}}, \bibinfo {author} {\bibfnamefont {T.}~\bibnamefont {Meunier}},
  \bibinfo {author} {\bibfnamefont {F.}~\bibnamefont {Portier}}, \bibinfo
  {author} {\bibfnamefont {P.}~\bibnamefont {Roche}}, \bibinfo {author}
  {\bibfnamefont {P.}~\bibnamefont {Roulleau}}, \bibinfo {author}
  {\bibfnamefont {S.}~\bibnamefont {Takada}},\ and\ \bibinfo {author}
  {\bibfnamefont {X.}~\bibnamefont {Waintal}},\ }\href
  {https://doi.org/10.1088/1361-6633/aaa98a} {\bibfield  {journal} {\bibinfo
  {journal} {Rep. Prog. Phys.}\ }\textbf {\bibinfo {volume}
  {81}},\ \bibinfo {pages} {056503} (\bibinfo {year} {2018})}\BibitemShut
  {NoStop}%
\bibitem [{\citenamefont {Samuelsson}\ and\ \citenamefont
  {B{\"u}ttiker}(2006)}]{samuelsson-2006-quant-state}%
  \BibitemOpen
  \bibfield  {author} {\bibinfo {author} {\bibfnamefont {P.}~\bibnamefont
  {Samuelsson}}\ and\ \bibinfo {author} {\bibfnamefont {M.}~\bibnamefont
  {B{\"u}ttiker}},\ }\href {https://doi.org/10.1103/physrevb.73.041305}
  {\bibfield  {journal} {\bibinfo  {journal} {Phys. Rev. B}\ }\textbf
  {\bibinfo {volume} {73}},\ \bibinfo {pages} {041305(R)} (\bibinfo {year}
  {2006})}\BibitemShut {NoStop}%
\bibitem [{\citenamefont {Grenier}\ \emph {et~al.}(2011)\citenamefont
  {Grenier}, \citenamefont {Herv{\'e}}, \citenamefont {Bocquillon},
  \citenamefont {Parmentier}, \citenamefont {Pla{\c{c}}ais}, \citenamefont
  {Berroir}, \citenamefont {F{\`e}ve},\ and\ \citenamefont
  {Degiovanni}}]{grenier-2011-singl-elect}%
  \BibitemOpen
  \bibfield  {author} {\bibinfo {author} {\bibfnamefont {C.}~\bibnamefont
  {Grenier}}, \bibinfo {author} {\bibfnamefont {R.}~\bibnamefont {Herv{\'e}}},
  \bibinfo {author} {\bibfnamefont {E.}~\bibnamefont {Bocquillon}}, \bibinfo
  {author} {\bibfnamefont {F.~D.}\ \bibnamefont {Parmentier}}, \bibinfo
  {author} {\bibfnamefont {B.}~\bibnamefont {Pla{\c{c}}ais}}, \bibinfo {author}
  {\bibfnamefont {J.~M.}\ \bibnamefont {Berroir}}, \bibinfo {author}
  {\bibfnamefont {G.}~\bibnamefont {F{\`e}ve}},\ and\ \bibinfo {author}
  {\bibfnamefont {P.}~\bibnamefont {Degiovanni}},\ }\href
  {https://doi.org/10.1088/1367-2630/13/9/093007} {\bibfield  {journal}
  {\bibinfo  {journal} {New J. Phys.}\ }\textbf {\bibinfo {volume}
  {13}},\ \bibinfo {pages} {093007} (\bibinfo {year} {2011})}\BibitemShut
  {NoStop}%
\bibitem [{\citenamefont {Jullien}\ \emph {et~al.}(2014)\citenamefont
  {Jullien}, \citenamefont {Roulleau}, \citenamefont {Roche}, \citenamefont
  {Cavanna}, \citenamefont {Jin},\ and\ \citenamefont
  {Glattli}}]{jullien-2014-quant-tomog-elect}%
  \BibitemOpen
  \bibfield  {author} {\bibinfo {author} {\bibfnamefont {T.}~\bibnamefont
  {Jullien}}, \bibinfo {author} {\bibfnamefont {P.}~\bibnamefont {Roulleau}},
  \bibinfo {author} {\bibfnamefont {B.}~\bibnamefont {Roche}}, \bibinfo
  {author} {\bibfnamefont {A.}~\bibnamefont {Cavanna}}, \bibinfo {author}
  {\bibfnamefont {Y.}~\bibnamefont {Jin}},\ and\ \bibinfo {author}
  {\bibfnamefont {D.~C.}\ \bibnamefont {Glattli}},\ }\href
  {http://dx.doi.org/10.1038/nature13821} {\bibfield  {journal} {\bibinfo
  {journal} {Nature}\ }\textbf {\bibinfo {volume} {514}},\ \bibinfo {pages}
  {603} (\bibinfo {year} {2014})}\BibitemShut {NoStop}%
\bibitem [{\citenamefont {Bisognin}\ \emph {et~al.}(2019)\citenamefont
  {Bisognin}, \citenamefont {Marguerite}, \citenamefont {Roussel},
  \citenamefont {Kumar}, \citenamefont {Cabart}, \citenamefont {Chapdelaine},
  \citenamefont {Mohammad-Djafari}, \citenamefont {Berroir}, \citenamefont
  {Bocquillon}, \citenamefont {Pla{\c{c}}ais}, \citenamefont {Cavanna},
  \citenamefont {Gennser}, \citenamefont {Jin}, \citenamefont {Degiovanni},\
  and\ \citenamefont {F{\`e}ve}}]{bisognin-2019-quant-tomog}%
  \BibitemOpen
  \bibfield  {author} {\bibinfo {author} {\bibfnamefont {R.}~\bibnamefont
  {Bisognin}}, \bibinfo {author} {\bibfnamefont {A.}~\bibnamefont
  {Marguerite}}, \bibinfo {author} {\bibfnamefont {B.}~\bibnamefont {Roussel}},
  \bibinfo {author} {\bibfnamefont {M.}~\bibnamefont {Kumar}}, \bibinfo
  {author} {\bibfnamefont {C.}~\bibnamefont {Cabart}}, \bibinfo {author}
  {\bibfnamefont {C.}~\bibnamefont {Chapdelaine}}, \bibinfo {author}
  {\bibfnamefont {A.}~\bibnamefont {Mohammad-Djafari}}, \bibinfo {author}
  {\bibfnamefont {J.-M.}\ \bibnamefont {Berroir}}, \bibinfo {author}
  {\bibfnamefont {E.}~\bibnamefont {Bocquillon}}, \bibinfo {author}
  {\bibfnamefont {B.}~\bibnamefont {Pla{\c{c}}ais}}, \bibinfo {author}
  {\bibfnamefont {A.}~\bibnamefont {Cavanna}}, \bibinfo {author} {\bibfnamefont
  {U.}~\bibnamefont {Gennser}}, \bibinfo {author} {\bibfnamefont
  {Y.}~\bibnamefont {Jin}}, \bibinfo {author} {\bibfnamefont {P.}~\bibnamefont
  {Degiovanni}},\ and\ \bibinfo {author} {\bibfnamefont {G.}~\bibnamefont
  {F{\`e}ve}},\ }\href {https://doi.org/10.1038/s41467-019-11369-5} {\bibfield
  {journal} {\bibinfo  {journal} {Nat. Commun.}\ }\textbf {\bibinfo
  {volume} {10}},\ \bibinfo {pages} {3379} (\bibinfo {year}
  {2019})}\BibitemShut {NoStop}%
\bibitem [{\citenamefont {Roussel}\ \emph {et~al.}(2021)\citenamefont
  {Roussel}, \citenamefont {Cabart}, \citenamefont {F{\`e}ve},\ and\
  \citenamefont
  {Degiovanni}}]{roussel21_proces_quant_signal_carried_by_elect_curren}%
  \BibitemOpen
  \bibfield  {author} {\bibinfo {author} {\bibfnamefont {B.}~\bibnamefont
  {Roussel}}, \bibinfo {author} {\bibfnamefont {C.}~\bibnamefont {Cabart}},
  \bibinfo {author} {\bibfnamefont {G.}~\bibnamefont {F{\`e}ve}},\ and\
  \bibinfo {author} {\bibfnamefont {P.}~\bibnamefont {Degiovanni}},\ }\href
  {https://doi.org/10.1103/prxquantum.2.020314} {\bibfield  {journal} {\bibinfo
   {journal} {PRX Quantum}\ }\textbf {\bibinfo {volume} {2}},\ \bibinfo {pages}
  {020314} (\bibinfo {year} {2021})}\BibitemShut {NoStop}%
\bibitem [{\citenamefont {Keeling}\ \emph {et~al.}(2006)\citenamefont
  {Keeling}, \citenamefont {Klich},\ and\ \citenamefont
  {Levitov}}]{keeling-2006-minim-excit}%
  \BibitemOpen
  \bibfield  {author} {\bibinfo {author} {\bibfnamefont {J.}~\bibnamefont
  {Keeling}}, \bibinfo {author} {\bibfnamefont {I.}~\bibnamefont {Klich}},\
  and\ \bibinfo {author} {\bibfnamefont {L.~S.}\ \bibnamefont {Levitov}},\
  }\href {https://doi.org/10.1103/PhysRevLett.97.116403} {\bibfield  {journal}
  {\bibinfo  {journal} {Phys. Rev. Lett.}\ }\textbf {\bibinfo {volume}
  {97}},\ \bibinfo {pages} {116403} (\bibinfo {year} {2006})}\BibitemShut
  {NoStop}%
\bibitem [{\citenamefont {Dubois}\ \emph
  {et~al.}(2013{\natexlab{a}})\citenamefont {Dubois}, \citenamefont {Jullien},
  \citenamefont {Portier}, \citenamefont {Roche}, \citenamefont {Cavanna},
  \citenamefont {Jin}, \citenamefont {Wegscheider}, \citenamefont {Roulleau},\
  and\ \citenamefont {Glattli}}]{dubois-2013-minim-excit}%
  \BibitemOpen
  \bibfield  {author} {\bibinfo {author} {\bibfnamefont {J.}~\bibnamefont
  {Dubois}}, \bibinfo {author} {\bibfnamefont {T.}~\bibnamefont {Jullien}},
  \bibinfo {author} {\bibfnamefont {F.}~\bibnamefont {Portier}}, \bibinfo
  {author} {\bibfnamefont {P.}~\bibnamefont {Roche}}, \bibinfo {author}
  {\bibfnamefont {A.}~\bibnamefont {Cavanna}}, \bibinfo {author} {\bibfnamefont
  {Y.}~\bibnamefont {Jin}}, \bibinfo {author} {\bibfnamefont {W.}~\bibnamefont
  {Wegscheider}}, \bibinfo {author} {\bibfnamefont {P.}~\bibnamefont
  {Roulleau}},\ and\ \bibinfo {author} {\bibfnamefont {D.~C.}\ \bibnamefont
  {Glattli}},\ }\href {http://dx.doi.org/10.1038/nature12713} {\bibfield
  {journal} {\bibinfo  {journal} {Nature}\ }\textbf {\bibinfo {volume} {502}},\
  \bibinfo {pages} {659} (\bibinfo {year} {2013}{\natexlab{a}})}\BibitemShut
  {NoStop}%
\bibitem [{\citenamefont {Moskalets}\ and\ \citenamefont
  {Haack}(2016)}]{moskalets16_singl_elect_coher}%
  \BibitemOpen
  \bibfield  {author} {\bibinfo {author} {\bibfnamefont {M.}~\bibnamefont
  {Moskalets}}\ and\ \bibinfo {author} {\bibfnamefont {G.}~\bibnamefont
  {Haack}},\ }\href
  {https://doi.org/https://doi.org/10.1016/j.physe.2015.09.002} {\bibfield
  {journal} {\bibinfo  {journal} {Phys. E (Amsterdam, Neth.)}\ }\textbf {\bibinfo {volume} {75}},\ \bibinfo {pages} {358}
  (\bibinfo {year} {2016})}\BibitemShut {NoStop}%
\bibitem [{\citenamefont
  {Moskalets}(2018{\natexlab{a}})}]{moskalets18_high_temper_fusion_multiel_levit}%
  \BibitemOpen
  \bibfield  {author} {\bibinfo {author} {\bibfnamefont {M.}~\bibnamefont
  {Moskalets}},\ }\href {https://doi.org/10.1103/physrevb.97.155411} {\bibfield
   {journal} {\bibinfo  {journal} {Phys. Rev. B}\ }\textbf {\bibinfo
  {volume} {97}},\ \bibinfo {pages} {155411} (\bibinfo {year}
  {2018}{\natexlab{a}})}\BibitemShut {NoStop}%
\bibitem [{\citenamefont {Glattli}\ and\ \citenamefont
  {Roulleau}(2016{\natexlab{b}})}]{glattli16_hanbur_brown_twiss_noise_correl}%
  \BibitemOpen
  \bibfield  {author} {\bibinfo {author} {\bibfnamefont {D.}~\bibnamefont
  {Glattli}}\ and\ \bibinfo {author} {\bibfnamefont {P.}~\bibnamefont
  {Roulleau}},\ }\href {https://doi.org/10.1016/j.physe.2015.10.034} {\bibfield
   {journal} {\bibinfo  {journal} {Phys. E (Amsterdam, Neth.)}\ }\textbf {\bibinfo {volume} {76}},\ \bibinfo {pages} {216}
  (\bibinfo {year} {2016}{\natexlab{b}})}\BibitemShut {NoStop}%
\bibitem [{\citenamefont {Bocquillon}\ \emph {et~al.}(2012)\citenamefont
  {Bocquillon}, \citenamefont {Parmentier}, \citenamefont {Grenier},
  \citenamefont {Berroir}, \citenamefont {Degiovanni}, \citenamefont {Glattli},
  \citenamefont {Pla{\c{c}}ais}, \citenamefont {Cavanna}, \citenamefont {Jin},\
  and\ \citenamefont {F{\`e}ve}}]{bocquillon12_elect_quant_optic}%
  \BibitemOpen
  \bibfield  {author} {\bibinfo {author} {\bibfnamefont {E.}~\bibnamefont
  {Bocquillon}}, \bibinfo {author} {\bibfnamefont {F.~D.}\ \bibnamefont
  {Parmentier}}, \bibinfo {author} {\bibfnamefont {C.}~\bibnamefont {Grenier}},
  \bibinfo {author} {\bibfnamefont {J.-M.}\ \bibnamefont {Berroir}}, \bibinfo
  {author} {\bibfnamefont {P.}~\bibnamefont {Degiovanni}}, \bibinfo {author}
  {\bibfnamefont {D.~C.}\ \bibnamefont {Glattli}}, \bibinfo {author}
  {\bibfnamefont {B.}~\bibnamefont {Pla{\c{c}}ais}}, \bibinfo {author}
  {\bibfnamefont {A.}~\bibnamefont {Cavanna}}, \bibinfo {author} {\bibfnamefont
  {Y.}~\bibnamefont {Jin}},\ and\ \bibinfo {author} {\bibfnamefont
  {G.}~\bibnamefont {F{\`e}ve}},\ }\href
  {https://doi.org/10.1103/physrevlett.108.196803} {\bibfield  {journal}
  {\bibinfo  {journal} {Phys. Rev. Lett.}\ }\textbf {\bibinfo {volume}
  {108}},\ \bibinfo {pages} {196803} (\bibinfo {year} {2012})}\BibitemShut
  {NoStop}%
\bibitem [{\citenamefont {Dubois}\ \emph
  {et~al.}(2013{\natexlab{b}})\citenamefont {Dubois}, \citenamefont {Jullien},
  \citenamefont {Grenier}, \citenamefont {Degiovanni}, \citenamefont
  {Roulleau},\ and\ \citenamefont {Glattli}}]{dubois-2013-integ-fract}%
  \BibitemOpen
  \bibfield  {author} {\bibinfo {author} {\bibfnamefont {J.}~\bibnamefont
  {Dubois}}, \bibinfo {author} {\bibfnamefont {T.}~\bibnamefont {Jullien}},
  \bibinfo {author} {\bibfnamefont {C.}~\bibnamefont {Grenier}}, \bibinfo
  {author} {\bibfnamefont {P.}~\bibnamefont {Degiovanni}}, \bibinfo {author}
  {\bibfnamefont {P.}~\bibnamefont {Roulleau}},\ and\ \bibinfo {author}
  {\bibfnamefont {D.~C.}\ \bibnamefont {Glattli}},\ }\href
  {https://doi.org/10.1103/physrevb.88.085301} {\bibfield  {journal} {\bibinfo
  {journal} {Phys. Rev. B}\ }\textbf {\bibinfo {volume} {88}},\ \bibinfo
  {pages} {085301} (\bibinfo {year} {2013}{\natexlab{b}})}\BibitemShut
  {NoStop}%
\bibitem [{\citenamefont
  {Brandes}(2008)}]{brandes08_waitin_times_noise_singl_partic_trans}%
  \BibitemOpen
  \bibfield  {author} {\bibinfo {author} {\bibfnamefont {T.}~\bibnamefont
  {Brandes}},\ }\href {https://doi.org/10.1002/andp.200810306} {\bibfield
  {journal} {\bibinfo  {journal} {Ann. Phys. (Berlin, Ger.)}\ }\textbf {\bibinfo
  {volume} {17}},\ \bibinfo {pages} {477} (\bibinfo {year} {2008})}\BibitemShut
  {NoStop}%
\bibitem [{\citenamefont {Albert}\ \emph {et~al.}(2012)\citenamefont {Albert},
  \citenamefont {Haack}, \citenamefont {Flindt},\ and\ \citenamefont
  {B{\"u}ttiker}}]{albert-2012-elect-waitin}%
  \BibitemOpen
  \bibfield  {author} {\bibinfo {author} {\bibfnamefont {M.}~\bibnamefont
  {Albert}}, \bibinfo {author} {\bibfnamefont {G.}~\bibnamefont {Haack}},
  \bibinfo {author} {\bibfnamefont {C.}~\bibnamefont {Flindt}},\ and\ \bibinfo
  {author} {\bibfnamefont {M.}~\bibnamefont {B{\"u}ttiker}},\ }\href
  {https://doi.org/10.1103/PhysRevLett.108.186806} {\bibfield  {journal}
  {\bibinfo  {journal} {Phys. Rev. Lett.}\ }\textbf {\bibinfo {volume}
  {108}},\ \bibinfo {pages} {186806} (\bibinfo {year} {2012})}\BibitemShut
  {NoStop}%
\bibitem [{\citenamefont {Haack}\ \emph {et~al.}(2014)\citenamefont {Haack},
  \citenamefont {Albert},\ and\ \citenamefont
  {Flindt}}]{haack14_distr_elect_waitin_times_quant_coher_conduc}%
  \BibitemOpen
  \bibfield  {author} {\bibinfo {author} {\bibfnamefont {G.}~\bibnamefont
  {Haack}}, \bibinfo {author} {\bibfnamefont {M.}~\bibnamefont {Albert}},\ and\
  \bibinfo {author} {\bibfnamefont {C.}~\bibnamefont {Flindt}},\ }\href
  {https://doi.org/10.1103/PhysRevB.90.205429} {\bibfield  {journal} {\bibinfo
  {journal} {Phys. Rev. B}\ }\textbf {\bibinfo {volume} {90}},\ \bibinfo
  {pages} {205429} (\bibinfo {year} {2014})}\BibitemShut {NoStop}%
\bibitem [{\citenamefont {Albert}\ and\ \citenamefont
  {Devillard}(2014)}]{albert14_waitin_time_distr_train_quant_elect_pulses}%
  \BibitemOpen
  \bibfield  {author} {\bibinfo {author} {\bibfnamefont {M.}~\bibnamefont
  {Albert}}\ and\ \bibinfo {author} {\bibfnamefont {P.}~\bibnamefont
  {Devillard}},\ }\href {https://doi.org/10.1103/PhysRevB.90.035431} {\bibfield
   {journal} {\bibinfo  {journal} {Phys. Rev. B}\ }\textbf {\bibinfo
  {volume} {90}},\ \bibinfo {pages} {035431} (\bibinfo {year}
  {2014})}\BibitemShut {NoStop}%
\bibitem [{\citenamefont {Dasenbrook}\ \emph {et~al.}(2014)\citenamefont
  {Dasenbrook}, \citenamefont {Flindt},\ and\ \citenamefont
  {B{\"u}ttiker}}]{dasenbrook14_floquet_theor_elect_waitin_times}%
  \BibitemOpen
  \bibfield  {author} {\bibinfo {author} {\bibfnamefont {D.}~\bibnamefont
  {Dasenbrook}}, \bibinfo {author} {\bibfnamefont {C.}~\bibnamefont {Flindt}},\
  and\ \bibinfo {author} {\bibfnamefont {M.}~\bibnamefont {B{\"u}ttiker}},\
  }\href {https://doi.org/10.1103/physrevlett.112.146801} {\bibfield  {journal}
  {\bibinfo  {journal} {Phys. Rev. Lett.}\ }\textbf {\bibinfo {volume}
  {112}},\ \bibinfo {pages} {146801} (\bibinfo {year} {2014})}\BibitemShut
  {NoStop}%
\bibitem [{\citenamefont {Dasenbrook}\ \emph {et~al.}(2015)\citenamefont
  {Dasenbrook}, \citenamefont {Hofer},\ and\ \citenamefont
  {Flindt}}]{dasenbrook15_elect_waitin_times_coher_conduc_are_correl}%
  \BibitemOpen
  \bibfield  {author} {\bibinfo {author} {\bibfnamefont {D.}~\bibnamefont
  {Dasenbrook}}, \bibinfo {author} {\bibfnamefont {P.~P.}\ \bibnamefont
  {Hofer}},\ and\ \bibinfo {author} {\bibfnamefont {C.}~\bibnamefont
  {Flindt}},\ }\href {https://doi.org/10.1103/PhysRevB.91.195420} {\bibfield
  {journal} {\bibinfo  {journal} {Phys. Rev. B}\ }\textbf {\bibinfo
  {volume} {91}},\ \bibinfo {pages} {195420} (\bibinfo {year}
  {2015})}\BibitemShut {NoStop}%
\bibitem [{\citenamefont {Haack}\ \emph {et~al.}(2012)\citenamefont {Haack},
  \citenamefont {Moskalets},\ and\ \citenamefont
  {B{\"u}ttiker}}]{haack12_glaub_coher_singl_elect_sourc}%
  \BibitemOpen
  \bibfield  {author} {\bibinfo {author} {\bibfnamefont {G.}~\bibnamefont
  {Haack}}, \bibinfo {author} {\bibfnamefont {M.}~\bibnamefont {Moskalets}},\
  and\ \bibinfo {author} {\bibfnamefont {M.}~\bibnamefont {B{\"u}ttiker}},\
  }\href {https://doi.org/10.1103/PhysRevB.87.201302} {\bibfield  {journal}
  {\bibinfo  {journal} {Phys. Rev. B}\ }\textbf {\bibinfo {volume} {87}},\
  \bibinfo {pages} {201302(R)} (\bibinfo {year} {2013})}\BibitemShut {NoStop}%
\bibitem [{\citenamefont
  {Glauber}(1963{\natexlab{a}})}]{glauber63_coher_incoh_states_radiat_field}%
  \BibitemOpen
  \bibfield  {author} {\bibinfo {author} {\bibfnamefont {R.~J.}\ \bibnamefont
  {Glauber}},\ }\href {https://doi.org/10.1103/PhysRev.131.2766} {\bibfield
  {journal} {\bibinfo  {journal} {Phys. Rev.}\ }\textbf {\bibinfo {volume}
  {131}},\ \bibinfo {pages} {2766} (\bibinfo {year}
  {1963}{\natexlab{a}})}\BibitemShut {NoStop}%
\bibitem [{\citenamefont
  {Glauber}(1963{\natexlab{b}})}]{glauber63_photon_correl}%
  \BibitemOpen
  \bibfield  {author} {\bibinfo {author} {\bibfnamefont {R.~J.}\ \bibnamefont
  {Glauber}},\ }\href {https://doi.org/10.1103/PhysRevLett.10.84} {\bibfield
  {journal} {\bibinfo  {journal} {Phys. Rev. Lett.}\ }\textbf {\bibinfo
  {volume} {10}},\ \bibinfo {pages} {84} (\bibinfo {year}
  {1963}{\natexlab{b}})}\BibitemShut {NoStop}%
\bibitem [{\citenamefont {Kelley}\ and\ \citenamefont
  {Kleiner}(1964)}]{kelley64_theor_elect_field_measur_photoel_count}%
  \BibitemOpen
  \bibfield  {author} {\bibinfo {author} {\bibfnamefont {P.~L.}\ \bibnamefont
  {Kelley}}\ and\ \bibinfo {author} {\bibfnamefont {W.~H.}\ \bibnamefont
  {Kleiner}},\ }\href {https://doi.org/10.1103/physrev.136.a316} {\bibfield
  {journal} {\bibinfo  {journal} {Phys. Rev.}\ }\textbf {\bibinfo {volume}
  {136}},\ \bibinfo {pages} {A316} (\bibinfo {year} {1964})}\BibitemShut
  {NoStop}%
\bibitem [{\citenamefont {Cahill}\ and\ \citenamefont
  {Glauber}(1999)}]{cahill99_densit_operat_fermion}%
  \BibitemOpen
  \bibfield  {author} {\bibinfo {author} {\bibfnamefont {K.~E.}\ \bibnamefont
  {Cahill}}\ and\ \bibinfo {author} {\bibfnamefont {R.~J.}\ \bibnamefont
  {Glauber}},\ }\href {https://doi.org/10.1103/PhysRevA.59.1538} {\bibfield
  {journal} {\bibinfo  {journal} {Phys. Rev. A}\ }\textbf {\bibinfo {volume}
  {59}},\ \bibinfo {pages} {1538} (\bibinfo {year} {1999})}\BibitemShut
  {NoStop}%
\bibitem [{\citenamefont
  {Glauber}(2006{\natexlab{a}})}]{glauber06_quant_theor_optic_coher}%
  \BibitemOpen
  \bibfield  {author} {\bibinfo {author} {\bibfnamefont {R.~J.}\ \bibnamefont
  {Glauber}},\ }\href {https://doi.org/10.1002/9783527610075} {\emph {\bibinfo
  {title} {Quantum Theory of Optical Coherence}}}\ (\bibinfo  {publisher}
  {Wiley},\ \bibinfo {year} {2006})\BibitemShut {NoStop}%
\bibitem [{\citenamefont
  {Glauber}(2006{\natexlab{b}})}]{glauber06_nobel_lectur}%
  \BibitemOpen
  \bibfield  {author} {\bibinfo {author} {\bibfnamefont {R.~J.}\ \bibnamefont
  {Glauber}},\ }\href {https://doi.org/10.1103/RevModPhys.78.1267} {\bibfield
  {journal} {\bibinfo  {journal} {Rev. Mod. Phys.}\ }\textbf {\bibinfo {volume}
  {78}},\ \bibinfo {pages} {1267} (\bibinfo {year}
  {2006}{\natexlab{b}})}\BibitemShut {NoStop}%
\bibitem [{\citenamefont {Mah{\'e}}\ \emph {et~al.}(2010)\citenamefont
  {Mah{\'e}}, \citenamefont {Parmentier}, \citenamefont {Bocquillon},
  \citenamefont {Berroir}, \citenamefont {Glattli}, \citenamefont {Kontos},
  \citenamefont {Pla{\c{c}}ais}, \citenamefont {F{\`e}ve}, \citenamefont
  {Cavanna},\ and\ \citenamefont
  {Jin}}]{mahe10_curren_correl_deman_elect_sourc}%
  \BibitemOpen
  \bibfield  {author} {\bibinfo {author} {\bibfnamefont {A.}~\bibnamefont
  {Mah{\'e}}}, \bibinfo {author} {\bibfnamefont {F.~D.}\ \bibnamefont
  {Parmentier}}, \bibinfo {author} {\bibfnamefont {E.}~\bibnamefont
  {Bocquillon}}, \bibinfo {author} {\bibfnamefont {J.~M.}\ \bibnamefont
  {Berroir}}, \bibinfo {author} {\bibfnamefont {D.~C.}\ \bibnamefont
  {Glattli}}, \bibinfo {author} {\bibfnamefont {T.}~\bibnamefont {Kontos}},
  \bibinfo {author} {\bibfnamefont {B.}~\bibnamefont {Pla{\c{c}}ais}}, \bibinfo
  {author} {\bibfnamefont {G.}~\bibnamefont {F{\`e}ve}}, \bibinfo {author}
  {\bibfnamefont {A.}~\bibnamefont {Cavanna}},\ and\ \bibinfo {author}
  {\bibfnamefont {Y.}~\bibnamefont {Jin}},\ }\href
  {https://doi.org/10.1103/PhysRevB.82.201309} {\bibfield  {journal} {\bibinfo
  {journal} {Phys. Rev. B}\ }\textbf {\bibinfo {volume} {82}},\ \bibinfo
  {pages} {201309(R)} (\bibinfo {year} {2010})}\BibitemShut {NoStop}%
\bibitem [{\citenamefont {Gustavsson}\ \emph {et~al.}(2009)\citenamefont
  {Gustavsson}, \citenamefont {Leturcq}, \citenamefont {Studer}, \citenamefont
  {Shorubalko}, \citenamefont {Ihn}, \citenamefont {Ensslin}, \citenamefont
  {Driscoll},\ and\ \citenamefont
  {Gossard}}]{gustavsson09_elect_count_quant_dots}%
  \BibitemOpen
  \bibfield  {author} {\bibinfo {author} {\bibfnamefont {S.}~\bibnamefont
  {Gustavsson}}, \bibinfo {author} {\bibfnamefont {R.}~\bibnamefont {Leturcq}},
  \bibinfo {author} {\bibfnamefont {M.}~\bibnamefont {Studer}}, \bibinfo
  {author} {\bibfnamefont {I.}~\bibnamefont {Shorubalko}}, \bibinfo {author}
  {\bibfnamefont {T.}~\bibnamefont {Ihn}}, \bibinfo {author} {\bibfnamefont
  {K.}~\bibnamefont {Ensslin}}, \bibinfo {author} {\bibfnamefont
  {D.}~\bibnamefont {Driscoll}},\ and\ \bibinfo {author} {\bibfnamefont
  {A.}~\bibnamefont {Gossard}},\ }\href
  {https://doi.org/10.1016/j.surfrep.2009.02.001} {\bibfield  {journal}
  {\bibinfo  {journal} {Surf. Sci. Rep.}\ }\textbf {\bibinfo {volume}
  {64}},\ \bibinfo {pages} {191} (\bibinfo {year} {2009})}\BibitemShut
  {NoStop}%
\bibitem [{\citenamefont {Maisi}\ \emph {et~al.}(2011)\citenamefont {Maisi},
  \citenamefont {Saira}, \citenamefont {Pashkin}, \citenamefont {Tsai},
  \citenamefont {Averin},\ and\ \citenamefont
  {Pekola}}]{maisi11_real_time_obser_discr_andreev_tunnel_event}%
  \BibitemOpen
  \bibfield  {author} {\bibinfo {author} {\bibfnamefont {V.~F.}\ \bibnamefont
  {Maisi}}, \bibinfo {author} {\bibfnamefont {O.~P.}\ \bibnamefont {Saira}},
  \bibinfo {author} {\bibfnamefont {Y.~A.}\ \bibnamefont {Pashkin}}, \bibinfo
  {author} {\bibfnamefont {J.~S.}\ \bibnamefont {Tsai}}, \bibinfo {author}
  {\bibfnamefont {D.~V.}\ \bibnamefont {Averin}},\ and\ \bibinfo {author}
  {\bibfnamefont {J.~P.}\ \bibnamefont {Pekola}},\ }\href@noop {} {\bibfield
  {journal} {\bibinfo  {journal} {Phys. Rev. Lett.}\ }\textbf {\bibinfo
  {volume} {106}},\ \bibinfo {pages} {217003} (\bibinfo {year}
  {2011})}\BibitemShut {NoStop}%
\bibitem [{\citenamefont {Kurzmann}\ \emph {et~al.}(2019)\citenamefont
  {Kurzmann}, \citenamefont {Stegmann}, \citenamefont {Kerski}, \citenamefont
  {Schott}, \citenamefont {Ludwig}, \citenamefont {Wieck}, \citenamefont
  {K{\"o}nig}, \citenamefont {Lorke},\ and\ \citenamefont
  {Geller}}]{kurzmann19_optic_detec_singl_elect_trans_dynam}%
  \BibitemOpen
  \bibfield  {author} {\bibinfo {author} {\bibfnamefont {A.}~\bibnamefont
  {Kurzmann}}, \bibinfo {author} {\bibfnamefont {P.}~\bibnamefont {Stegmann}},
  \bibinfo {author} {\bibfnamefont {J.}~\bibnamefont {Kerski}}, \bibinfo
  {author} {\bibfnamefont {R.}~\bibnamefont {Schott}}, \bibinfo {author}
  {\bibfnamefont {A.}~\bibnamefont {Ludwig}}, \bibinfo {author} {\bibfnamefont
  {A.~D.}\ \bibnamefont {Wieck}}, \bibinfo {author} {\bibfnamefont
  {J.}~\bibnamefont {K{\"o}nig}}, \bibinfo {author} {\bibfnamefont
  {A.}~\bibnamefont {Lorke}},\ and\ \bibinfo {author} {\bibfnamefont
  {M.}~\bibnamefont {Geller}},\ }\href@noop {} {\bibfield  {journal} {\bibinfo
  {journal} {Phys. Rev. Lett.}\ }\textbf {\bibinfo {volume} {122}},\
  \bibinfo {pages} {247403} (\bibinfo {year} {2019})}\BibitemShut {NoStop}%
\bibitem [{\citenamefont {Ranni}\ \emph {et~al.}(2021)\citenamefont {Ranni},
  \citenamefont {Brange}, \citenamefont {Mannila}, \citenamefont {Flindt},\
  and\ \citenamefont {Maisi}}]{ranni21_real_time_obser_cooper_pair}%
  \BibitemOpen
  \bibfield  {author} {\bibinfo {author} {\bibfnamefont {A.}~\bibnamefont
  {Ranni}}, \bibinfo {author} {\bibfnamefont {F.}~\bibnamefont {Brange}},
  \bibinfo {author} {\bibfnamefont {E.~T.}\ \bibnamefont {Mannila}}, \bibinfo
  {author} {\bibfnamefont {C.}~\bibnamefont {Flindt}},\ and\ \bibinfo {author}
  {\bibfnamefont {V.~F.}\ \bibnamefont {Maisi}},\ }\href@noop {} {\bibfield
  {journal} {\bibinfo  {journal} {Nat. Commun.}\ }\textbf {\bibinfo
  {volume} {12}},\ \bibinfo {pages} {6358} (\bibinfo {year}
  {2021})}\BibitemShut {NoStop}%
\bibitem [{\citenamefont {Brange}\ \emph {et~al.}(2021)\citenamefont {Brange},
  \citenamefont {Schmidt}, \citenamefont {Bayer}, \citenamefont {Wagner},
  \citenamefont {Flindt},\ and\ \citenamefont
  {Haug}}]{brange21_contr_emiss_time_statis_dynam}%
  \BibitemOpen
  \bibfield  {author} {\bibinfo {author} {\bibfnamefont {F.}~\bibnamefont
  {Brange}}, \bibinfo {author} {\bibfnamefont {A.}~\bibnamefont {Schmidt}},
  \bibinfo {author} {\bibfnamefont {J.~C.}\ \bibnamefont {Bayer}}, \bibinfo
  {author} {\bibfnamefont {T.}~\bibnamefont {Wagner}}, \bibinfo {author}
  {\bibfnamefont {C.}~\bibnamefont {Flindt}},\ and\ \bibinfo {author}
  {\bibfnamefont {R.~J.}\ \bibnamefont {Haug}},\ }\href
  {https://doi.org/10.1126/sciadv.abe0793} {\bibfield  {journal} {\bibinfo
  {journal} {Sci. Adv.}\ }\textbf {\bibinfo {volume} {7}},\ \bibinfo
  {pages} {793} (\bibinfo {year} {2021})}\BibitemShut {NoStop}%
\bibitem [{\citenamefont {Mandel}\ and\ \citenamefont
  {Wolf}(1965)}]{mandel65_coher_proper_optic_field}%
  \BibitemOpen
  \bibfield  {author} {\bibinfo {author} {\bibfnamefont {L.}~\bibnamefont
  {Mandel}}\ and\ \bibinfo {author} {\bibfnamefont {E.}~\bibnamefont {Wolf}},\
  }\href {https://doi.org/10.1103/revmodphys.37.231} {\bibfield  {journal}
  {\bibinfo  {journal} {Rev. Mod. Phys.}\ }\textbf {\bibinfo {volume}
  {37}},\ \bibinfo {pages} {231} (\bibinfo {year} {1965})}\BibitemShut
  {NoStop}%
\bibitem [{\citenamefont
  {Macchi}(1975)}]{macchi75_coinc_approac_to_stoch_point_proces}%
  \BibitemOpen
  \bibfield  {author} {\bibinfo {author} {\bibfnamefont {O.}~\bibnamefont
  {Macchi}},\ }\href {https://doi.org/10.1017/s0001867800040313} {\bibfield
  {journal} {\bibinfo  {journal} {Adv. Appl. Probab.}\ }\textbf
  {\bibinfo {volume} {7}},\ \bibinfo {pages} {83} (\bibinfo {year}
  {1975})}\BibitemShut {NoStop}%
\bibitem [{\citenamefont
  {Iwankiewicz}(1995)}]{iwankiewicz95_dynam_mechan_system_under_random_impul}%
  \BibitemOpen
  \bibfield  {author} {\bibinfo {author} {\bibfnamefont {R.}~\bibnamefont
  {Iwankiewicz}},\ }\href
  {https://www.ebook.de/de/product/26321209/radoslaw_iwankiewicz_dynamical_mechanical_systems_under_random_impulses.html}
  {\emph {\bibinfo {title} {Dynamical Mechanical Systems Under Random
  Impulses}}}\ (\bibinfo  {publisher} {World Scientific Publishing London},\
  \bibinfo {year} {1995})\BibitemShut {NoStop}%
\bibitem [{\citenamefont {Levitov}\ \emph {et~al.}(1996)\citenamefont
  {Levitov}, \citenamefont {Lee},\ and\ \citenamefont
  {Lesovik}}]{levitov96_elect_count_statis_coher_states_elect_curren}%
  \BibitemOpen
  \bibfield  {author} {\bibinfo {author} {\bibfnamefont {L.~S.}\ \bibnamefont
  {Levitov}}, \bibinfo {author} {\bibfnamefont {H.}~\bibnamefont {Lee}},\ and\
  \bibinfo {author} {\bibfnamefont {G.~B.}\ \bibnamefont {Lesovik}},\ }\href
  {https://doi.org/10.1063/1.531672} {\bibfield  {journal} {\bibinfo  {journal}
  {J. Math. Phys. (Melville, NY, U. S.)}\ }\textbf {\bibinfo {volume} {37}},\
  \bibinfo {pages} {4845} (\bibinfo {year} {1996})}\BibitemShut {NoStop}%
\bibitem [{\citenamefont {D.~J.~Daley}(2003)}]{D.J.Daley2003}%
  \BibitemOpen
  \bibfield  {author} {\bibinfo {author} {\bibfnamefont {D.~V.-J.}\
  \bibnamefont {D.~J.~Daley}},\ }\href
  {https://www.ebook.de/de/product/3671300/d_j_daley_d_vere_jones_an_introduction_to_the_theory_of_point_processes.html}
  {\emph {\bibinfo {title} {An Introduction to the Theory of Point
  Processes}}}\ (\bibinfo  {publisher} {Springer New York},\ \bibinfo {year}
  {2003})\BibitemShut {NoStop}%
\bibitem [{\citenamefont {Snyder}\ and\ \citenamefont
  {Miller}(1991)}]{snyder91_random_point_proces_time_space}%
  \BibitemOpen
  \bibfield  {author} {\bibinfo {author} {\bibfnamefont {D.~L.}\ \bibnamefont
  {Snyder}}\ and\ \bibinfo {author} {\bibfnamefont {M.~I.}\ \bibnamefont
  {Miller}},\ }\href {https://doi.org/10.1007/978-1-4612-3166-0} {\emph
  {\bibinfo {title} {Random Point Processes in Time and Space}}}\ (\bibinfo
  {publisher} {Springer New York},\ \bibinfo {year} {1991})\BibitemShut
  {NoStop}%
\bibitem [{\citenamefont {Beenakker}\ \emph {et~al.}(2005)\citenamefont
  {Beenakker}, \citenamefont {Titov},\ and\ \citenamefont
  {Trauzettel}}]{beenakker05_optim_spin_entan_elect_hole_pair_pump}%
  \BibitemOpen
  \bibfield  {author} {\bibinfo {author} {\bibfnamefont {C.~W.~J.}\
  \bibnamefont {Beenakker}}, \bibinfo {author} {\bibfnamefont {M.}~\bibnamefont
  {Titov}},\ and\ \bibinfo {author} {\bibfnamefont {B.}~\bibnamefont
  {Trauzettel}},\ }\href {https://doi.org/10.1103/physrevlett.94.186804}
  {\bibfield  {journal} {\bibinfo  {journal} {Phys. Rev. Lett.}\
  }\textbf {\bibinfo {volume} {94}},\ \bibinfo {pages} {186804} (\bibinfo
  {year} {2005})}\BibitemShut {NoStop}%
\bibitem [{\citenamefont {Cheong}\ and\ \citenamefont
  {Henley}(2004)}]{cheong04_many_body_densit_matric_free_fermion}%
  \BibitemOpen
  \bibfield  {author} {\bibinfo {author} {\bibfnamefont {S.-A.}\ \bibnamefont
  {Cheong}}\ and\ \bibinfo {author} {\bibfnamefont {C.~L.}\ \bibnamefont
  {Henley}},\ }\href {https://doi.org/10.1103/PhysRevB.69.075111} {\bibfield
  {journal} {\bibinfo  {journal} {Phys. Rev. B}\ }\textbf {\bibinfo {volume}
  {69}},\ \bibinfo {pages} {075111} (\bibinfo {year} {2004})}\BibitemShut
  {NoStop}%
\bibitem [{\citenamefont {Corney}\ and\ \citenamefont
  {Drummond}(2006)}]{corney06_gauss_phase_space_repres_fermion}%
  \BibitemOpen
  \bibfield  {author} {\bibinfo {author} {\bibfnamefont {J.~F.}\ \bibnamefont
  {Corney}}\ and\ \bibinfo {author} {\bibfnamefont {P.~D.}\ \bibnamefont
  {Drummond}},\ }\href {https://doi.org/10.1103/PhysRevB.73.125112} {\bibfield
  {journal} {\bibinfo  {journal} {Phys. Rev. B}\ }\textbf {\bibinfo {volume}
  {73}},\ \bibinfo {pages} {125112} (\bibinfo {year} {2006})}\BibitemShut
  {NoStop}%
\bibitem [{\citenamefont {Yin}(2019)}]{yin-2019-quasip-states}%
  \BibitemOpen
  \bibfield  {author} {\bibinfo {author} {\bibfnamefont {Y.}~\bibnamefont
  {Yin}},\ }\href {https://doi.org/10.1088/1361-648x/ab0fc4} {\bibfield
  {journal} {\bibinfo  {journal} {J. Phys.: Condens. Matter}\
  }\textbf {\bibinfo {volume} {31}},\ \bibinfo {pages} {245301} (\bibinfo
  {year} {2019})}\BibitemShut {NoStop}%
\bibitem [{\citenamefont {Yue}\ and\ \citenamefont
  {Yin}(2021)}]{yue21_quasip_states_integ_fract_charg}%
  \BibitemOpen
  \bibfield  {author} {\bibinfo {author} {\bibfnamefont {X.~K.}\ \bibnamefont
  {Yue}}\ and\ \bibinfo {author} {\bibfnamefont {Y.}~\bibnamefont {Yin}},\
  }\href@noop {} {\bibfield  {journal} {\bibinfo  {journal} {Phys. Rev. B}\ }\textbf {\bibinfo {volume} {103}},\ \bibinfo {pages} {245429} (\bibinfo
  {year} {2021})}\BibitemShut {NoStop}%
\bibitem [{Note1()}]{Note1}%
  \BibitemOpen
  \bibinfo {note} {Alternatively, it can also be obtained from Eq.~\protect
  \textup {\hbox {\mathsurround \z@ \protect \normalfont (\ignorespaces \ref
  {s2:eq20}\unskip \@@italiccorr )}}, which can be calculated more easily in
  this case.}\BibitemShut {Stop}%
\bibitem [{\citenamefont
  {Moskalets}(2018{\natexlab{b}})}]{moskalets18_singl_elect_secon_order_correl}%
  \BibitemOpen
  \bibfield  {author} {\bibinfo {author} {\bibfnamefont {M.}~\bibnamefont
  {Moskalets}},\ }\href {https://doi.org/10.1103/physrevb.98.115421} {\bibfield
   {journal} {\bibinfo  {journal} {Phys. Rev. B}\ }\textbf {\bibinfo
  {volume} {98}},\ \bibinfo {pages} {115421} (\bibinfo {year}
  {2018}{\natexlab{b}})}\BibitemShut {NoStop}%
\bibitem [{\citenamefont {Kashcheyevs}\ and\ \citenamefont
  {Samuelsson}(2017)}]{kashcheyevs17_class_to_quant_cross_elect_deman_emiss}%
  \BibitemOpen
  \bibfield  {author} {\bibinfo {author} {\bibfnamefont {V.}~\bibnamefont
  {Kashcheyevs}}\ and\ \bibinfo {author} {\bibfnamefont {P.}~\bibnamefont
  {Samuelsson}},\ }\href {https://doi.org/10.1103/physrevb.95.245424}
  {\bibfield  {journal} {\bibinfo  {journal} {Phys. Rev. B}\ }\textbf
  {\bibinfo {volume} {95}},\ \bibinfo {pages} {245424} (\bibinfo {year}
  {2017})}\BibitemShut {NoStop}%
\bibitem [{\citenamefont {F{\`e}ve}\ \emph {et~al.}(2007)\citenamefont
  {F{\`e}ve}, \citenamefont {Mah{\'e}}, \citenamefont {Berroir}, \citenamefont
  {Kontos}, \citenamefont {Pla{\c c}ais}, \citenamefont {Glattli},
  \citenamefont {Cavanna}, \citenamefont {Etienne},\ and\ \citenamefont
  {Jin}}]{feve-2007-deman-coher}%
  \BibitemOpen
  \bibfield  {author} {\bibinfo {author} {\bibfnamefont {G.}~\bibnamefont
  {F{\`e}ve}}, \bibinfo {author} {\bibfnamefont {A.}~\bibnamefont {Mah{\'e}}},
  \bibinfo {author} {\bibfnamefont {J.~M.}\ \bibnamefont {Berroir}}, \bibinfo
  {author} {\bibfnamefont {T.}~\bibnamefont {Kontos}}, \bibinfo {author}
  {\bibfnamefont {B.}~\bibnamefont {Pla{\c c}ais}}, \bibinfo {author}
  {\bibfnamefont {D.~C.}\ \bibnamefont {Glattli}}, \bibinfo {author}
  {\bibfnamefont {A.}~\bibnamefont {Cavanna}}, \bibinfo {author} {\bibfnamefont
  {B.}~\bibnamefont {Etienne}},\ and\ \bibinfo {author} {\bibfnamefont
  {Y.}~\bibnamefont {Jin}},\ }\href
  {http://science.sciencemag.org/content/316/5828/1169.abstract} {\bibfield
  {journal} {\bibinfo  {journal} {Science}\ }\textbf {\bibinfo {volume}
  {316}},\ \bibinfo {pages} {1169} (\bibinfo {year} {2007})}\BibitemShut
  {NoStop}%
\bibitem [{\citenamefont {Fletcher}\ \emph {et~al.}(2012)\citenamefont
  {Fletcher}, \citenamefont {Kataoka}, \citenamefont {Howe}, \citenamefont
  {Pepper}, \citenamefont {See}, \citenamefont {Giblin}, \citenamefont
  {Griffiths}, \citenamefont {Jones}, \citenamefont {Farrer}, \citenamefont
  {Ritchie},\ and\ \citenamefont
  {Janssen}}]{fletcher12_clock_contr_emiss_singl_elect}%
  \BibitemOpen
  \bibfield  {author} {\bibinfo {author} {\bibfnamefont {J.~D.}\ \bibnamefont
  {Fletcher}}, \bibinfo {author} {\bibfnamefont {M.}~\bibnamefont {Kataoka}},
  \bibinfo {author} {\bibfnamefont {H.}~\bibnamefont {Howe}}, \bibinfo {author}
  {\bibfnamefont {M.}~\bibnamefont {Pepper}}, \bibinfo {author} {\bibfnamefont
  {P.}~\bibnamefont {See}}, \bibinfo {author} {\bibfnamefont {S.~P.}\
  \bibnamefont {Giblin}}, \bibinfo {author} {\bibfnamefont {J.~P.}\
  \bibnamefont {Griffiths}}, \bibinfo {author} {\bibfnamefont {G.~A.~C.}\
  \bibnamefont {Jones}}, \bibinfo {author} {\bibfnamefont {I.}~\bibnamefont
  {Farrer}}, \bibinfo {author} {\bibfnamefont {D.~A.}\ \bibnamefont
  {Ritchie}},\ and\ \bibinfo {author} {\bibfnamefont {T.~J. B.~M.}\
  \bibnamefont {Janssen}},\ }\href
  {https://doi.org/10.1103/PhysRevLett.111.216807} {\bibfield  {journal}
  {\bibinfo  {journal} {Phys. Rev. Lett.}\ }\textbf {\bibinfo {volume}
  {111}},\ \bibinfo {pages} {216807} (\bibinfo {year} {2013})}\BibitemShut
  {NoStop}%
\bibitem [{\citenamefont {Ubbelohde}\ \emph {et~al.}(2014)\citenamefont
  {Ubbelohde}, \citenamefont {Hohls}, \citenamefont {Kashcheyevs},
  \citenamefont {Wagner}, \citenamefont {Fricke}, \citenamefont {K{\"a}stner},
  \citenamefont {Pierz}, \citenamefont {Schumacher},\ and\ \citenamefont
  {Haug}}]{ubbelohde14_partit_deman_elect_pairs}%
  \BibitemOpen
  \bibfield  {author} {\bibinfo {author} {\bibfnamefont {N.}~\bibnamefont
  {Ubbelohde}}, \bibinfo {author} {\bibfnamefont {F.}~\bibnamefont {Hohls}},
  \bibinfo {author} {\bibfnamefont {V.}~\bibnamefont {Kashcheyevs}}, \bibinfo
  {author} {\bibfnamefont {T.}~\bibnamefont {Wagner}}, \bibinfo {author}
  {\bibfnamefont {L.}~\bibnamefont {Fricke}}, \bibinfo {author} {\bibfnamefont
  {B.}~\bibnamefont {K{\"a}stner}}, \bibinfo {author} {\bibfnamefont
  {K.}~\bibnamefont {Pierz}}, \bibinfo {author} {\bibfnamefont {H.~W.}\
  \bibnamefont {Schumacher}},\ and\ \bibinfo {author} {\bibfnamefont {R.~J.}\
  \bibnamefont {Haug}},\ }\href {https://doi.org/10.1038/nnano.2014.275}
  {\bibfield  {journal} {\bibinfo  {journal} {Nat. Nanotechnol.}\ }\textbf
  {\bibinfo {volume} {10}},\ \bibinfo {pages} {46} (\bibinfo {year}
  {2014})}\BibitemShut {NoStop}%
\bibitem [{\citenamefont {Wigner}(1932)}]{wigner32_quant_correc_therm_equil}%
  \BibitemOpen
  \bibfield  {author} {\bibinfo {author} {\bibfnamefont {E.}~\bibnamefont
  {Wigner}},\ }\href {https://doi.org/10.1103/PhysRev.40.749} {\bibfield
  {journal} {\bibinfo  {journal} {Phys. Rev.}\ }\textbf {\bibinfo {volume}
  {40}},\ \bibinfo {pages} {749} (\bibinfo {year} {1932})}\BibitemShut
  {NoStop}%
\bibitem [{\citenamefont {Ferraro}\ \emph {et~al.}(2013)\citenamefont
  {Ferraro}, \citenamefont {Feller}, \citenamefont {Ghibaudo}, \citenamefont
  {Thibierge}, \citenamefont {Bocquillon}, \citenamefont {F{\`e}ve},
  \citenamefont {Grenier},\ and\ \citenamefont
  {Degiovanni}}]{ferraro-2013-wigner-funct}%
  \BibitemOpen
  \bibfield  {author} {\bibinfo {author} {\bibfnamefont {D.}~\bibnamefont
  {Ferraro}}, \bibinfo {author} {\bibfnamefont {A.}~\bibnamefont {Feller}},
  \bibinfo {author} {\bibfnamefont {A.}~\bibnamefont {Ghibaudo}}, \bibinfo
  {author} {\bibfnamefont {E.}~\bibnamefont {Thibierge}}, \bibinfo {author}
  {\bibfnamefont {E.}~\bibnamefont {Bocquillon}}, \bibinfo {author}
  {\bibfnamefont {G.}~\bibnamefont {F{\`e}ve}}, \bibinfo {author}
  {\bibfnamefont {C.}~\bibnamefont {Grenier}},\ and\ \bibinfo {author}
  {\bibfnamefont {P.}~\bibnamefont {Degiovanni}},\ }\href
  {https://doi.org/10.1103/physrevb.88.205303} {\bibfield  {journal} {\bibinfo
  {journal} {Phys. Rev. B}\ }\textbf {\bibinfo {volume} {88}},\ \bibinfo
  {pages} {205303} (\bibinfo {year} {2013})}\BibitemShut {NoStop}%
\bibitem [{\citenamefont {Fletcher}\ \emph {et~al.}(2019)\citenamefont
  {Fletcher}, \citenamefont {Johnson}, \citenamefont {Locane}, \citenamefont
  {See}, \citenamefont {Griffiths}, \citenamefont {Farrer}, \citenamefont
  {Ritchie}, \citenamefont {Brouwer}, \citenamefont {Kashcheyevs},\ and\
  \citenamefont {Kataoka}}]{fletcher19_contin_variab_tomog_solit_elect}%
  \BibitemOpen
  \bibfield  {author} {\bibinfo {author} {\bibfnamefont {J.~D.}\ \bibnamefont
  {Fletcher}}, \bibinfo {author} {\bibfnamefont {N.}~\bibnamefont {Johnson}},
  \bibinfo {author} {\bibfnamefont {E.}~\bibnamefont {Locane}}, \bibinfo
  {author} {\bibfnamefont {P.}~\bibnamefont {See}}, \bibinfo {author}
  {\bibfnamefont {J.~P.}\ \bibnamefont {Griffiths}}, \bibinfo {author}
  {\bibfnamefont {I.}~\bibnamefont {Farrer}}, \bibinfo {author} {\bibfnamefont
  {D.~A.}\ \bibnamefont {Ritchie}}, \bibinfo {author} {\bibfnamefont {P.~W.}\
  \bibnamefont {Brouwer}}, \bibinfo {author} {\bibfnamefont {V.}~\bibnamefont
  {Kashcheyevs}},\ and\ \bibinfo {author} {\bibfnamefont {M.}~\bibnamefont
  {Kataoka}},\ }\href {https://doi.org/10.1038/s41467-019-13222-1} {\bibfield
  {journal} {\bibinfo  {journal} {Nat. Commun.}\ }\textbf {\bibinfo
  {volume} {10}},\ \bibinfo {pages} {5298} (\bibinfo {year}
  {2019})}\BibitemShut {NoStop}%
\end{thebibliography}

\end{document}